\documentclass{article}
\usepackage{preambles}

\begin{document}
%========================
% Title Page
%========================
\title{
	Complex Saddles of Truncated String Amplitudes
}
\author{
	\Large
	Takuya Yoda$^{1}$ \footnote{t.yoda(at)gauge.scphys.kyoto-u.ac.jp}%, \quad
	%Coauthor Name$^{2}$ \footnote{coauthor(at)e-mail}
	\vspace{1em} \\
	{$^{1}$ \small{\it Department of Physics, Kyoto University, Kyoto 606-8502, Japan }} \\
	%{$^{2}$ \small{\it Department of Physics, University, Address Postal-code, Country }}
}
\date{\small{ \today }}
\maketitle
\thispagestyle{empty}

% Abstract
\begin{abstract}
	String scattering amplitudes in the high energy asymptotic region have been studied by saddle point approximation.
	Recently, it was pointed out that infinitely many complex saddles contribute to string amplitudes even at tree-level after redefining the original formal integration contour so that the new contour analytically continues amplitudes appropriately.
	It is a challenging problem to identify which saddles contribute to higher genus corrections of string amplitudes.
	In this paper, we construct QFT toy models which have the same infinite mass tower as string amplitudes but ignoring degeneracies.
	Their higher loop Feynman diagrams are evaluated by identifying their contributing complex saddles.
	We find that the saddles associated with infinitely many stringy excitations provide highly oscillatory terms to the amplitudes.
	We conjecture that string amplitudes, as functions of momenta, approach multi-fractal functions in the high energy asymptotic regions if higher genus contributions are fully included.
	Their fractal dimensions should be determined purely by the type of string theory and the spacetime dimension where string scatterings occur.
\end{abstract}
\vfill
\noindent
% Preprint number
KUNS-3018

%========================
% Contents Page
%========================
\newpage
\thispagestyle{empty}
\tableofcontents
\thispagestyle{empty}

%================================================================================
%========================
% Main Sections
%========================
\newpage
\renewcommand{\thefootnote}{\arabic{footnote}}
\setcounter{footnote}{0}
\pagenumbering{arabic}
\setcounter{page}{1}

%========================
\section{Introduction}
\label{sec:intro}
%========================

%----------------------------------
\subsection{Amplitude poles/zeros and oscillations}
\begin{figure}[t]
	\centering
	\includegraphics[height=0.4\textwidth]{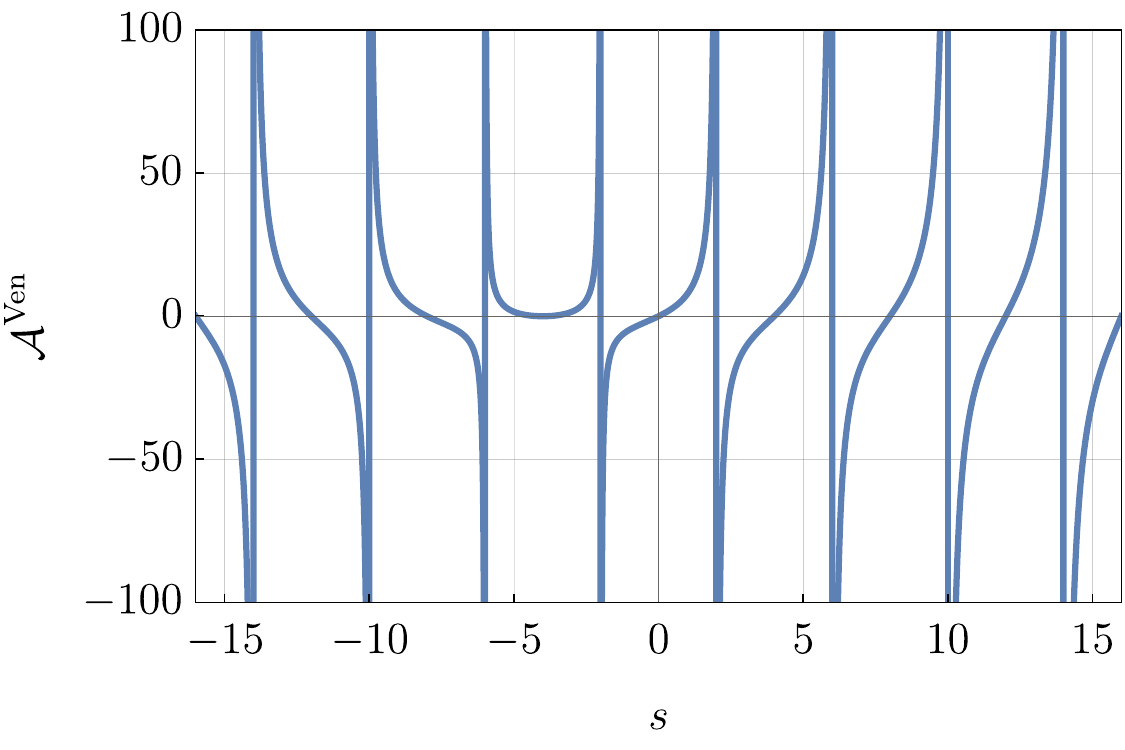}
	\caption{
		Typical behavior of the Veneziano amplitude.
		The plot is when $t=0$.
		A region $s\geq-8$ is a physical region where momentum $\abs{\mathbf{p}}$ is real,
		while the other region $s<-8$ is an unphysical region where momentum $\abs{\mathbf{p}}$ is imaginary.
	}
	\label{fig:erratic_behaviors}
\end{figure}

%----------------------------------
String amplitudes have rich analytic structure of poles and zeros.
For example, the Veneziano Amplitude is expressed as
\begin{align}
	\mathcal{A}^{\text{Ven}}
	&=
	\sqrt{\pi}
	\prod_{x=s,t,u}
	\dfrac{ \Gamma(-\frac{\alpha(x)}{2}) }{ \Gamma(\frac{1+\alpha(x)}{2}) },
	\quad \alpha(x)=x/2 +1%\alpha(x)=\alpha' x +1
\end{align}
by collecting all channel contributions\cite{Freund:1987ck}.
From this expression, we find that there are infinitely many poles and zeros at
\begin{align}
	\begin{array}{lr}
		\text{Poles:} & \frac{\alpha(x)}{2} = +n \\
		\text{Zeros:} & \frac{1+\alpha(x)}{2} = -n
	\end{array}
\end{align}
where $n=0,1,2,\dots$.
These non-negative integers $n$ are associated with stringy excitations.
Since poles or zeros appear periodically with respect to $\alpha(x)$,
its plot shows oscillating patterns in generic parameter regions as Fig.~\ref{fig:erratic_behaviors} shows\footnote{
	Conventions in this paper is
	\begin{align}
		s &= -(p_1+p_2)^2=+4(\abs{\mathbf{p}}^2-2), \notag\\
		t &= -(p_1+p_3)^2=-4\abs{\mathbf{p}}^2\sin^2\theta, \notag\\
		u &= -(p_1+p_4)^2=-4\abs{\mathbf{p}}^2\cos^2\theta.
	\end{align}
}.
It is in clear contrast to typical perturbative QFT amplitudes which do not have infinitely many excitations.
More generally the Koba-Nielsen formula says that tachyon $n$-point amplitude is
\begin{align}
	\mathcal{A}_n^{\text{KN}}
	= g^{n-2}\int
	\dd{\mu_n}
	\prod_{i<j} (y_i-y_j)^{+p_ip_j}.
\end{align}
It shows more erratic behaviors.
%If $n$ is sufficiently small, it is expressed by the hypergeometric functions.
%It is expanded by a combination of gamma functions.
%\mycomm{Check this!}
Furthermore, considering amplitudes with higher stringy excitations,
we find that the amplitudes become more and more erratic.
These raise following questions:
\begin{itemize}
	\item Where are the poles and zeros of string amplitudes, particularly when higher genus corrections are included?
	\item What is the statistics of the location of poles and zeros?
	\item Can we quantify the erraticity of their distributions?
\end{itemize}

These questions are also motivated by black hole physics.
We can regard black hole as an ensemble of highly excited strings
by the black hole-string correspondence \cite{Horowitz:1996nw,Horowitz:1997jc},
by thermal radiation from highly excited strings \cite{Amati:1999fv},
and by series of related works
\cite{Gross:2021gsj,Rosenhaus:2021xhm}.
%\cite{Gross:2021gsj,Rosenhaus:2021xhm,Firrotta:2022cku,Bianchi:2022mhs,Firrotta:2023wem,Bianchi:2023uby,10.1093/ptep/ptad045,Hashimoto:2022bll}.
%In \cite{Hashimoto:2022bll}.
These results motivate us to expect that
scatterings by a black hole should be reproduced by string scattering amplitudes.
Black holes as scatterers have at least two characteristic features:
thermal radiation and time-delay due to redshift near horizon.
Their counterparts in string side, if they exist,
should be statistics of energy of emitted wave packets and their time-delay.
It is essential to evaluate highly erratic string amplitudes.
In this paper, we focus on studying the high energy asymptotic behaviors of stringy amplitudes.
Challenging studies on statistics and time-delay of wave packets are left for future works\footnote{
The author discussed time-delay in tree-level string amplitudes in a previous work \cite{Yoda:2024pie}.
}.

%----------------------------------
\subsection{Toward higher genus}
%----------------------------------
The asymptotic behaviors of string scattering amplitudes have been studied by a series of works using saddle point analysis\cite{Gross:1987kza,Gross:1987ar,Gross:1988ue,Gross:1988sj,Gross:1989ge,Mende:1989wt}.
Each perturbative order contribution was evaluated and Borel resummed.
The result was interpreted as an improvement of the locality of strings while maintaining ultraviolet finiteness \cite{Mende:1989wt}.
One caveat of their analysis was that the original integration contour was a formal one which diverges for typical 
physical kinematic regions.

This problem can be cured by redefining the integration contour\cite{Mimachi2003,Mimachi2004,Witten:2013pra}.
Deforming the new integration contour to a combination of thimbles, it is found that infinitely many complex saddles contribute to string amplitudes even at tree-level\cite{Mizera:2019vvs,Yoda:2024pie}.
All of these complex saddles are essential to reproduce the appropriate location of poles and zeros of string amplitudes.
This method of refining integration contour has been further developed to evaluate more general tree-level amplitudes and one-loop amplitudes\cite{Eberhardt:2024twy,Eberhardt:2023xck}.

%----------------------------------
\subsection{Ideas of truncation}
%----------------------------------
It is worthwhile to develop this method to improve saddle point approximation in\cite{Gross:1987kza,Gross:1987ar,Gross:1988ue,Gross:1988sj,Gross:1989ge,Mende:1989wt}
and to estimate the location of poles and zeros of higher genus amplitudes.
However, it is still a hard task since moduli integrals of string amplitudes are highly non-trivial\footnote{
	See reviews or textbooks e.g.\cite{DHoker:1988pdl,Kaku:1999yd,DhokerLec}
}.
In this paper, we avoid this problem by considering QFT toy models which have the same mass tower as string amplitudes but ignoring degeneracies.
These models are also interpreted as the ``pinching limit'' of string amplitudes.
Later, we will find that the toy amplitudes have infinitely many complex saddles associated with stringy mass tower
and that the saddles provide highly oscillatory terms to amplitudes.
We will discuss that such behaviors should be robust although we sacrificed moduli invariance.
Finally, the toy amplitudes will lead us to our conjecture that string amplitudes should approach fractal functions.

\begin{figure}[t]
	\centering
	\includegraphics[width=0.45\textwidth]{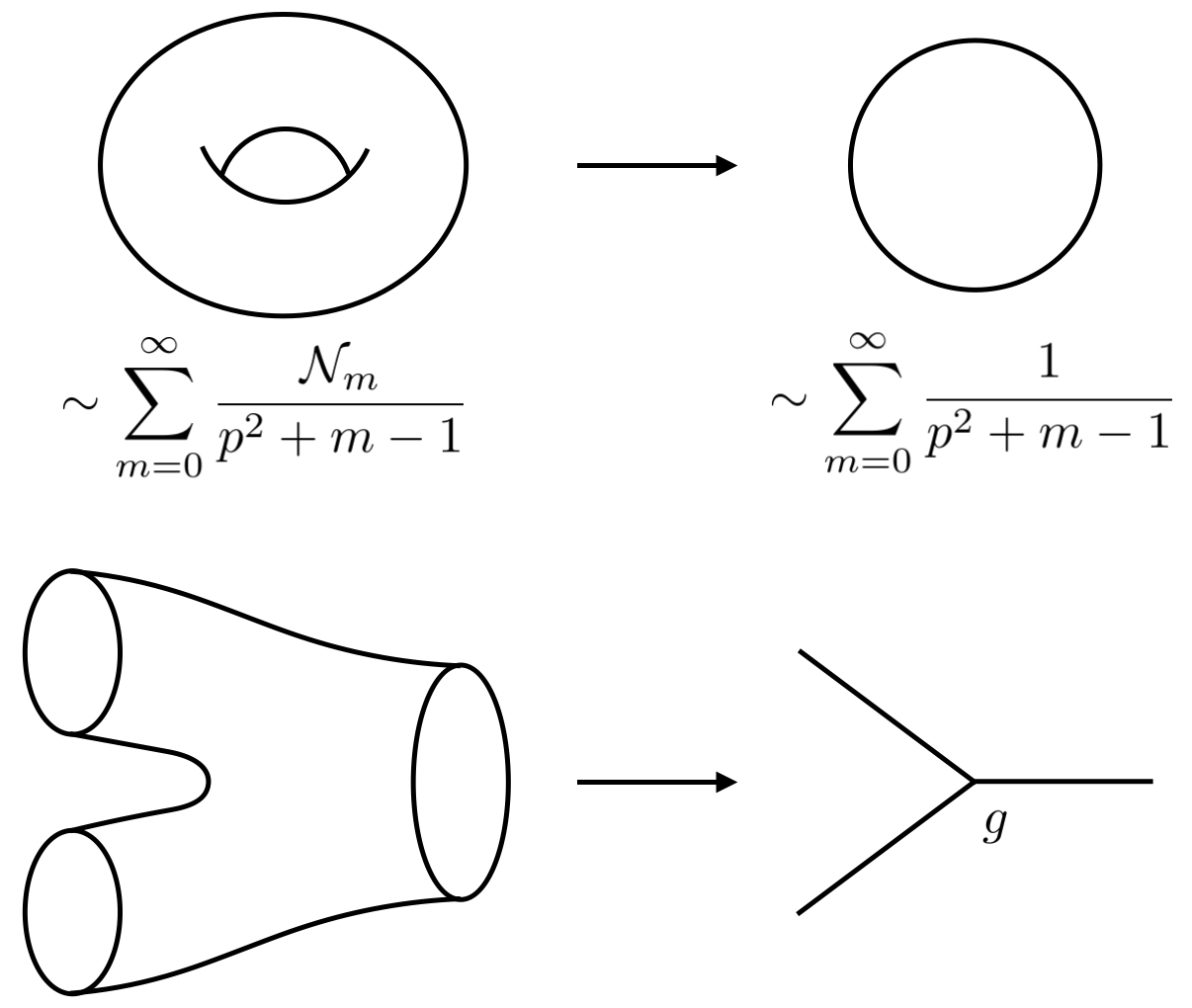}
	\caption{
		Ideas of truncating string amplitudes.
		A family of stringy excitations are removed and interacting vertices are replaced by a constant coupling.
		Although this truncation breaks modular invariance, it preserves typical oscillatory behaviors of amplitudes due to infinitely many stringy excitations.
	}
	\label{fig:idea_truncation}
\end{figure}

Our toy amplitudes are obtained by truncating stringy excitations based on the ideas summarized in Fig.~\ref{fig:idea_truncation}.
More precise rules will be explained in Sec.~\ref{sec:truncate_string_amp}.
For example, torus amplitude is thought to be the sum of all propagators of string excitations with degeneracies.
We will truncate a family of infinitely many string excitations,
leaving only a single excitation mode for each mass level $\alpha'M^2 = m-1, m\in\integernum_{\geq}$.
This truncation sacrifices modular invariance of string amplitudes.
However, it preserves infinitely many poles at $p^2+m-1=0$.
As explained in Sec.~\ref{sec:evaluate_amp},
such truncated amplitudes show highly oscillatory behaviors due to these infinitely many mass excitations.
It allows us to discuss the location of poles/zeros and erraticity of stringy amplitudes.

Since we broke modular invariance, we need to fix interaction vertices by hand.
Motivated by the forward limit of string amplitudes and higher spin models,
we will introduce three-point interaction with a constant coupling.
The truncated model is understood also as a model of infinitely many scalars with mass $\alpha'M^2 = m-1, m\in\integernum_{\geq}$.
In other words
\begin{align}
	\label{eq:scalar_action}
	S[\phi]
	&= \int \dd[d]{x} \left[
	\sum_{I=0}^{\infty}
	-\phi_I (-\partial^2 + \alpha'M^2(I)) \phi_I
	-g \sum_{I,J,K=0}^{\infty} \phi_I\phi_{J}\phi_{K}
	\right].
\end{align}

%\begin{align}
%	\sim \sum_{m=0}^{\infty} \frac{\mathcal{N}_m}{p^2 + m-1} \\
%	\sim \sum_{m=0}^{\infty} \frac{1}{p^2 + m-1}  \\
%	g
%\end{align}

Evaluating the amplitudes of this truncated model by saddle point approximation,
we will find that the amplitudes show highly oscillatory behaviors as a function of momentum due to infinitely many stringy mass tower.
As we go to higher loop corrections, we find smaller oscillatory structures in the amplitudes.

If we can go further to infinitely many loop corrections and high energy asymptotic limit,
the amplitudes have infinitely small structures.
In the asymptotic limit, the model no longer has typical mass scale
since momentum $\alpha'p^2$ is sent to infinity and mass gap goes to relatively zero.
Thus, there should be a fair chance that all oscillatory modes contribute to the amplitudes equivalently
and the graph of amplitudes become fractal.

We will check that this expectation is reasonable by studying an example whose self-similar structure becomes clean.
We will estimate the fractal dimension $D_{\text{F}}$ its amplitude as a function of momentum.
Our observations indicate that there exists a sweet parameter region
where the fractal dimension fits into an interval $1<D_{\text{F}}<2$ as the spacetime dimension is sent to $d\rightarrow6$ so that the three-point scalar interaction coupling becomes dimensionless.

We admit that there are caveats in these observations.
One of the caveats is modular invariance breaking.
It is possible that highly oscillatory behaviors we have found cancels after recovering modular invariance.
Second point is that our saddle point approximation method becomes worse as we approach the classically conformal point.
Our estimation of the fractal dimension is obtained by extrapolating our results in asymptotic regions.
Third point is about renormalization.
Our model is motivated by developing saddle point approximation method for string amplitudes,
thus, the truncated model is evaluated in higher dimension $d>6$.
The dimensional regularization factor, say $\Gamma(2-d/2)$ for one-loop, is just factored out.
Lastly, we note that our saddle point approximation is valid for $\alpha'p^2 \gg1$ where string momentum is imaginary.
If we are purely interested in the zeros of amplitudes, we can claim that our results give direct answers.
However, if we try to answer the original motivation, black hole physics, our results do not give complete answer at this stage.
We may need to find relations between zeros and poles,
or to give up our saddle point method and seek other powerful tools to analyze amplitudes.

Nevertheless, our truncated model detects oscillatory behaviors associated with infinitely many stringy excitations.
The form of string one-loop integral, as explained in Sec.~\ref{sec:truncate_string_amp} and the beginning of Sec.~\ref{sec:evaluate_amp}, tells us that
relation between oscillatory factors and stringy mass excitations should be robust even if we change the level of truncation.
Thus, in this paper, we conjecture that amplitudes become fractal in the asymptotic region,
intending to further motivate asymptotic analysis of string amplitudes and discussion about correspondence between black holes and strings as scatterers.

%----------------------------------
\subsection{Organization of this paper}
%----------------------------------

This paper is organized as follows.
In Sec.~\ref{sec:string_amp}, we review string scattering amplitudes and redefinition of integration contours.
Using the new integration contours, integrals converge and we can find appropriate saddles contributing to amplitudes.
In Sec.~\ref{sec:truncate_string_amp}, we will truncate the integral expressions of string amplitudes and construct our models.
Three-point interaction is introduced by hand, motivated by the forward limit of string amplitudes.
In Sec.~\ref{sec:barnes}, we will take a detour to prepare for saddle point approximation.
In Sec.~\ref{sec:evaluate_amp}, we will evaluate the truncated models of string amplitudes.
We will show that the amplitudes show oscillatory behaviors due to infinitely many stringy excitations.
In Sec.~\ref{sec:conjec}, we will compare the oscillatory behaviors with a typical fractal function, the Weierstrass function,
and estimate fractal dimension of our model in the asymptotic region.
Finally, we will state our conjecture that string amplitudes approach fractal functions in the higher genus and in the high energy asymptotic region.
In Sec.~\ref{sec:conc}, we summarize this paper and discuss future prospects.
Appendix App.~\ref{sec:ac_gamma} is devoted to reviewing the idea of redefining integration contours.
We will explain how redefinition of integration contour analytically continues the original integral, using the Euler gamma function as an example.

%========================
\section{String Amplitudes with Deformed Contour}
\label{sec:string_amp}
%========================

In this section, we review string scattering amplitudes.
Their formal integration contours are redefined to cure divergence.
The following integral expressions will be used later in Sec.~\ref{sec:truncate_string_amp} to construct our truncated models.

%----------------------------------
\subsection{Free energy}
\begin{figure}[t]
	\centering
	\includegraphics[width=.45\textwidth]{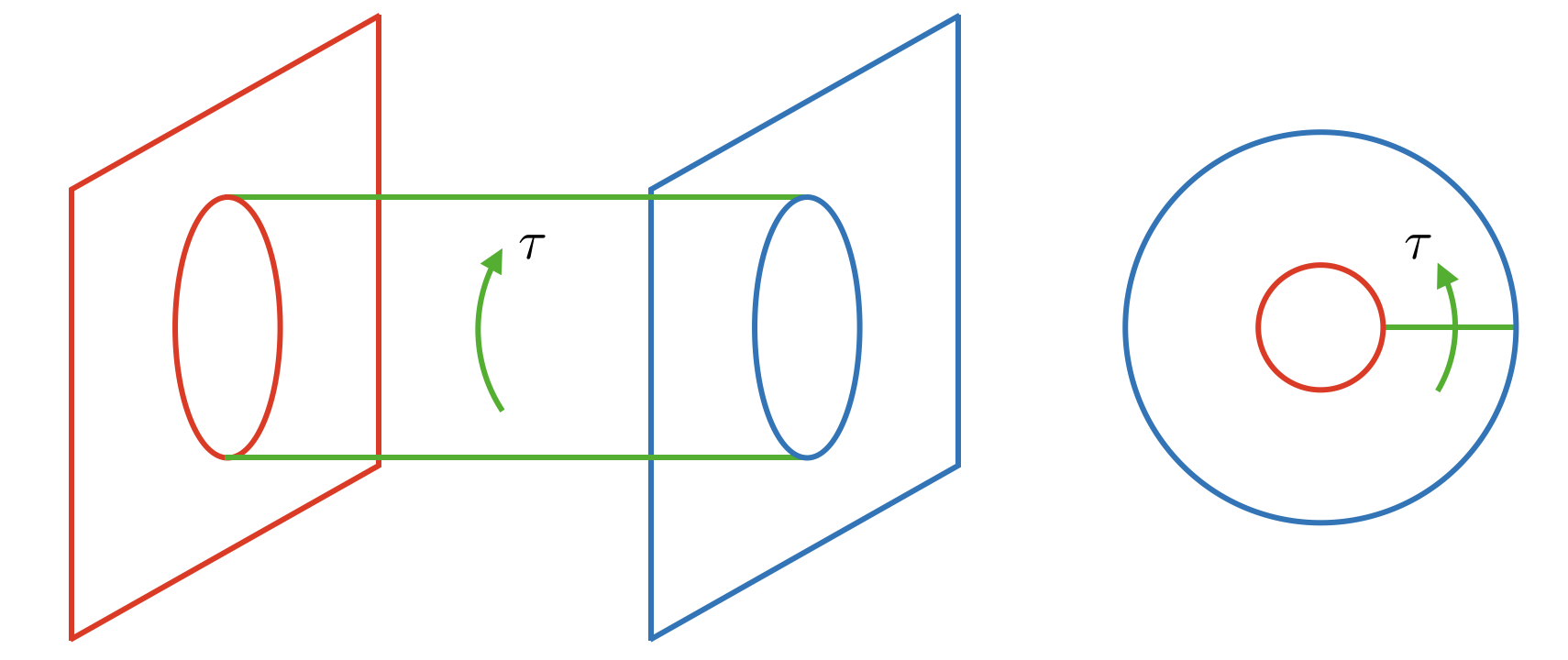}
	\caption{
		Open string attached to $\text{D}p$-branes.
		String travels in $d=p+1$ spacetime dimension.
		Its Schwinger proper time $\tau$ is along the direction of allows in the figures.
	}
	\label{fig:annulus}
\end{figure}

%----------------------------------
Let us consider open bosonic string theory in $D=26$ dimension.
Its open string excitations are
\begin{align}
	\alpha' M^2
	&= \sum_{n=1}^{\infty} \sum_{i=1}^{24} \alpha_{-n}^i\alpha_{n}^i -1 \notag\\
	&= \sum_{n=1}^{\infty} \sum_{i=1}^{24} n N_n^i -1
\end{align}
where
\begin{align}
	[\alpha_{m}^{\mu},\alpha_{n}^{\nu}]
	= m\delta_{m+n}\eta^{\mu\nu}.
\end{align}
The eigenvalues $\{N_n^i\}$ count occupation numbers of $n$-th excitation level and $i$ is an index for polarizations.
For example,
\begin{align}
	\begin{array}{ll}
		\{N_n\} = \{0,0,\dots\} & \alpha'M^2 = -1, \\
		\{N_n\} = \{1,0,\dots\} & \alpha'M^2 = 0, \\
		\{N_n\} = \{0,1,\dots\} & \alpha'M^2 = 1, \\
		\hspace{3em} \vdots &
	\end{array}
\end{align}
Now suppose that an open string is attached to $\text{D}p$-branes as Fig.~\ref{fig:annulus} shows.
Strings travel in $d=p+1$ spacetime dimension.
%\begin{align}
%	\tau
%\end{align}
Its cylinder amplitude, or equivalently annulus amplitude (see e.g. \cite{Johnson:2023onr}) is
\begin{align}
	\mathcal{A}^{\text{ann}}
	&= \int \frac{\dd[d]{\ell}}{(2\pi)^{d}}
	\int_0^{\infty} \frac{\dd{\tau}}{2\tau}
	\Tr \left[
	e^{ -2\pi\alpha'(\ell^2+M^2)\tau }
	\right].
\end{align}
Here $\tau$ is understood as the Euclidean proper time for each excitation with mass $\alpha'M^2$.
We have a factor $1/2\tau$ in the integrand to compute free energy of bosonic theory.
The trace over all string excitations becomes
\begin{align}
	\label{eq:trace_annulus}
	\Tr\left[
	e^{ -2\pi\alpha' M^2 \tau }
	\right]
	&= e^{ +2\pi\tau }
	\prod_{n=1}^{\infty} \prod_{i=1}^{D-2}
	\sum_{N_n^i=0}^{\infty} e^{ -2\pi n N_n^i \tau } \notag\\
	&=  e^{ +2\pi\tau }
	\prod_{n=1}^{\infty} \prod_{i=1}^{D-2}
	\frac{1}{ 1-e^{-2\pi n\tau} }
	= \dfrac{ e^{+2\pi\tau} }
	{ \left[ \prod_{n=1}^{\infty}(1-e^{-2\pi n\tau}) \right]^{D-2} } \notag\\
	&= \frac{1}{\eta(i\tau)^{D-2}} = \frac{e^{+2\pi\tau}}{f(i\tau)^{D-2}}.
\end{align}
Here we used the Dedekind eta function and the Euler function such that
\begin{align}
	\eta(\tau)
	&= e^{ \frac{\pi i\tau}{12} }
	\prod_{n=1}^{\infty} (1-e^{ 2\pi in\tau })
	= e^{ \frac{\pi i\tau}{12} } f(\tau).
\end{align}
Substituting this expression into the original integral expression, we obtain
\begin{align}
	\label{eq:annulus_amp}
	\mathcal{A}^{\text{ann}}
	&= \int \frac{\dd[d]{\ell}}{(2\pi)^{d}}
	\int_0^{\infty} \frac{\dd{\tau}}{2\tau}
	\frac{ e^{ -2\pi\tau(\alpha'\ell^2-1) } }{ f(i\tau)^{D-2} } \notag\\
	&= \left( \frac{1}{8\pi^2\alpha'} \right)^{d/2}
	\int_0^{\infty} \frac{\dd{\tau}}{2\tau}\:
	\tau^{-d/2}
	\frac{ e^{+2\pi\tau} }{ f(i\tau)^{D-2} } \notag\\
	&= \left( \frac{1}{8\pi^2\alpha'} \right)^{d/2}
	\int_0^{\infty} \frac{\dd{\tau}}{2\tau}\:
	\tau^{-d/2}
	\frac{ 1 }{ \eta(i\tau)^{D-2} }.
\end{align}
We have arrived at a simpler expression, an integral of the Dedekind eta function.
However, this is a formal integral.
%The Dedekind eta function $\eta(\tau)$ is singular on the real axis $\Im\tau=0$.
%In particular around the origin, its asymptotic expansion is \mycomm{Check This!}
The asymptotic form of the Dedekind eta function $\eta(\tau)$ around the origin is
\begin{align}
	\eta(\tau)
	&\sim
	\text{const.} \frac{q^{1/24}}{ (1-q)^{1/2} } \: e^{ -\frac{\pi^2}{6(1-q)} }, \quad
	q = e^{2\pi i\tau} \sim 1.
\end{align}
Thus the integral diverges around the origin $\tau \rightarrow +0$.
Also noting a functional relation
\begin{align}
	%\eta(\tau+1) &= e^{\pi i/12} \: \eta(\tau) \\
	\eta(-1/\tau) &= \sqrt{-i\tau} \: \eta(\tau),
\end{align}
similarly the integral diverges around infinity $\tau \rightarrow +\infty$.

\begin{figure}[H]
	\centering
	\includegraphics[height=.4\textwidth]{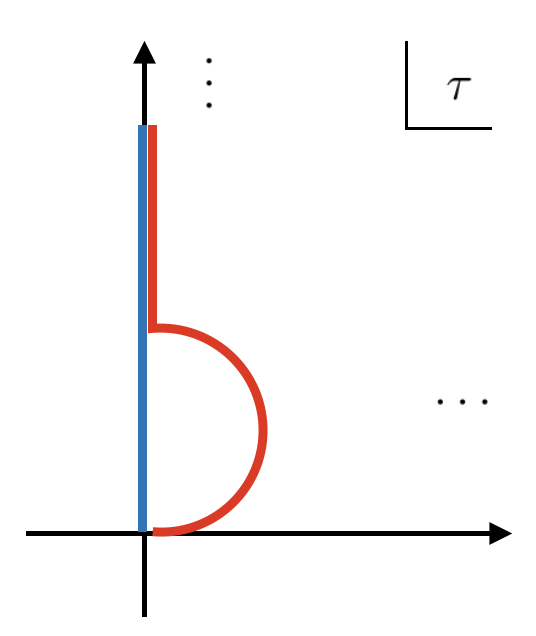}
	\caption{
		Redefined integration contour for an open string amplitude.
		The original integration contour (blue) diverges around the origin,
		while the new contour (red) converges.
		The bent contour is understood as Lorentzian proper time of a string.
	}
	\label{fig:eta_contour}
\end{figure}

This problem was cured in \cite{Eberhardt:2023xck} by redefining the integration contour.
In particular, around the origin, the new integration contour is bent as Fig.~\ref{fig:eta_contour} shows.
Recalling that the Schwinger parameter $\tau$ was originally Euclidean proper time of a traveling string,
the bent contour is understood as Lorentzian time of a string in the asymptotic region.
%\begin{align}
%	i\tau \quad \cdots
%\end{align}
%\begin{align}
%	\mathcal{A}(D,d) = ???
%\end{align}
%Left for future works

If we combine annulus and M\"{o}bius strip contributions,
the integral is divided into the residue around infinity
and the Rademacher contour\cite{Eberhardt:2023xck}.
Their contributions can be evaluated explicitly.
However, we pose here and proceed to next example.

%----------------------------------
\subsection{Two vertices}
%----------------------------------
Next let us consider annulus amplitude with $N=2$ vertices inserted.
We denote external momenta by $k_i, i=1,\dots,N$
and internal loop momentum by $\ell$.
From momentum conservation
\begin{align}
	k_1 + k_2 = 0.
\end{align}
We also write
\begin{align}
	&k=k_1=-k_2, \\
	&\ell_1 = \ell+k_1+k_2 = \ell, \quad \ell_2 = \ell+k_2.
\end{align}
Its amplitude is (see e.g.\cite{Kaku:1999yd,Green:1987sp,Green:1987mn})
\begin{align}
	\mathcal{A}^{\text{ann}}_N
	&= \int \frac{\dd[d]{\ell}}{(2\pi)^{d}}
	\int_0^{1} \prod_{i=1}^{N} \dd{x_i}
	x_i^{ \alpha'\ell_i^2-2 } \;
	\frac{1}{f(w)^{D-2}}
	\prod_{i<j}\prod_{n=1}^{\infty}
	\left[
	\frac{ (1-w^{n-1}c_{ji})(1-w^nc^{-1}_{ji}) }{ (1-w^n)^2 }
	\right]^{+\alpha'k_ik_j}
\end{align}
where
\begin{align}
	w &= x_1\cdots x_N, \\
	c_{ji} &= \rho_j/\rho_i = (x_1\cdots x_j)/(x_1\cdots x_i).
\end{align}
We change the variables as
\begin{align}
	\tau_i = \frac{-1}{2\pi}\ln x_i, \quad
	\tau = \tau_1+\cdots+\tau_N
\end{align}
so that the integral is over the Schwinger proper time
\begin{align}
	\mathcal{A}^{\text{ann}}_N
	&= \int \frac{\dd[d]{\ell}}{(2\pi)^{d}}
	\int_0^{\infty} \prod_{i=1}^{N} \dd{\tau_i} \:
	e^{ -2\pi(\alpha'\ell_i^2-1)\tau_i }
	\frac{1}{ f(i\tau)^{D-2} } \notag\\
	&\qquad \times
	\prod_{i<j} \prod_{n=1}^{\infty}
	\left[
	\frac{ (1-e^{-2\pi n\tau}e^{+2\pi(\tau_1+\cdots+\tau_i+\tau_{j+1}+\cdots_+\tau_N)})(1-e^{-2\pi n\tau}e^{+2\pi(\tau_{i+1}+\cdots+\tau_j)}) }{ (1-e^{-2\pi n\tau})^2 }
	\right]^{+\alpha'k_ik_j}
\end{align}
Since we set $N=2$, it reduces to
\begin{align}
	\mathcal{A}^{\text{ann}}_2
	&= \left( \frac{1}{8\pi^2\alpha'} \right)^{d/2}
	\int_{0}^{\infty} \dd{\tau_1}\dd{\tau_2} \:
	(\tau_1+\tau_2)^{-d/2} \:
	\frac{ e^{-2\pi \frac{\tau_1\tau_2}{\tau_1+\tau_2}\alpha'k^2+2\pi(\tau_1+\tau_2)} }
	{ f(i\tau)^{D-2} } \notag\\
	&\qquad\times
	\prod_{n=1}^{\infty}
	\left[
	\frac{ (1-e^{-2\pi n\tau}e^{+2\pi \tau_1})(1-e^{-2\pi n\tau}e^{+2\pi\tau_2}) }
	{ (1-e^{-2\pi n\tau})^2 }
	\right]^{-\alpha'k^2}.
\end{align}
%Evaluating
%\begin{align}
%	\mathcal{A}_2(D,d; k^2) = ???
%\end{align}
%Left for future works

This is also an formal integral since the integrand diverges around $\tau=0,\infty$.
The problem of the divergence is similarly cured by redefining integration contour.

%========================
\section{Truncating String Amplitudes}
\label{sec:truncate_string_amp}
%========================

In this section, we truncate string amplitudes reviewed in the previous section.
This truncation remove a family of infinitely many string excitations.
Since modular invariance is broken by this truncation,
we need to introduce interactions by hand.
In this paper, we introduce three-point interaction with a constant coupling,
motivated by the forward limit of string scattering and higher spin models.

%----------------------------------
\subsection{Truncation of mass tower}
%----------------------------------
In the previous examples in Sec.~\ref{sec:string_amp},
the trace is taken over all possible combinations of $\{N_n\}$.
In order to construct a simplified model with stringy excitations,
we restrict the range of the trace.

For example, we may restrict only on the first excitation level $n=1$,
\begin{align}
	N_1 = 0,1,2,\dots, \quad
	N_{n\geq2} &= 0.
\end{align}
Then the trace is truncated as
\begin{align}
	\left. \Tr\left[
	e^{ -2\pi\alpha' M^2 \tau }
	\right] \right|_{n=1}
	&= e^{ +2\pi\tau }
	\prod_{i=1}^{D-2}
	\sum_{N_1^i=0}^{\infty} e^{ -2\pi N_1^i \tau } \notag\\
	&=  e^{ +2\pi\tau }
	\prod_{i=1}^{D-2}
	\frac{1}{ 1-e^{-2\pi \tau} }
	= \dfrac{ e^{+2\pi\tau} }
	{ (1-e^{-2\pi \tau})^{D-2} }.
\end{align}
This is also understood as \eqref{eq:trace_annulus} with $\prod_{n\geq2}$ factors truncated.
The result is also equivalent to summation over
\begin{align}
	\{N_n\}
	= \{0,0,0,\dots\},
	\{1,0,0,\dots\},
	\{0,1,0,\dots\},
	\{0,0,1,\dots\},
	\dots
\end{align}

We may consider other types of truncation.
Restricting only on the second excitation level $n=2$,
\begin{align}
	N_1 = 0, \quad
	N_{2} = 0,1,2,\dots, \quad
	N_{n\geq3} = 0,
\end{align}
the trace is truncated as
\begin{align}
	\left. \Tr\left[
	e^{ -2\pi\alpha' M^2 \tau }
	\right] \right|_{n=2}
	&= e^{ +2\pi\tau }
	\prod_{i=1}^{D-2}
	\sum_{N_2^i=0}^{\infty} e^{ -4\pi N_2^i \tau } \notag\\
	&=  e^{ +2\pi\tau }
	\prod_{i=1}^{D-2}
	\frac{1}{ 1-e^{-4\pi \tau} }
	= \dfrac{ e^{+2\pi\tau} }
	{ (1-e^{-4\pi \tau})^{D-2} }.
\end{align}

Further we may consider truncation on the first and second excitation levels $n=1,2$,
\begin{align}
	N_1 = 0,1,2,\dots, \quad
	N_{2} = 0,1,2,\dots, \quad
	N_{n\geq3} = 0.
\end{align}
Then
\begin{align}
	\left. \Tr\left[
	e^{ -2\pi\alpha' M^2 \tau }
	\right] \right|_{n=1,2}
	&= e^{ +2\pi\tau }
	\prod_{n=1,2}
	\prod_{i=1}^{D-2}
	\sum_{N_1^i,N_2^i=0}^{\infty} e^{ -2\pi n N_n^i \tau } \notag\\
	&=  e^{ +2\pi\tau }
	\prod_{i=1}^{D-2}
	\frac{1}{ (1-e^{-2\pi\tau})(1-e^{-4\pi \tau}) }
	= \dfrac{ e^{+2\pi\tau} }
	{ \left[ (1-e^{-2\pi\tau})(1-e^{-4\pi \tau}) \right]^{D-2} }.
\end{align}

%----------------------------------
\subsection{Free energy}
%----------------------------------

Now we define truncated amplitude models.
For example in the $n=1$ truncation case,
\begin{align}
	\label{eq:ann_amp_n1}
	\left. \mathcal{A}^{\text{ann}} \right|_{n=1}
	&= \int \frac{\dd[d]{\ell}}{(2\pi)^{d}}
	\int_0^{\infty} \frac{\dd{\tau}}{2\tau}\:
	e^{-2\pi\tau \alpha'\ell^2}
	\left. \Tr\left[
	e^{ -2\pi\alpha' M^2 \tau }
	\right] \right|_{n=1} \notag\\
	&= \int \frac{\dd[d]{\ell}}{(2\pi)^{d}}
	\int_0^{\infty} \frac{\dd{\tau}}{2\tau}\:
	\frac{ e^{-2\pi\tau(\alpha'\ell^2-1)} }
	{ (1-e^{-2\pi\tau})^{D-2} } \notag\\
	&= \left( \frac{1}{8\pi^2\alpha'} \right)^{d/2}
	\int_0^{\infty} \frac{\dd{\tau}}{2\tau}\:
	\tau^{-d/2}
	\frac{ e^{+2\pi\tau} }{ (1-e^{-2\pi\tau})^{D-2} }
%	&= \left( \frac{1}{8\pi^2\alpha'} \right)^{d/2}
%	\frac{\Gamma(-d/2)}{2}
%	\zeta_{D-2}(-(p+1)/2,-2\pi; 2\pi,\dots,2\pi)
\end{align}
This is almost identical with the original string amplitude \eqref{eq:annulus_amp} except $\prod_{n\geq2}$ factors of the integrand.
Since the structure of the integrand is the same, they share similar oscillatory behaviors as discussed later in Sec.~\ref{sec:barnes} and Sec.~\ref{sec:evaluate_amp}.

Note that the integral is also an formal integral since the integrand diverges around $\tau=0,\infty$.
One way to cure this problem is to analytically continue the integral from an unphysical region to a physical region as we did in the original string amplitude, or in dimensional regularization.
Consider a function defined as
\begin{align}
	\zeta_N(s,x; \mathbf{a})
	= \frac{1}{\Gamma(s)} \int_0^{\infty} \dd{\tau} \:
	\tau^{s-1} \frac{e^{-x\tau}}{ \prod_{i=1}^{N}(1-e^{-a_i\tau}) }.
\end{align}
The reason why we denoted it by $\zeta$ will be explained later in Sec.~\ref{sec:barnes}.
If we choose parameters appropriately, this integral converges.
Later we may analytically continue this function to a physical region
$N\rightarrow D-2,\: s\rightarrow-d/2,\: x\rightarrow-2\pi,\: a_i\rightarrow+2\pi$.
Then, we obtain well-defined finite result
\begin{align}
	\left. \mathcal{A}^{\text{ann}} \right|_{n=1}
	&= \left( \frac{1}{8\pi^2\alpha'} \right)^{d/2}
	2\Gamma(-d/2)\: 
	\zeta_{D-2}(-d/2,-2\pi; 2\pi,\dots,2\pi).
\end{align}

In standard QFT computations, the gamma function $\Gamma(-d/2)$ may cause divergence.
It should be renormalized to other physical parameters.
However, in this paper, we are interested in high energy asymptotic behaviors of string amplitudes where $d/2$ is not necessarily non-negative integers.
Later in Sec.~\ref{sec:evaluate_amp}, we will show that saddle point approximation of the function $\zeta$ gives oscillatory factors $e^{-(\text{const.})x}$,
which essentially comes from the factor inside the integrand $e^{-x\tau}$.
This oscillatory behaviors are robust even if we increase the truncation level.
Indeed, the one-loop results in \cite{Eberhardt:2023xck} consists of oscillatory factors.
Thus, as long as we are interested in such oscillatory behaviors,
the divergence factor $\Gamma(-d/2)$ should be ignored.

Such a treatment of analytical continuation is equivalent to redefining the integration contour.
Some simple examples are the Hankel integration contour and the Pochhammer integration contour to analytically continue the Euler gamma function and beta function respectively.
In the context of string theory a new integration contour is introduced in \cite{Witten:2013pra}.
Generalizing the idea of Pochhammer integration contour, it was shown that a spiral integration contour analytically continues string amplitudes in a consistent way with the Feynman $i\epsilon$ prescription.
Such a method is reviewed in App.~\ref{sec:ac_gamma} using the Euler gamma function as an example.

\begin{figure}[t]
	\centering
	\includegraphics[height=.45\textwidth]{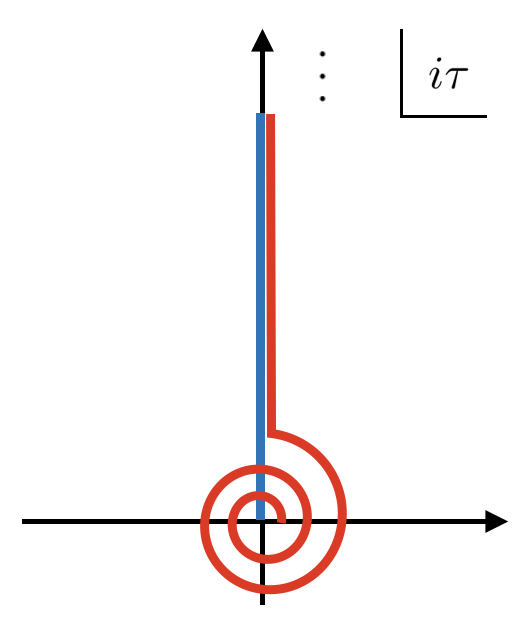}
	\caption{
		Redefined integration contour for truncated string amplitude model.
		The original integration contour (red) diverges around the origin,
		while the new contour (red) converges.
		The bent contour is understood as Lorentzian proper time of each stringy excitation mode.
	}
	\label{fig:barnes_contour}
\end{figure}

In our case, the integration contour is redefined as Fig.~\ref{fig:barnes_contour}.
This is a straightforward generalization of techniques reviewed in App.~\ref{sec:ac_gamma}.
The spiral contour around the origin $\tau=0$ absorbs the divergence of the integrand.
Another divergence from $\tau=\infty$ may require more explanations.
It will be cured in Sec.~\ref{sec:barnes}.

%Evaluating
%\begin{align}
%	\mathcal{A}(D,d) = (???)
%\end{align}
%Left for future works

%Ex. $n=1,2$ truncation.
%Similarly
%\begin{align}
%	\mathcal{A}
%	&= \left( \frac{1}{8\pi^2\alpha'} \right)^{(p+1)/2}
%	\int_0^{\infty} \frac{\dd{\tau}}{2\tau}\:
%	\tau^{-(p+1)/2}
%	\frac{ e^{+2\pi\tau} }
%	{ \left[ (1-e^{-2\pi\tau})(1-e^{-4\pi\tau}) \right]^{D-2} }
%\end{align}
%This is also written with the Barnes multiple zeta.

We can generalize this method straightforwardly to other truncation types such as $n=2,\: n=1,2$.

%----------------------------------
\subsection{Two vertices}
%----------------------------------

Next let us truncate two vertices amplitudes.
For example in $n=1$ truncation
\begin{align}
	\left. \mathcal{A}_2^{\text{ann}} \right|_{n=1}
	&= \left( \frac{1}{8\pi^2\alpha'} \right)^{d/2}
	\int_{0}^{\infty} \dd{\tau_1}\dd{\tau_2} \:
	(\tau_1+\tau_2)^{-d/2} \:
	\frac{ e^{-2\pi \frac{\tau_1\tau_2}{\tau_1+\tau_2}\alpha'k^2+2\pi(\tau_1+\tau_2)} }
	{ (1-e^{-2\pi \tau})^{D-2} } \:
	\left[
	\frac{ (1-e^{-2\pi \tau_1})(1-e^{-2\pi \tau_2}) }
	{ (1-e^{-2\pi \tau})^2 }
	\right]^{-\alpha'k^2}.
\end{align}
Changing the variables as
\begin{align}
	\tau_1 = x_1\tau, \quad
	\tau_2 = x_2\tau,
\end{align}
with a constraint $x_1+x_2=1$, we obtain
\begin{align}
	\label{eq:ann_amp_n2}
	\left. \mathcal{A}_2^{\text{ann}} \right|_{n=1}
	&= \left( \frac{1}{8\pi^2\alpha'} \right)^{d/2}
	\int_0^1 \dd{x_1}\dd{x_2} \delta(x_1+x_2-1)
	\left( \frac{1}{x_1+x_2} \right)^{d/2} \notag\\
	&\qquad\times
	\int_0^{\infty} \dd{\tau}\:
	\tau^{-d/2}\:
	\frac{ e^{ +\frac{x_1x_2}{x_1+x_2}(-2\pi\alpha'k^2)\tau +2\pi(x_1+x_2)\tau } }
	{ (1-e^{-2\pi\tau})^{D-2} }
	\left[
	\frac{ (1-e^{-2\pi x_1\tau})(1-e^{-2\pi x_2\tau}) }
	{ (1-e^{-2\pi \tau})^2 }
	\right]^{-\alpha'k^2}.
\end{align}
Here $\tau$ is understood as string proper time along the loop,
and $x_1,x_2$ are ratios of two propagators between two vertices.
The $\tau$ integral in \eqref{eq:ann_amp_n2} is a generalization of \eqref{eq:ann_amp_n1} with extra factors.
This formal integral is also cured by the same method.

%Evaluating
%\begin{align}
%	\mathcal{A}(D,d) = (???)
%\end{align}
%Left for future works

%----------------------------------
\subsection{Introducing interactions}
%----------------------------------
We would like to construct similar truncation models for higher genus amplitudes.
In \cite{Gross:1987kza,Gross:1987ar,Gross:1988ue,Gross:1988sj,Gross:1989ge,Mende:1989wt},
they used modular invariance to do some guess works.
Their method was useful to focus on asymptotic dumping factors $\sim e^{-(1/4\#\text{genus})(s\ln s+t\ln t+u\ln u)}$.
In our case, rather, we are interested in oscillatory behaviors of string amplitudes.
We sacrifice the modular invariance to simplify integral expressions.
Instead, we need to introduce interaction vertices of strings by hand.
We introduce three-point interaction with a constant coupling motivated by the following features of string amplitude.
Although we may consider more general types of interactions,
we focus on the simplest example in this paper.

The $st$-part of the Veneziano amplitude is
\begin{align}
	\mathcal{A}^{\text{Ven}}_{st}
	&= \frac{ \Gamma(-s-1)\Gamma(-t-1) }{ \Gamma(-s-t-1) }.
\end{align}
Around the $s$-channel pole, it becomes
\begin{align}
	\mathcal{A}^{\text{Ven}}_{st}
	\sim
	\sum_{n\geq-1}
	\frac{ \Gamma(n+t+2) }{ \Gamma(n+2)\Gamma(t+2) }
	\frac{1}{s-n},
\end{align}
If we regard this as QFT diagram,
the pre-factor is interpreted as a momentum dependent coupling
\begin{align}
	\mathcal{A}^{\text{Ven}}_{st}
	\sim
	\sum_{n\geq-1}
	\frac{ g_n^2(t) }{ s-n }, \quad
	g_n^2(t) = \frac{ \Gamma(n+t+2) }{ \Gamma(n+2)\Gamma(t+2) },
\end{align}
and $1/(s-n)$ is interpreted as a propagator.
In the forward limit $t\rightarrow0$, the coupling goes to a constant value $g_n(t)\rightarrow1$.
%\begin{align}
%	\mathcal{A}^{\text{VEN}}_{st}
%	\sim
%	\sum_{n\geq-1}
%	\frac{ 1 }{ s-n }.
%\end{align}

Motivated by this limit, we set rules to construct higher loop diagrams in our truncated models.
\begin{itemize}
	\item Propagator: sum of propagators with the $n=1$ truncated string excitations
	\item Interaction vertex: three-point interaction with a constant coupling
\end{itemize}
See also Fig.~\ref{fig:idea_truncation} for graphical expression of these rules.
%\begin{figure}[t]
%	\centering
%	\includegraphics[width=0.6\textwidth]{fig/toy_rule.pdf}
%	%\caption{}
%	\label{fig:toy_rule}
%\end{figure}
These rules are also understood as a model with infinitely many scalars \eqref{eq:scalar_action} with constant three-point interaction coupling.
We will claim that, at the end of Sec.~\ref{sec:evaluate_amp}, the effect of ignoring the momentum dependence of a coupling should be negligible as long as we are interested in oscillatory behaviors of amplitudes.

%--------------------------------
\subsubsection{Propagator}
%--------------------------------
Following the rule summarized above, we define our truncated model amplitudes.
The simplest example is a propagator
\begin{align}
	\mathcal{A}_0(p^2)
	&=
	\frac{1}{2\pi\alpha'}
	\sum_{\alpha'M^2=-1,0,1,\dots} \frac{1}{p^2+M^2}.
\end{align}
For later convenience, it is divided by $2\pi\alpha'$.
Introducing the Schwinger parameter, we obtain
\begin{align}
	\mathcal{A}_0(p^2)
	&= \int_0^{\infty} \dd{\tau}\:
	\frac{ e^{ +2\pi(-\alpha'p^2+1)\tau } }{ 1-e^{-2\pi\tau} }.
\end{align}

%\begin{figure}[t]
%	\centering
%	\includegraphics[height=30mm]{fig/depth-1_mom.pdf}
%	\caption{
%		Examples of diagrams,
%		propagator and one-loop
%	}
%	\label{fig:depth-1_mom}
%\end{figure}

%--------------------------------
\subsubsection{One-loop diagram}
%--------------------------------
Next example is a one-loop diagram shown in Fig.~\ref{fig:diagrams_example}.
It is defined by
\begin{align}
	\mathcal{A}_1(p^2)
	&=
	\frac{ (2\pi\alpha')^{d/2-2} }{ \Gamma(-d/2+2) }
	\int \dd[d]{\ell_1}
	\sum_{\alpha'M_1^2, \alpha'M_2^2}
	\frac{1}{\ell_1^2+M_1^2}
	\frac{1}{(p-\ell_1)^2+M_2^2}.
\end{align}
Introducing the Feynman parameter and the Schwinger parameter in a standard way
\begin{align}
	\mathcal{A}_1(p^2)
	&=
	(2\pi\alpha')^{d/2}
	\int \dd[d]{\ell_1}
	\sum_{\alpha'M_1^2, \alpha'M_2^2}
	\int_0^1 \dd{x_1}\dd{x_2} \delta(x_1+x_2-1) \notag\\
	&\qquad\times
	\frac{1}{\Gamma(-d/2+2)}
	\int_0^{\infty} \dd{\tau} \tau^{2-1}
	e^{ -2\pi\alpha'(\ell_1^2+M_1^2)x_1\tau -2\pi\alpha'((p-\ell_1)^2+M_2^2)x_2\tau }.
\end{align}
Completing the square for the momentum integral
\begin{align}
	\label{eq:depth-1_comp_sqr}
	&x_1(\ell_1^2+M_1^2)+ x_2((p-\ell_1)^2+M_2^2) \notag\\
	&= (x_1+x_2) \ell_1^2 -2x_2 p\ell_1 +x_2p^2 + x_1M_1^2 +x_2M_2^2 \notag\\
	&= (x_1+x_2)\left( \ell_1 - \frac{x_2p}{x_1+x_2} \right)^2
	+\frac{x_1x_2}{x_1+x_2}p^2 +x_1M_1^2+x_2M_2^2,
\end{align}
we have
\begin{align}
	\label{eq:trunc_oneloop}
	\mathcal{A}_1(p^2)
	&=
	\sum_{\alpha'M_1^2, \alpha'M_2^2}
	\int_0^1 \dd{x_1}\dd{x_2} \delta(x_1+x_2-1)
	\left( \frac{1}{x_1+x_2} \right)^{d/2} \notag\\
	&\qquad\times
	\frac{1}{\Gamma(-d/2+2)}
	\int_0^{\infty} \dd{\tau} \tau^{-d/2+2-1}
	e^{ -2\pi\alpha'\tau\left( \frac{x_1x_2}{x_1+x_2}p^2+x_1M_1^2+x_2M_2^2 \right) } \notag\\
	&=
	\int_0^1 \dd{x_1}\dd{x_2} \delta(x_1+x_2-1)
	\left( \frac{1}{x_1+x_2} \right)^{d/2} \notag\\
	&\qquad\times
	\frac{1}{\Gamma(-d/2+2)}
	\int_0^{\infty} \dd{\tau} \tau^{-d/2+2-1}\:
	\frac{ e^{ +2\pi\tau\left( \frac{x_1x_2}{x_1+x_2}(-\alpha'p^2)+x_1+x_2 \right) } }
	{ (1-e^{-2\pi x_1\tau})(1-e^{-2\pi x_2\tau}) }.
\end{align}
This is identical with \eqref{eq:ann_amp_n2} except the polarization and the momentum dependent coupling part.

%--------------------------------
\subsubsection{Two-loop diagrams}
\begin{figure}[t]
	\centering
	\includegraphics[width=0.8\textwidth]{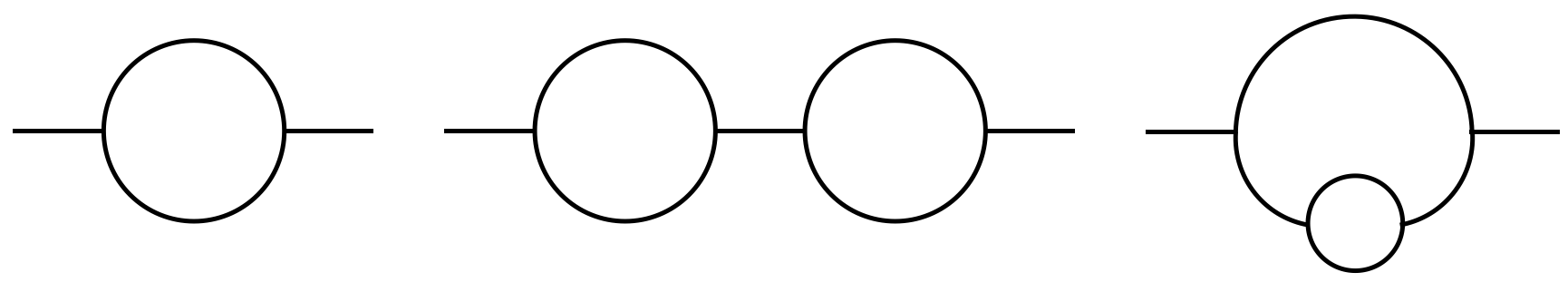}
	\caption{
		Some examples of loop diagrams for truncated string model.
		One-loop, two-loop reducible, and two-loop irreducible, from the left.
	}
	\label{fig:diagrams_example}
\end{figure}

%--------------------------------
There are two types of two-loop diagrams.
One of them is reducible to one-loop diagram as shown in Fig.~\ref{fig:diagrams_example}.
In the same way as the previous case, we have
\begin{align}
	\label{eq:trunc_twoloop_red}
	\mathcal{A}_2(p^2)
	&=
	\frac{ (2\pi\alpha')^{d-5} }{ \Gamma(-d+5) }
	\int \dd[d]{\ell_1} \dd[d]{\ell_2}
	\sum_{\alpha'M_1^2,\dots,\alpha'M_5^2} \notag\\
	&\qquad\qquad
	\frac{1}{ \ell_1^2+M_1^2 } \frac{1}{(p-\ell_1)^2+M_2^2}
	\frac{1}{ p^2+M_3^2 }
	\frac{1}{ \ell_2^2+M_4^2 } \frac{1}{(p-\ell_2)^2+M_5^2} \notag\\
	&=
	\int_0^1 \dd{x_1}\dots\dd{x_5} \delta(x_1+\cdots+x_5-1)
	\left(\frac{1}{x_1+x_2}\right)^{d/2}
	\left(\frac{1}{x_4+x_5}\right)^{d/2} \notag\\
	&\qquad\qquad\times
	\frac{1}{\Gamma(-d+5)}
	\int_0^{\infty} \dd{\tau}\: \tau^{-d+5-1}\:
	\frac{ \exp2\pi\tau \left[ \left(\frac{x_1x_2}{x_1+x_2}+x_3+\frac{x_4x_5}{x_4+x_5}\right)(-\alpha'p^2)+x_1+\cdots+x_5 \right] }
	{ (1-e^{-2\pi x_1\tau}) \cdots (1-e^{-2\pi x_5\tau}) }.
\end{align}
Another is irreducible one.
Similarly
\begin{align}
	\label{eq:trunc_twoloop_irred}
	\mathcal{A}_2(p^2)
	&=
	\frac{ (2\pi\alpha')^{d-5} }{ \Gamma(-d+5) }
	\int \dd[d]{\ell_1} \dd[d]{\ell_2}
	\sum_{\alpha'M_1^2,\dots,\alpha'M_5^2} \notag\\
	&\qquad\qquad
	\frac{1}{ \ell_1^2+M_1^2 } \frac{1}{(p-\ell_1)^2+M_2^2}
	\frac{1}{ \ell_2^2+M_3^2 }
	\frac{1}{ (p-\ell_1-\ell_2)^2+M_4^2 }
	\frac{1}{(p-\ell_1)^2+M_5^2} \notag\\
	&=
	\int_0^1 \dd{x_1}\dots\dd{x_5} \delta(x_1+\cdots+x_5-1)
	\left(\frac{1}{x_1+x_2+\frac{x_3x_4}{x_3+x_4}+x_5}\right)^{d/2}
	\left(\frac{1}{x_3+x_4}\right)^{d/2} \notag\\
	&\qquad\qquad\times
	\frac{1}{\Gamma(-d+5)}
	\int_0^{\infty} \dd{\tau}\: \tau^{-d+5-1}\:
	\frac{ \exp2\pi\tau \left[ \frac{ x_1\left(x_2+\frac{x_3x_4}{x_3+x_4}+x_5\right) }{ x_1+\left(x_2+\frac{x_3x_4}{x_3+x_4}+x_5\right)  }(-\alpha'p^2)+x_1+\cdots+x_5 \right] }
	{ (1-e^{-2\pi x_1\tau}) \cdots (1-e^{-2\pi x_5\tau}) }.
\end{align}

%\begin{figure}[t]
%	\centering
%	\includegraphics[height=30mm]{fig/depth-2_mom.pdf}
%	\caption{
%		Examples of diagrams,
%		two-loop diagrams
%	}
%	\label{fig:depth-2_mom}
%\end{figure}

%--------------------------------
\subsubsection{Nested melon diagrams}
\begin{figure}[t]
	\centering
	\includegraphics[width=0.8\textwidth]{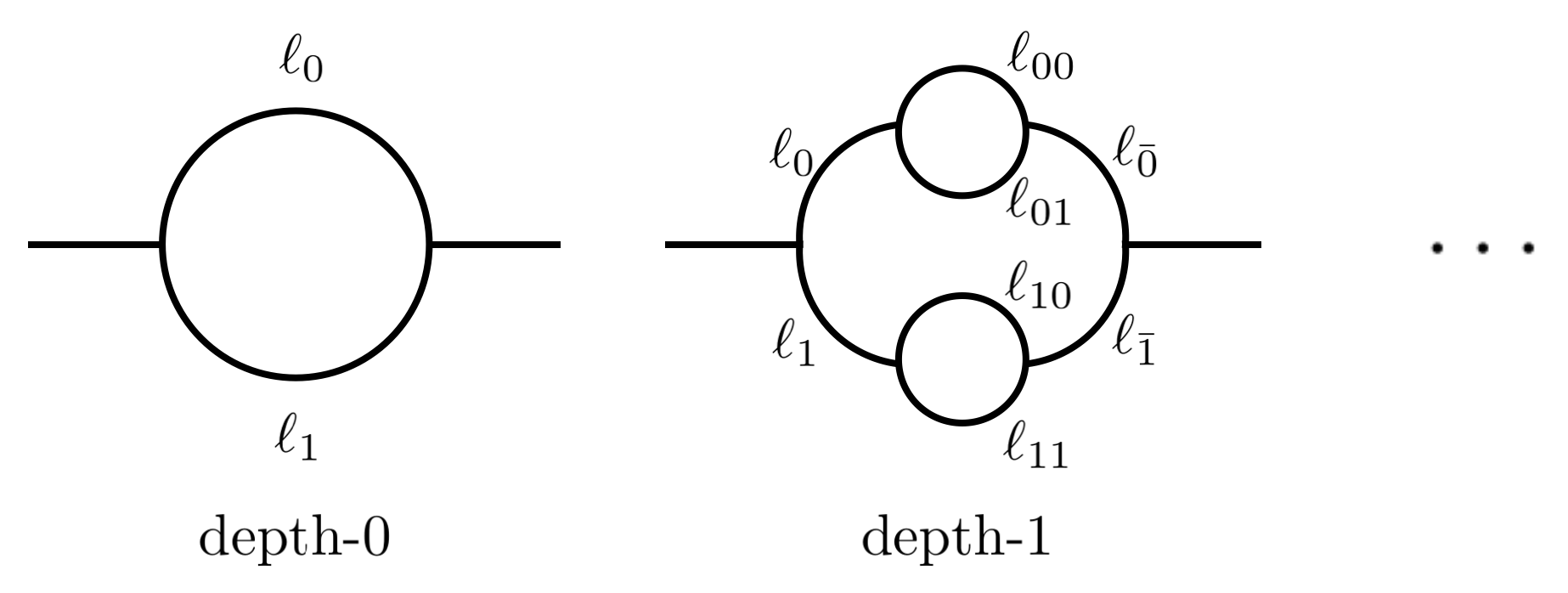}
	\caption{
		Examples of nested melon diagrams.
		In the depth-$n$ diagram, a single propagator bifurcates $(n+1)$-times into $2^{n+1}$ propagators
		and they merge again $(n+1)$-times into a single propagator.
	}
	\label{fig:nested_loops}
\end{figure}

%--------------------------------
We can generalize previous amplitudes in a straightforward way.
Here, we introduce a special class of multi-loop diagrams shown in Fig.~\ref{fig:nested_loops}.
%\begin{align}
%	\text{depth-}0 \; \text{depth-}1 \\
%	\ell_0 \; \ell_1 \; \ell_{00} \; \ell_{01} \; \ell_{10} \; \ell_{11} \; \ell_{\bar{0}} \; \ell_{\bar{1}}
%\end{align}
This class of diagrams will be compared in later Sec.~\ref{sec:conjec} to a clean example of fractal function.
We will see that observations from this example is straightforwardly generalized to other multi-loop diagrams.

For later convenience,
we assign a binary number to each propagator as described in Fig.~\ref{fig:nested_loops}.
We distinguish binary numbers with the same value but with different digits.
We also distinguish binary numbers with a bar symbol.
For example, $0$, $00$, $\overline{0}$, and $\overline{00}$ are all distinguished.

We also introduce a set of all $n$-digit binary numbers with/without a bar symbol on them, defined by
\begin{subequations}
	\begin{align}
		\mathbb{B}_n
		&= \left\{
		\underbrace{0\cdots0}_{n}, \:
		\underbrace{0\cdots1}_{n}, \:\dots, \:
		\underbrace{1\cdots1}_{n}
		\right\}, \\
		\overline{\mathbb{B}}_n
		&= \left\{
		\underbrace{\overline{0\cdots0}}_{n}, \:
		\underbrace{\overline{0\cdots1}}_{n}, \:\dots, \:
		\underbrace{\overline{1\cdots1}}_{n}
		\right\},
	\end{align}
\end{subequations}
for $n>0$.

We use these elements to label momenta and mass of propagator propagators.
For example, as $\ell_a, M_a,\: a \in \mathbb{B}_n$,
and as $\ell_{\overline{a}}, M_{\overline{a}},\: \overline{a} \in \overline{\mathbb{B}}_n$.
Particularly when $n=0$,
the set $\mathbb{B}_0$ is understood as the null set.
Accordingly, we define $\ell_{\text{null}}=p, M_{\text{null}}=0$.

We denote unions of $\mathbb{B}_n, \overline{\mathbb{B}}_n$ by
\begin{subequations}
	\begin{align}
		\mathbb{B}_n^m
		&= \mathbb{B}_n \cup \mathbb{B}_{n+1} \cup \cdots \cup \mathbb{B}_m, \\
		\overline{\mathbb{B}}_n^m
		&= \overline{\mathbb{B}}_n \cup \overline{\mathbb{B}}_{n+1} \cup \cdots \cup \overline{\mathbb{B}}_m,
	\end{align}
\end{subequations}
for $m \geq n \geq 0$.

Finally, the size of these sets are denoted by $\abs{\cdot}$.
For example,
$\abs{\mathbb{B}_{n+1}}=2^{n+1},\: \abs{\mathbb{B}_1^n}=\abs{\overline{\mathbb{B}}_1^n}=2^{n+1}-2$.

Now we are ready to define depth-$n$ melon diagram.
\begin{align}
	\label{eq:amp_depn}
	\mathcal{A}^{(n)}(p^2)
	&\propto \sum_{M^2_{a \in \mathbb{B}_1^n} }
	\int \prod_{ a \in \mathbb{B}_1^n } \frac{\dd[d]{\ell_a}}{(2\pi)^d}
	\prod_{ a \in \mathbb{B}_0^{n-1} }
	\delta^d ( \ell_a - \ell_{a0} -\ell_{a1} )\:
	\frac{1}{ \ell_{a0}^2+M_{a0}^2 }
	\frac{1}{ \ell_{a1}^2+M_{a1}^2 } \notag\\
	&\times
	\sum_{M^2_{ \overline{a} \in \mathbb{\overline{B}}_1^n }}
	\int \prod_{ \overline{a} \in \mathbb{\overline{B}}_1^n } \frac{\dd[d]{\ell_{\overline{a}}}}{(2\pi)^d}
	\prod_{ \overline{a} \in \mathbb{\overline{B}}_0^{n-1} }
	\delta^d ( \ell_{\overline{a}} - \ell_{\overline{a0}} -\ell_{\overline{a1}} )\:
	\frac{1}{ \ell_{\overline{a0}}^2+M_{\overline{a0}}^2 }
	\frac{1}{ \ell_{\overline{a1}}^2+M_{\overline{a1}}^2 } \notag\\
	&\times
	\sum_{ M^2_{ b\in\mathbb{B}_{n+1} } }
	\int \prod_{ b\in\mathbb{B}_{n+1} } \frac{\dd[d]{\ell_b}}{(2\pi)^d}
	\prod_{ b\in\mathbb{B}_{n} }
	\delta^d( \ell_b-\ell_{b0}-\ell_{b1} )
	\frac{1}{ \ell_{b0}^2+M_{b0}^2 }
	\frac{1}{ \ell_{b1}^2+M_{b1}^2 }
	\delta^d( \ell_{\overline{b}}-\ell_{b0}-\ell_{b1} ).
\end{align}
%Here the integrals over momenta $\ell_{a}, \ell_{\overline{a}}, \ell_b$ are regularized by the ordinal Feynman-$i\varepsilon$ prescription and the dimensional regularization.
%The summations over $m_a^2, m_{\overline{a}}^2, m_b^2$ are understood as the sum over all non-negative integers,
%\begin{subequations}
%	\begin{align}
%		m_a^2 = n_a-1, \: n_a\in\mathbb{Z}_{\geq}, \\
%		m_{\overline{a}}^2 = n_{\overline{a}}-1, \: n_{\overline{a}}\in\mathbb{Z}_{\geq}, \\
%		m_b^2 = n_b-1, \: n_b\in\mathbb{Z}_{\geq}.
%	\end{align}
%\end{subequations}

In the same way as the previous cases,
we introduce Feynman parameters,
$x_a,x_{\overline{a}},x_b$ with $a\in\mathbb{B}_1^n,\: \overline{a}\in\overline{\mathbb{B}}_1^n,\: b\in\mathbb{B}_{n+1}$,
and Schwinger parameter $\tau$.
The amplitude becomes
\begin{align}
	\label{eq:amp_depn_int}
	\mathcal{A}^{(n)}
	&\propto
	\sum_{ M^2_{a}, M^2_{\overline{a}}, M^2_b }
	\int_0^1 \prod_{a,\overline{a},b} \dd{x_a}\dd{x_{\overline{a}}}\dd{x_b}\:
	\delta \left( \sum_ax_a + \sum_{\overline{a}}x_{\overline{a}}+\sum_bx_b-1 \right) \notag\\
	&\times
	\int_0^\infty \dd{\tau} \tau^{ (3\cdot2^{n+1}-4)-1 }
	\int \prod_{a,\overline{a},b}
	\frac{\dd[d]{\ell_a}}{(2\pi)^d}
	\frac{\dd[d]{\ell_{\overline{a}}}}{(2\pi)^d}
	\frac{\dd[d]{\ell_b}}{(2\pi)^d} \notag\\
	&\times
	\prod_{
		a\in\mathbb{B}_0^{n-1},\:
		\overline{a}\in\overline{\mathbb{B}}_{0}^{n-1},\:
		b\in\mathbb{B}_n
	}
	\delta^d ( \ell_a - \ell_{a0} -\ell_{a1} )\:
	\delta^d ( \ell_{\overline{a}} - \ell_{\overline{a0}} -\ell_{\overline{a1}} ) \notag\\
	&\hspace{32mm}\times
	\delta^d( \ell_b-\ell_{b0}-\ell_{b1} )\:
	\delta^d( \ell_{\overline{b}}-\ell_{b0}-\ell_{b1} ) \notag\\
	&\times
	\exp\left[
	-2\pi\alpha'\tau\left(
	\sum_a x_a(\ell_a^2+M_a^2)
	+\sum_{\overline{a}} x_{\overline{a}}(\ell_{\overline{a}}+M_{\overline{a}}^2)
	+\sum_b x_b(\ell_b^2+M_b^2)
	\right)
	\right].
\end{align}
Here the indices $a,\overline{a},b$ run over $\mathbb{B}_1^n, \overline{\mathbb{B}}_1^n, \mathbb{B}_{n+1}$ respectively except the ones in the delta functions.

We can systematically perform the Gaussian integrals over momenta $\ell$ step-by-step from the loops in the middle of the diagram toward the left/right-hand side of the diagram.
It is convenient to define new variables iteratively by
\begin{subequations}
	\begin{align}
		x^{(0)}_a
		&= x_a, \\
		x^{(n+1)}_a
		&= x_a
		+\frac{x^{(n)}_{a0}x^{(n)}_{a1}}{x^{(n)}_{a0}+x^{(n)}_{a1}}
		+x_{\overline{a}}.
	\end{align}
\end{subequations}
Then, recursively we can show that
\begin{align}
	\label{eq:n-loop_int}
	\mathcal{A}^{(n)}
	&\propto
	\sum_{ M^2_{a}, M^2_{\overline{a}}, M^2_b }
	\int_0^1 \prod_{a,\overline{a},b} \dd{x_a}\dd{x_{\overline{a}}}\dd{x_b}\:
	\delta \left( \sum_ax_a + \sum_{\overline{a}}x_{\overline{a}}+\sum_bx_b-1 \right) \notag\\
	&\times
	\prod_{i=0}^{n} \prod_{a\in\mathbb{B}_{n-i}}
	\left( \frac{1}{x^{(i)}_{a0}+x^{(i)}_{a1}} \right)^{d/2}
	\int_0^\infty \dd{\tau} \tau^{ (3\cdot2^{n+1}-4)-(2^{n+1}-1)\frac{d}{2}-1 } \: \notag\\
	&\times
	\exp\left[
	- 2\pi\alpha'\tau \left(
	\frac{x^{(n)}_0x^{(n)}_1}{x^{(n)}_0+x^{(n)}_1}p^2
	+ \sum_a x_a M_a^2
	+ \sum_{\overline{a}} x_{\overline{a}}M_{\overline{a}}^2
	+ \sum_{b} x_b M_b^2
	\right)
	\right].
\end{align}
Finally performing the summation over stringy excitations, we arrive at
\begin{align}
	\label{eq:trunc_nested}
	\mathcal{A}^{(n)}
	&\propto
	\int_0^1 \prod_{a,\overline{a},b} \dd{x_a}\dd{x_{\overline{a}}}\dd{x_b}\:
	\delta \left( \sum_ax_a + \sum_{\overline{a}}x_{\overline{a}}+\sum_bx_b-1 \right) \notag\\
	&\times
	\prod_{i=0}^{n} \prod_{a\in\mathbb{B}_{n-i}}
	\left( \frac{1}{x^{(i)}_{a0}+x^{(i)}_{a1}} \right)^{d/2} \notag\\
	&\times
	\frac{1}{\Gamma(-k^{(n)})}
	\int_0^{\infty} \dd{\tau}\:
	\tau^{-k^{(n)}-1}\:
	\frac{ \exp 2\pi\tau\left[ \frac{x^{(n)}_0x^{(n)}_1}{x^{(n)}_0+x^{(n)}_1}(-\alpha'p^2) +\sum_ax_a+\sum_{\bar{a}}x_{\bar{a}}+\sum_bx_b \right] }
	{ \prod_{a}(1-e^{-2\pi x_a\tau}) \prod_{\bar{a}}(1-e^{-2\pi x_{\bar{a}}\tau}) \prod_{b}(1-e^{-2\pi x_b\tau}) }
\end{align}
where
\begin{align}
	k^{(n)}
	&= (2^{n+1}-1)\frac{d}{2} - (3\cdot2^{n+1}-4).
\end{align}

%========================
\section{Barnes Multiple Zeta Function}
\label{sec:barnes}
%========================

So far, we have obtained integral expressions for truncated string amplitudes.
In this section, we take a detour to a special function, called the Barnes multiple zeta function,
in order to prepare for evaluating the diagrams.
We review this special function and obtain its asymptotic expansion by saddle point method.
Techniques reviewed in this part will be used in the next Sec.~\ref{sec:evaluate_amp}.

%--------------------------------
\subsection{Definitions}
%--------------------------------
The Barnes multiple zeta function\cite{Barnes1904}\footnote{
	See also \cite{Barnes1899,Barnes1900,Barnes1901}
	and a textbook e.g. \cite{Ibukiyama2014}.
	The Barnes multiple zeta function is also called the multiple Hurwitz zeta function since it is understood as a generalization of the Hurwitz zeta function.
}
is defined by
\begin{align}
	\label{eq:bzeta_def}
	\zeta_N(s,x;\mathbf{a})
	= \sum_{n_1,\dots,n_{N}=0}^{\infty}
	\frac{1}{(x+a_1n_1+\cdots+a_{N}n_{N})^s}, \quad
	x>0, \Re s>N.
\end{align}
This infinite series absolutely converges for $\Re s>N$.
Particularly when $N=1, x=1, a_1=1$,
the Barnes multiple zeta function reduces to the Riemann zeta function.
Also when $N=1, a_1=1$, it reduces to the Hurwitz zeta function.

Introducing the Schwinger parameter,
we can move to an integral expression
\begin{align}
	\label{eq:bzeta_int}
	\zeta_N(s,x; \mathbf{a})
	&= \frac{1}{\Gamma(s)} \int_0^{\infty} \dd{t}\:
	t^{s-1} \frac{e^{-xt}}{ (1-e^{-a_1t}) \cdots (1-e^{-a_{N}t}) }.
\end{align}
It has another expression
\begin{align}
	\label{eq:bzeta_hankel}
	\zeta_N(s,x;\mathbf{a})
	= \frac{1}{ \Gamma(s)(e^{2\pi is}-1) }
	\int_{\mathcal{H}_0} \dd{t}\:
	t^{s-1} \frac{e^{+(a_1+\cdots+a_N-x)t}}{(e^{a_1t}-1)\cdots(e^{a_Nt}-1)}.
\end{align}
Here $\mathcal{H}_0$ is the Hankel contour wrapping the positive real axis in the counter clockwise direction.
Through this contour integral representation, the Barnes multiple zeta function is analytically continued to the whole complex $s$-plane except simple poles at $s=1,\dots,N$.

From the definition of the Barnes multiple zeta function,
we can show that
\begin{align}
	\zeta_{N}(s,x+a_N; a_1,\dots,a_N)
	= \zeta_{N}(s,x;a_1,\dots,a_N) - \zeta_{N-1}(s,x;a_1,\dots,a_{N-1}).
\end{align}
Using this relation, we can analytically continue with respect to $x$.
Through these processes, we can cure the divergence we encountered in Sec.~\ref{sec:truncate_string_amp} around $\tau=\infty$.

%--------------------------------
\subsection{Special case $N=1$: Hurwitz zeta function}
%--------------------------------
A special case $N=1, a_1=1$ is known as the Hurwitz zeta function\footnote{
	See a textbook e.g. \cite{whittaker_watson_2021}.
}.
It is defined by an infinite series
\begin{align}
	\label{eq:hzeta_def}
	\zeta(s,x) = \sum_{n=0}^{\infty} \frac{1}{(x+n)^s}, \quad
	x>0,\: \Re s>1.
\end{align}
This infinite series absolutely converges for $\Re s>1$.
Particularly when $x=1$, it reduces to the Riemann zeta function.

Introducing the Schwinger parameter, we can move to an integral expression
\begin{align}
	\label{eq:hzeta_int}
	\zeta(s,x)
	&= \frac{1}{\Gamma(s)} \int_0^{\infty} \dd{t}\:
	t^{s-1} \frac{e^{-xt}}{1-e^{-t}}.
\end{align}
It has another expression
\begin{align}
	\label{eq:hzeta_hankel}
	\zeta(s,a)
	= \frac{1}{\Gamma(s)(e^{2\pi is}-1)}
	\int_{\mathcal{H}_{0}} \dd{t}\:
	t^{s-1} \frac{e^{(1-x)t}}{e^t-1}.
\end{align}
Here $\mathcal{H}_0$ is the Hankel contour wrapping the positive real axis in the counter clockwise direction.
Through this contour integral representation, the Hurwitz zeta function is analytically continued to the whole complex $s$-plane except the simple pole at $s=1$.

From the definition of the Hurwitz zeta function,
when $x>1$, the constant $x$ can be shifted by using
\begin{align}
	\label{eq:hzeta_xshift}
	\zeta(s,x)
	= \zeta(s,x-m) - \sum_{n=0}^{m-1}\frac{1}{(n+x-m )^s},
\end{align}
where $m$ is an integer such that $x-1\leq m<x$.
Using this relation, we can analytically continue in terms of $x$.
We may restrict $x$ to a region $0<x\leq 1$ if necessary.

\subsubsection{Integer argument}
Suppose that $s=-k$ is a non-positive integers.
Also let us restrict to a region $0<x\leq1$ for simplicity.
In this case, we can unwrap the Hankel contour $\mathcal{H}_0$ and deform to a horizontal line at negative infinity.
Through this deformation,
the integration contour crosses the poles at $t=2\pi i\mathbb{Z}_{\neq0}$.
Collecting the contributions from these poles, we can show that
\begin{align}
	\label{eq:hzeta_bernoulli}
	\zeta(-k,x)
	&= - \frac{B_{k+1}(x)}{k+1}, \quad
	k=0,1,2,\dots.
\end{align}
Here, $B_{n}(x)$ is the Bernoulli polynomial.
It is defined by the following generating function
\begin{align}
	\label{eq:bernoulli}
	\frac{ te^{xt} }{ e^{t}-1 }
	&= \sum_{k=0}^{\infty} B_{k}(x) \frac{t^k}{k!}.
\end{align}
From the definition of the Bernoulli polynomial, we find
\begin{align}
	\label{eq:bernoulli_fourier}
	B_k(x)
	&= \oint_{C_0} \frac{\dd{t}}{2\pi i} \frac{k!}{t^{k+1}}
	\frac{ te^{xt} }{ e^{t}-1 } \notag\\
	&= -k! \sum_{n\neq0}
	\oint_{C_n} \frac{\dd{t}}{2\pi i}
	\frac{ t^{-k} e^{xt} }{ e^{t}-1 } \notag\\
	&= -k! \sum_{n\neq0}
	\left( \frac{1}{2\pi in} \right)^k
	e^{2\pi in x}.
\end{align}
One may regard this relation as the Fourier expansion of the Bernoulli polynomial.
%It leads to the asymptotic form of the Bernoulli polynomial for the large order $k\gg1$
%\begin{align}
%	B_k(x)
%	&\sim
%	-\frac{k!}{(2\pi i)^k} \left( e^{2\pi ix} + (-1)^k e^{-2\pi ix} \right).
%\end{align}
Substituting \eqref{eq:bernoulli_fourier} into \eqref{eq:hzeta_bernoulli}, we obtain

\begin{align}
	\label{eq:hzeta_fourier}
	\zeta(-k,x)
	&= k! \sum_{n\neq0}
	\left( \frac{1}{2\pi in} \right)^{k+1}
	e^{2\pi inx}, \quad
	0<x\leq1.
	%&\sim
	%\frac{k!}{(2\pi i)^{k+1}}
	%\left( e^{2\pi ix} + (-1)^{k+1} e^{-2\pi ix} \right).
\end{align}
More generally, by using \eqref{eq:hzeta_xshift}, we obtain
\begin{align}
	\label{eq:hzeta_fourier_poly}
	\zeta(-k,x)
	&= 	k! \sum_{n\neq0}
	\left( \frac{1}{2\pi in} \right)^{k+1}
	e^{2\pi inx} 
	- \sum_{n=0}^{m-1}\frac{1}{(n+x-m )^{-k}}, \quad
	x>0.
\end{align}
In the asymptotic region $k\gg1$, the Hurwitz zeta function reduces to a simple $2\pi$-periodic function.
The size of the first term is $\sim (1/2\pi)^{k+1}k!$
while the second term is $\sim x^{k+1}$.
These two contributions become compatible when $k\sim x$.
The first oscillatory part dominates when $x$ is small $x \ll k$
while the second polynomial part dominates when $x$ is large $x\gg k$.

\begin{figure}[t]
	\centering
	\includegraphics[width=0.45\textwidth]{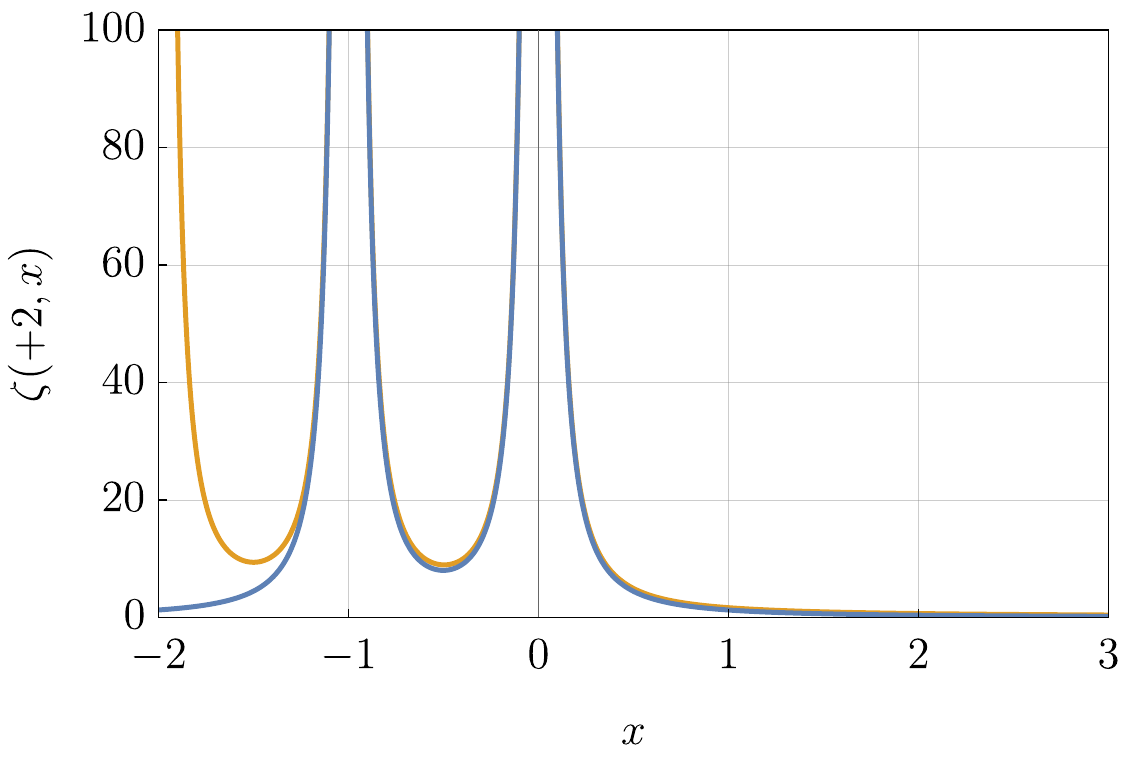} \hspace{5mm}
	\includegraphics[width=0.45\textwidth]{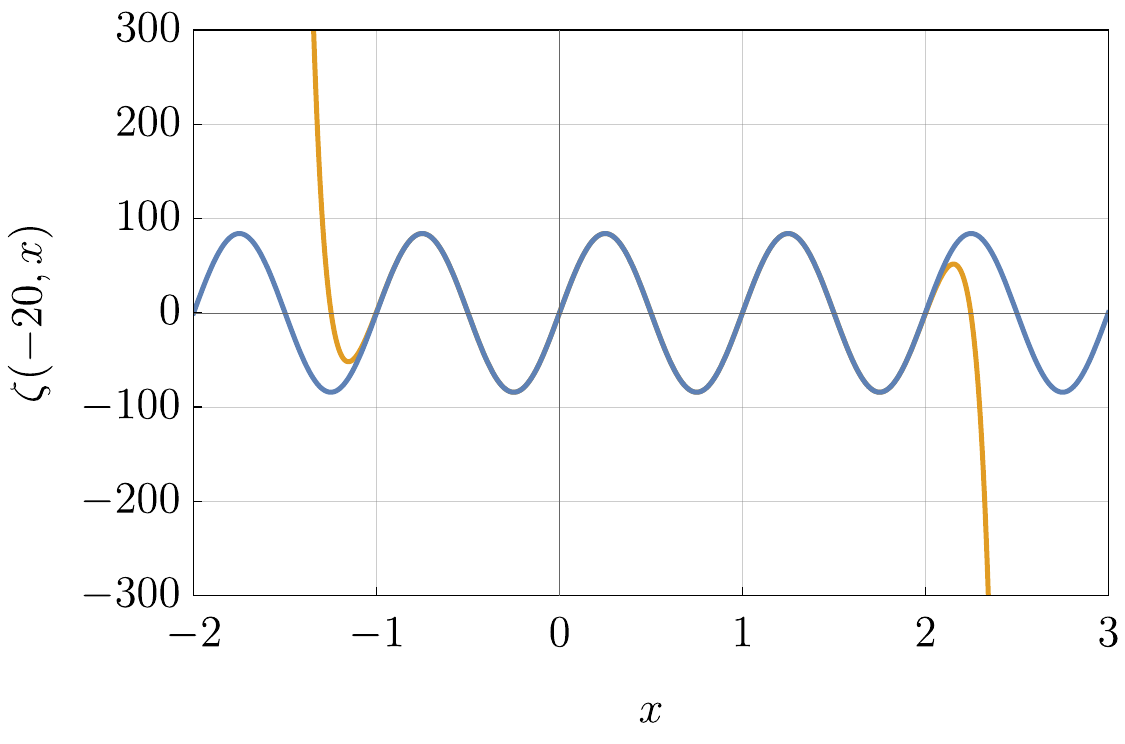}
	\caption{
		Plots of the Hurwitz zeta $\zeta(-k,x)$ in typical parameter regions.
		The left figure is for $s=-k=+2$ while
		the right figure is for $s=-k=-20$.
		The orange curves are full $\zeta(-k,x)$.
		The blue curves are its truncation of infinite series at the first term.
	}
	\label{fig:hurwitz_plot}
\end{figure}

The last expression \eqref{eq:hzeta_fourier_poly} converges only for $\Re s=-\Re k<0$.
However, we can analytically continue it.
From the definition of the Hurwitz zeta function,
we find that it has poles at $x\in\integernum_{\leq}$ when $\Re s>1$.
Some example plots of the Hurwitz zeta function are depicted in Fig.~\ref{fig:hurwitz_plot}.

\subsubsection{Saddle point approximation}
Saddle point approximation is also useful to obtain the expansion of the Hurwitz zeta function \eqref{eq:hzeta_fourier_poly}.
Let us start from the integral representation \eqref{eq:hzeta_int}.
When $s=-k\leq0$, the integral diverges around $t=0$.
In order to cure this problem, one may use the Hankel contour.
Instead we use a spiral integration contour following \cite{Witten:2013pra},
\begin{align}
	\zeta(-k,x)
	&= \frac{1}{\Gamma(-k-i\delta)}
	\int_{\mathcal{C}_{\delta}} \dd{t}\: t^{-k-i\delta-1}
	\frac{e^{-xt}}{1-e^{-t}}.
\end{align}
Here $\delta>0$ is a small regularization parameter to shift $k$, which is an analog of the small constant of the Feynman prescription small parameter.
In the subsequent discussion, we will omit the small regularization parameter $\delta$ as long as there is no risk of confusion.

If we write the integral in the form of
\begin{align}
	\zeta(-k,x)
	&= \frac{1}{\Gamma(-k)}
	\int_{\mathcal{C}} \dd{t}\: e^{S(t)},
\end{align}
the ``action'' is
\begin{align}
	S(t)
	&= -xt -\ln(1-e^{-t}) - (k+1)\ln t.
\end{align}
Its saddle points are determined by an equation
\begin{align}
	0 &= S'(t)
	= -x -\frac{1}{e^t-1} -\frac{k+1}{t}.
\end{align}
In the following we restrict in asymptotic region $k, x\gg 1$ where saddle point approximation is valid.

We can find a real saddle by setting ansatz $t = -1/\epsilon$.
The saddle point equation becomes
\begin{align}
	0 \sim
	x-1-\epsilon(k+1)
\end{align}
Its solution is
\begin{align}
	t_0 \sim -\frac{k+1}{x-1}.
\end{align}
Also we can also find infinitely many complex saddles.
By setting an ansatz $t=\epsilon+2\pi in,\: n\in\mathbb{Z}_{\neq0}$,
the saddle point equation becomes
\begin{align}
	0 \simeq
	%x(2\pi in+\epsilon) + \frac{2\pi in+\epsilon}{\epsilon} + (k+1).
	x + \frac{1}{\epsilon} + \frac{k+1}{2\pi n+\epsilon}.
\end{align}
Its solutions are
\begin{align}
	t_n \sim
	%2\pi in - \frac{1}{x+\frac{k}{2\pi in}}.
	2\pi in\left(
		1- \frac{1}{2\pi inx+k+2}
	\right).
\end{align}

\begin{figure}[t]
	\centering
	\includegraphics[width=0.6\textwidth]{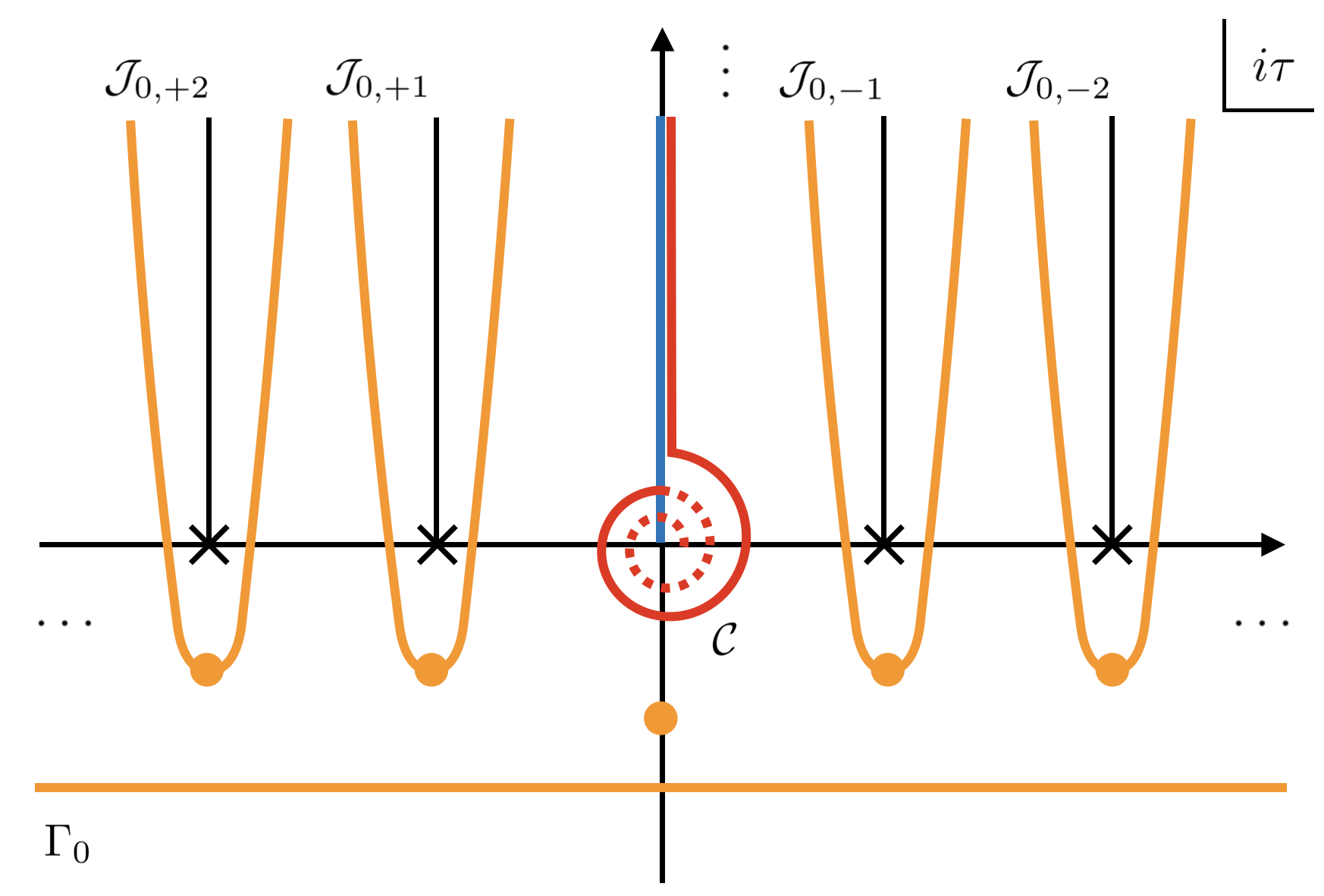}
	\caption{
		Thimble structure of the Hurwitz zeta function.
		The blue line is the original formal integration contour.
		The red spiral contour is the redefined integration contour $\mathcal{C}$ to analytically continue the Hurwitz zeta function. Its dotted part indicates that the contour flows into the $l(<0)$-th Riemann sheets.
		The black lines from cross symbols are logarithmic branch cuts.
		The integration contour $\mathcal{C}$ is deformed to a combination of thimbles (orange curves) associated with complex saddles (orange bulbs).
	}
	\label{fig:hurwitz_saddles}
\end{figure}

The thimbles, or the steepest descents, associated with these saddles are
\begin{align}
	\dv{\tau(s)}{s} = \overline{\pdv{S(\tau)}{\tau}}.
\end{align}
The saddles and thimbles are illustrated in Fig.~\ref{fig:hurwitz_saddles}.
%\begin{align}
%	\mathcal{J}_{0,-2} \; \mathcal{J}_{0,-1} \; \mathcal{J}_{0,+1} \; \mathcal{J}_{0,+2} \;
%	\Gamma_0 \\
%	\mathcal{C}
%\end{align}
On each $l(\leq0)$-th Riemann sheet, we have thimbles associated with complex saddles at $t\sim 2\pi in$.
We denote them by $\mathcal{J}_{l,n}$.
Also, we denote a horizontal line by $\Gamma_l$.
The original integration contour $\mathcal{C}$ is deformed to a combination of these curves.
\begin{align}
	\mathcal{C}
	\rightarrow
	\sum_{l=-\infty}^0
	\left[
	\sum_{n\neq0} (-1)^{\sigma_n}\mathcal{J}_{l,n}
	+ \Gamma_l
	\right]
\end{align}
where $(-1)^{\sigma_n}=1$ is the intersection number.
Note that the real saddle does not contribute to the integral.

At a complex saddle $t=t_n$ on the $l$-th Riemann sheet,
the action value is
\begin{align}
	e^{S(t_n)-2\pi i(k+1)l}
	&\sim
	%\left( -x-\frac{k}{2\pi in} \right)
	%%e^{ \frac{1}{1+\frac{k}{2\pi in x}} }
	%\left( \frac{1}{2\pi in} \right)^{k+1}
	%e^{-2\pi inx - 2\pi i(k+1)l}.
	\left( \frac{1}{2\pi in} \right)^{k+1}
	%\left( 1+\frac{k+1}{2\pi inx+k+2} \right)
	\frac{ e^{-2\pi in x-2\pi i(k+1)l} }{ 1-e^{ +\frac{2\pi in}{2\pi inx+k+2} } }.
\end{align}
The fluctuation part, or the second derivative, is
\begin{align}
	S''(t)
	&= \frac{e^t}{(e^t-1)^2} + \frac{k+1}{t^2},
\end{align}
thus
\begin{align}
	S''(t_n)
	&\sim
	%\left( -x-\frac{k}{2\pi n} \right)^2
	%+ \frac{k}{(2\pi in)^2}
	\left( -\frac{2\pi inx+k+2}{2\pi in} \right)^2.
\end{align}
Combining all contributions from the thimbles $\mathcal{J}_{l,n}$, we finally obtain
\begin{align}
	\label{eq:hzeta_fourier_saddle}
	\left. \zeta(-k,x) \right|_{\mathcal{J}}
	&\sim
	\frac{1}{\Gamma(-k)}
	\sum_{l=-\infty}^{0} \sum_{n\neq0}
	(-1)^{\sigma_{n}}
	\sqrt{ \frac{2\pi}{-S''(t_n)} } \: e^{S(t_n) -2\pi i(k+1)l} \notag\\
	&\sim
	k! %\textcolor{red}{\frac{1}{\sqrt{2\pi}}?}
	\sum_{n\neq0}
	%\frac{e^{\frac{1}{1-\frac{k}{2\pi inx}}}}{ \sqrt{1+\frac{k}{(k-2\pi inx)^2}} }
	\left( \frac{1}{2\pi in} \right)^{k+1} \: e^{2\pi in x}.
\end{align}
Here we used the gamma function reflection formula
to obtain $k!$ factor from $1/\Gamma(-k)$ and $e^{-2\pi i(k+1)l}$
and rewrote the label $n\rightarrow -n$.
This agrees with the oscillatory part of \eqref{eq:hzeta_fourier_poly}.
Another contribution from the horizontal line $\Gamma_l$ gives the polynomial part of \eqref{eq:hzeta_fourier_poly}.

%---------------------------------
\subsection{Special case $N=2$}
%---------------------------------

\subsubsection{Integer argument}
We can generalize what we found in the Hurwitz zeta function straightforwardly.
Suppose that $s=-k$ be non-positive integers.
In this case, the Barnes multiple zeta function is expressed as\cite{RUIJSENAARS2000107}
\begin{align}
	\label{eq:bzeta_bernoulli}
	\zeta_N(-k,x; \mathbf{a})
	&= \frac{ (-1)^N k! }{ (k+N)! }\:
	B_{k+N}(x;\mathbf{a}), \quad
	k=0,1,2,\dots
\end{align}
Here $B_{n}(x;\mathbf{a})$ is called the Barnes multiple Bernoulli polynomial.
It is defined by the following generating function
\begin{align}
	\label{eq:bmbernoulli}
	\frac{ t^Ne^{xt} }{ (e^{a_0t}-1) \cdots (e^{a_{N-1}t}-1) }
	&= \sum_{k=0}^{\infty} B_{k}(x;\mathbf{a}) \frac{t^k}{k!}.
\end{align}
%The expression \eqref{eq:bzeta_bernoulli} can be derived by deforming the Hankel contour $\mathcal{H}_0$ in \eqref{eq:bzeta_hankel} to a vertical line, and by collecting contributions from the poles at $t=2\pi in\mathbb{Z}_{\neq0}$.

Similarly to the Hurwitz zeta function,
we can show that the Barnes multiple zeta function has oscillatory part.
%\mycomm{no cancellation between remaining terms?}
In the following, we will explicitly show it in the case of $N=2,\mathbf{a}=(a_1,a_2)=(1/4,1)$.
From the definition of the Barnes multiple zeta function, we can shift the variable $x$ by
\begin{align}
	\label{eq:bzeta_ex_shift}
	\zeta_2(s,x;1/4,1)
	&= \zeta_2(s,x-m;1/4,1)
	- \sum_{n=1}^{m} \zeta_1(s,x-n;1/4),
\end{align}
where $m$ is an integer such that $x-1\leq m <x$.
Now we assume that $0<x<a_1+\cdots+a_n$.
From the definition of the Barnes multiple Bernoulli polynomial, we find
\begin{align}
	\label{eq:bmbernoulli_fourier}
	B_{k+N}(x;\mathbf{a})
	&= \oint_{C_0} \frac{\dd{t}}{2\pi i} \frac{(k+N)!}{t^{k+N+1}}
	\frac{ t^Ne^{xt} }{ (e^{a_1t}-1) \cdots (e^{a_{N}t}-1) } \notag\\
	&= -(k+N)! \sum_{\pm}
	\oint_{C_{\pm}} \frac{\dd{t}}{2\pi i}
	\frac{ t^{-k-1} e^{xt} }{ (e^{a_1t}-1) \cdots (e^{a_{N}t}-1) }.
\end{align}
Here the contours $C_{\pm}$ wraps the poles along the positive/negative imaginary axis.
The expression \eqref{eq:bmbernoulli_fourier} reduces to
\begin{align}
	B_{k+2}(x;1/4,1)
	&= -(k+2)! \sum_{n\neq0}
	\oint_{C_n} \frac{\dd{t}}{2\pi i}
	\frac{ t^{-k-1} e^{xt} }{ (e^{t/4}-1) (e^t-1) }.
\end{align}

Their poles at $t=2\pi i n,\: n\neq0$ contribute to the integral.
For $n\not\equiv0 \pmod{4}$, the poles are simple poles.
Their residues are
\begin{subequations}
	\begin{align}
		\left.
		(t-2\pi in)\frac{ t^{-k-1} e^{xt} }{ (e^{t/4}-1) (e^t-1) }
		\right|_{t=2\pi in}
		&= \frac{1}{e^{2\pi in/4}-1}
		\left( \frac{1}{2\pi in} \right)^{k+1}
		e^{2\pi in x}, \quad
		n\not\equiv0\pmod4.
	\end{align}
	For $n\equiv0 \pmod{4}$, the poles are 2nd order.
	Denoting $n=4m$, the residues are
	\begin{align}
		&\left.
		\frac{1}{1!} \dv[1]{t}
		(t-2\pi in)^2 \frac{ t^{-k-1} e^{xt} }{ (e^{t/4}-1) (e^t-1) }
		\right|_{t=2\pi in} \notag\\
		&= \left( 4x-\frac{5}{2} \right)
		\left( \frac{1/4}{2\pi im} \right)^{k+1}
		e^{2\pi im \cdot 4x}
		- 4(k+1) \left( \frac{1/4}{2\pi im} \right)^{k+2}
		e^{ 2\pi im \cdot 4x } \notag\\
		&=
		\left( -\frac{k+1}{2\pi im} + (4x-5/2) \right)
		\left( \frac{1/4}{2\pi im} \right)^{k+1}
		e^{2\pi im \cdot 4x},
		\quad m\neq0.
	\end{align}
\end{subequations}
Combining all of these residues, we find that
\begin{align}
	\label{eq:bmbernoulli_fourier_explicit}
	&B_{k+2}(x;1/4,1) \notag\\
	%&= -(k+2)! \left[
	%\sum_{n\not\equiv0\: \text{mod}4}
	%\frac{ 1 }{ e^{2\pi in/4}-1 }
	%\left( \frac{1}{2\pi in} \right)^{k+1}
	%e^{2\pi inx} \right.\notag\\
	%&\hspace{-20mm} \left.
	%+\sum_{n\neq0}
	%\left(
	%\left(4x-\frac{5}{2}\right)
	%\left( \frac{1/4}{2\pi in} \right)^{k+1}
	%e^{2\pi in \cdot 4x}
	%-4(k+1)
	%\left( \frac{1/4}{2\pi in} \right)^{k+2}
	%e^{2\pi in \cdot 4x}
	%\right)	
	%\right].
	&=
	-(k+2)! \left[
	\sum_{n\not\equiv0\: \text{mod}4}
	\frac{ 1 }{ e^{2\pi in/4}-1 }
	\left( \frac{1}{2\pi in} \right)^{k+1}
	e^{2\pi inx}
	+ \sum_{n\neq0}
	\left( -\frac{k+1}{2\pi in} + (4x-5/2) \right)
	\left( \frac{1/4}{2\pi in} \right)^{k+1}
	e^{2\pi in \cdot 4x}
	\right]
\end{align}
Substituting \eqref{eq:bmbernoulli_fourier_explicit} into \eqref{eq:bzeta_bernoulli}, we obtain
	\begin{align}
		\label{eq:bzeta_fourier}
		&\zeta_2(-k,x;1/4,1) \notag\\
		%&= -k! \left[
		%\sum_{n\not\equiv0 \: \text{mod}4}
		%\frac{ 1 }{ e^{2\pi in/4}-1 }
		%\left( \frac{1}{2\pi in} \right)^{k+1}
		%e^{2\pi inx} \right.\notag\\
		%&\hspace{-20mm} \left.
		%+\sum_{n\neq0}
		%\left(
		%\left(4x-\frac{5}{2}\right)
		%\left( \frac{1/4}{2\pi in} \right)^{k+1}
		%e^{2\pi in \cdot 4x}
		%-4(k+1)
		%\left( \frac{1/4}{2\pi in} \right)^{k+2}
		%e^{2\pi in \cdot 4x}
		%\right)	
		%\right].
		&= -k!
		\left[
		\sum_{n\not\equiv0\: \text{mod}4}
		\frac{ 1 }{ e^{2\pi in/4}-1 }
		\left( \frac{1}{2\pi in} \right)^{k+1}
		e^{2\pi inx}
		+ \sum_{n\neq0}
		\left( -\frac{k+1}{2\pi in} + (4x-5/2) \right)
		\left( \frac{1/4}{2\pi in} \right)^{k+1}
		e^{2\pi in \cdot 4x}
		\right]
	\end{align}
Now, it is clear that the Barnes multiple zeta function with a non-positive integer argument consists of periodic terms.
In the case of $\mathbf{a}=(a_1,a_2)=(1/4,1)$, the Barnes multiple zeta function $\zeta_2(-k,x;1/4,1)$ consists of two Fourier modes:
\begin{subequations}
	\begin{align}
		\label{eq:barnes_ex_n2_low}
		&\left\{ \left( \frac{1}{2\pi in} \right)^{k+1} e^{2\pi i nx} \right\}_{n\not\equiv0\: \text{mod}4} \\
		\label{eq:barnes_ex_n2_high}
		&\left\{ \left( \frac{1/4}{2\pi in} \right)^{k+1} e^{2\pi in \cdot 4x} \right\}_{n\neq0}
	\end{align}
\end{subequations}
The size of the oscillatory part is $\sim (a_i/2\pi)^{k+1} k!$
while the remaining part is $\sim x^{k+1}$.
These two parts become compatible when $k\sim x$.
The oscillatory part dominates when $x$ is small $x\ll k$
while the second part dominates when $x$ is large $x\gg k$.

%In the asymptotic region $k\gg1$, the Fourier expansion is dominated by the modes with $n=\pm1$.
%The Barnes multiple zeta function reduces to
%\begin{align}
%	\zeta_2(-k,x;1/4,1)
%	\sim
%	-k! \left( \frac{1}{2\pi i} \right)^{k+1}
%	\sum_{\pm}
%	\left[
%	c_{1,\pm} e^{\pm 2\pi ix}
%	+ c_{4,\pm} \frac{ k^1 e^{\pm2\pi i \cdot 4x }}{ 4^k }
%	\right],
%\end{align}
%where coefficients $c_{1,\pm}, c_{4,\pm}$ are some constants of order unity.
%The high frequency mode $e^{\pm 2\pi i \cdot 4x}$ is suppressed by a factor $\sim k^1/4^k$.

\begin{figure}[t]
	\centering
	\includegraphics[width=0.45\textwidth]{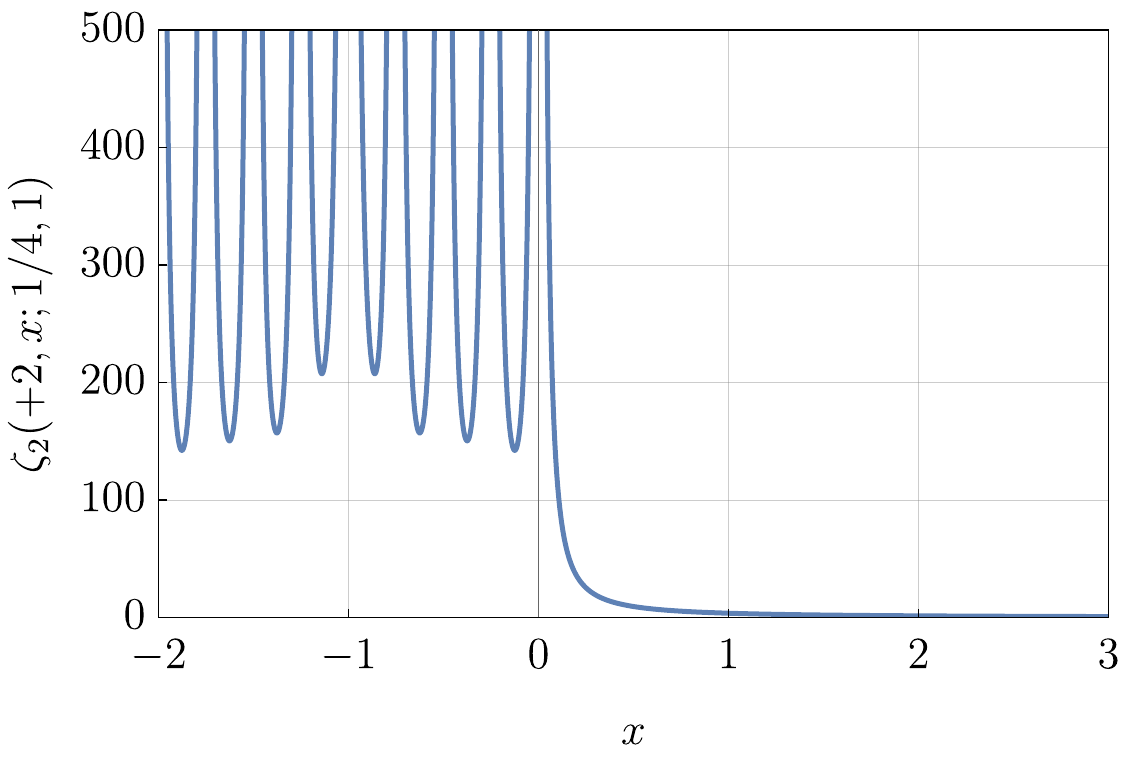} \hspace{5mm}
	\includegraphics[width=0.45\textwidth]{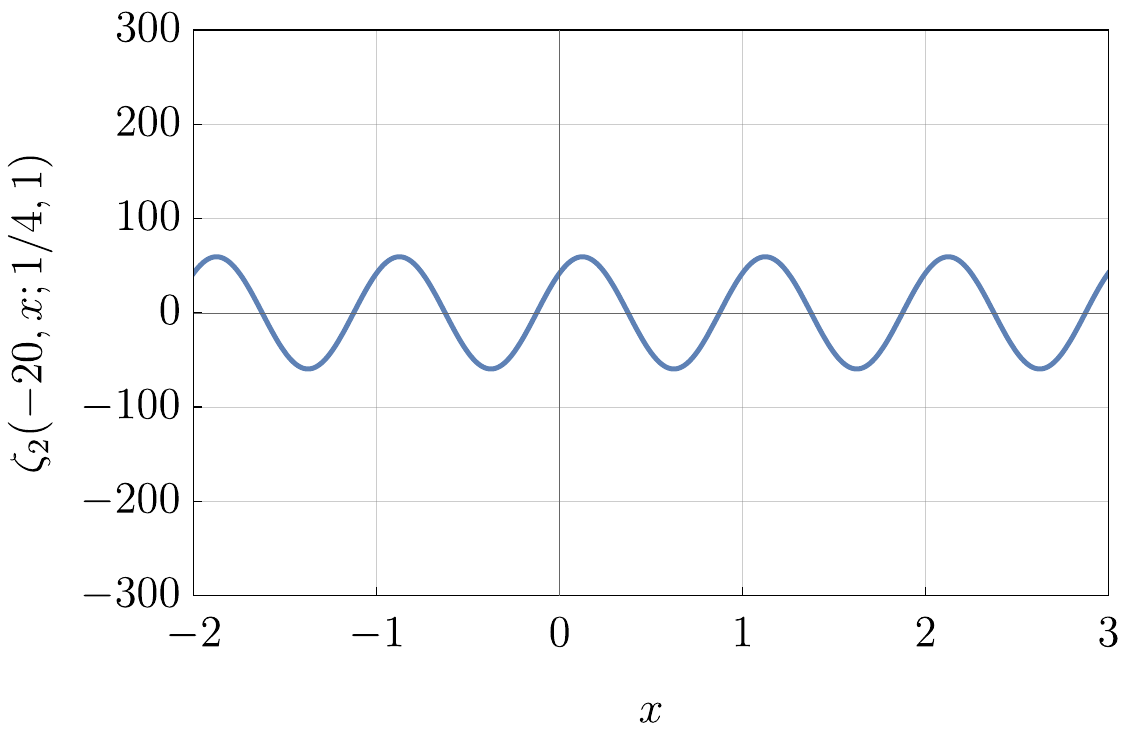}
	\caption{
		Plots of the Barnes zeta $\zeta_2(-k,x; 1/4,1)$ in typical parameter regions.
		The left figure is for $s=-k=2$
		while the right figure is for $s=-k=-20$.
		These blue curves are all truncation of infinite series expression.
		The left one is truncated at the pole $x=-2$
		while the right one is truncated at the first term of the high frequency mode $n_2=1$.
	}
	\label{fig:barnes_plot}
\end{figure}

The last expression	\eqref{eq:bzeta_fourier} converges only for $\Re s=-\Re k<0$.
However, we can analytically continue it.
From the definition of the Barnes multiple zeta function,
there are poles at $x\in a_1\integernum_{\leq}+a_2\integernum_{\leq}$ if $\Re s=-\Re k >1$.
Some example plots are depicted in Fig.~\ref{fig:barnes_plot}.

\begin{figure}[t]
	\centering
	\includegraphics[width=0.45\textwidth]{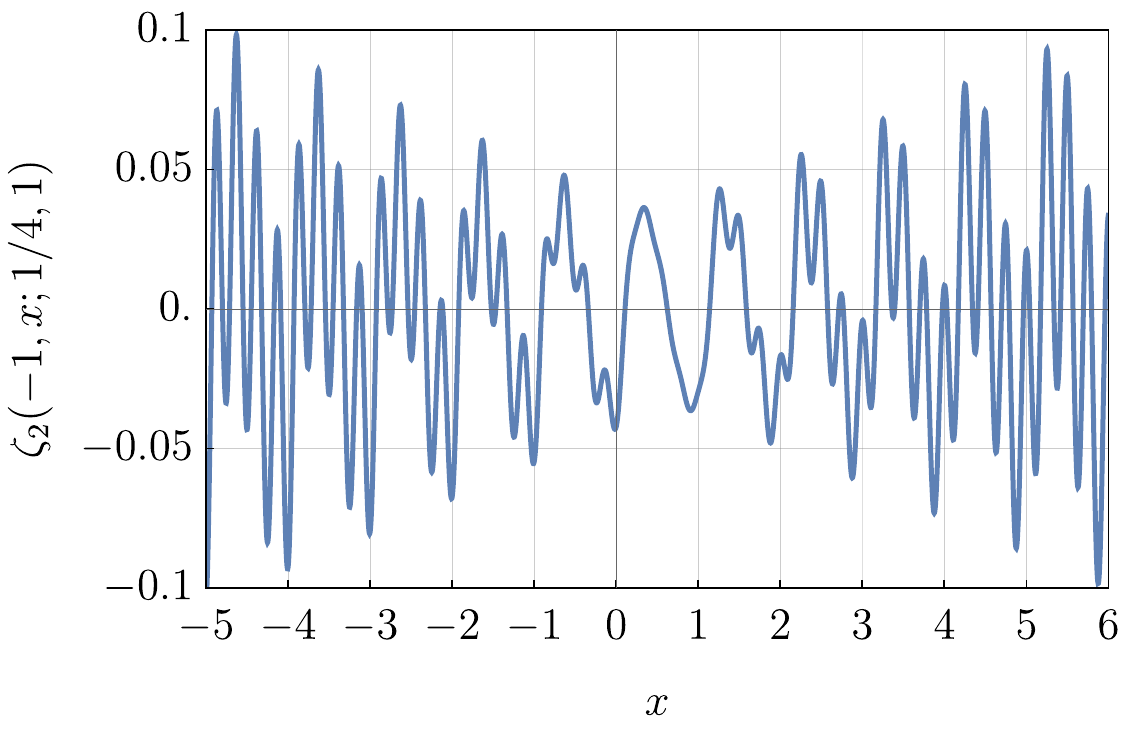}
	\caption{
		Plots of the Barnes zeta $\zeta_2(-k,x; 1/4,1)$ for smaller $s=-k=-1$.
		Truncation level is the same as Fig.~\ref{fig:barnes_plot}.
		In this case,
		since the size of the high frequency mode is larger than Fig.~\ref{fig:barnes_plot},
		the plot becomes more erratic.
	}
	\label{fig:barnes_plot3}
\end{figure}

As long as $k$ is sufficiently large, the lowest frequency mode \eqref{eq:barnes_ex_n2_low} gives the dominant contribution.
If $k$ becomes smaller, the higher frequency mode \eqref{eq:barnes_ex_n2_high} becomes compatible with other contributions.
Fig.~\ref{fig:barnes_plot3}.

\myspace
The residues can also be computed in a different way.
The following computation will be helpful for finding asymptotic expansion in terms of $k$.
The integrand is expanded around a simple pole as
\begin{align}
	\frac{ t^{-k-1} e^{xt} }{ (e^{a_0t}-1) (e^{a_1t}-1) }
	= \sum_{\ell=0}^{\infty}
	c_{\ell}
	(t-2\pi ina_1^{-1})^{-1+\ell}.
\end{align}
where $n\not\equiv0 \pmod{4}$.
Changing the variable as $w = t-2\pi ina_1^{-1}$, we obtain
\begin{align}
	&\sum_{\ell=0}^{\infty}
	c_{\ell} w^{-1+\ell} \notag\\
	&= \frac{ (w+2\pi ina_1^{-1})^{-k-1} e^{xw} }
	{ (e^{2\pi ina_0/a_1}e^{a_0w}-1) (e^{a_1w}-1) }\:
	e^{ 2\pi inx/a_1 } \notag\\
	&= \sum_{r=0}^{\infty}
	(2\pi ina_1^{-1})^{-k-1}
	\frac{ \Gamma(-k) }{ \Gamma(r+1)\Gamma(-k-r) }
	\left( \frac{w}{2\pi ina_1^{-1}} \right)^{r}
	w^{-1}
	\frac{ w^1 e^{xw} }{ e^{a_1w}-1 }
	\frac{1}{ e^{2\pi ina_0/a_1}e^{a_0w}-1 }\:
	e^{ 2\pi inx/a_1 }.
\end{align}
To proceed, we introduce a polynomial $\tilde{B}_{\ell}$ such that
\begin{align}
	\frac{ w^1 e^{xw} }{ e^{a_1w}-1 }
	\frac{1}{ e^{2\pi ina_0/a_1}e^{a_0w}-1 }
	&= \sum_{\ell=0}^{\infty}
	\tilde{B}_{\ell}(x;a_1;a_0)
	\frac{w^{\ell}}{\ell!}.
\end{align}
Particularly when $\ell=0$,
\begin{align}
	\tilde{B}_{0}(x;a_1;a_0)
	= B_0(x;a_1)\frac{1}{ e^{2\pi ina_0/a_1}-1 }.
\end{align}
Rewriting the labels as $\ell+r\rightarrow\ell, \ell\rightarrow r$,
\begin{align}
	&\sum_{\ell=0}^{\infty} c_{\ell} w^{-1-\ell} \notag\\
	&= \sum_{\ell=0}^{\infty} \sum_{r=0}^{\ell}
	\frac{ (2\pi ina_1^{-1})^{-k-1-\ell+r} \Gamma(-k) }
	{ \Gamma(\ell-r+1)\Gamma(-k-\ell+r) }
	\frac{\tilde{B}_{r}(x;a_1;a_0)}{r!}w^{-1+\ell}\:
	e^{2\pi inx/a_1} \notag\\
	&= \sum_{\ell=0}^{\infty} \sum_{r=0}^{\ell}
	\frac{ (2\pi ina_1^{-1})^{-k-1-\ell+r} (-1)^{\ell-r} \Gamma(k+1+\ell-r) }
	{ \Gamma(\ell-r+1)\Gamma(k+1) }
	\frac{\tilde{B}_{r}(x;a_1;a_0)}{r!}\:
	e^{2\pi inx/a_1}.
\end{align}
Thus
\begin{align}
	c_{\ell}
	&= \sum_{r=0}^{\ell}
	\frac{ (-1)^{\ell-r} \Gamma(k+1+\ell-r) }
	{ \Gamma(\ell-r+1)\Gamma(k+1) }
	\frac{\tilde{B}_r(x;a_1;a_0)}{r!}
	\left( \frac{a_1}{2\pi in} \right)^{k+1+\ell-r}
	e^{2\pi inx/a_1}.
\end{align}
Finally, we find that the residue around a simple pole is
\begin{align}
	c_0
	&= B_0(x;a_1)
	\frac{1}{ e^{2\pi ina_0/a_1}-1 }
	\left( \frac{a_1}{2\pi in} \right)^{k+1}
	e^{2\pi inx/a_1} \notag\\
	&= \frac{1}{ e^{2\pi in/4}-1 }
	\left( \frac{1}{2\pi in} \right)^{k+1}
	e^{2\pi inx}.
\end{align}
This agrees with the previous result.

Similarly, the residue around a pole of 2nd order is systematically computed as follows.
The integrand is expanded around a pole of 2nd order as
\begin{align}
	\frac{ t^{-k-1} e^{xt} }{ (e^{a_0t}-1) (e^{a_1t}-1) }
	= \sum_{\ell=0}^{\infty}
	d_{\ell}
	(t-2\pi ina_0^{-1})^{-2+\ell},
\end{align}
where $n$ is any non-zero integers $n\neq 0$.
Changing the variable as $w = t-2\pi ina_0^{-1}$, we obtain
\begin{align}
	&\sum_{\ell=0}^{\infty}
	d_{\ell} w^{-2+\ell} \notag\\
	&= \frac{ (w+2\pi ina_0^{-1})^{-k-1} e^{xw} }
	{ (e^{a_0w}-1) (e^{a_1w}-1) }\:
	e^{ 2\pi inx/a_0 } \notag\\
	&= \sum_{r=0}^{\infty}
	(2\pi ina_0^{-1})^{-k-1}
	\frac{ \Gamma(-k) }{ \Gamma(r+1)\Gamma(-k-r) }
	\left( \frac{w}{2\pi ina_0^{-1}} \right)^{r}
	w^{-2}
	\frac{ w^2 e^{xw} }{ (e^{a_0w}-1) (e^{a_1w}-1) }\:
	e^{ 2\pi inx/a_0 }.
\end{align}
In this case, the same Barnes multiple Bernoulli polynomial $B_{\ell}$ reappears.
Rewriting the labels as $\ell+r\rightarrow\ell, \ell\rightarrow r$,
\begin{align}
	&\sum_{\ell=0}^{\infty} d_{\ell} w^{-2-\ell} \notag\\
	&= \sum_{\ell=0}^{\infty} \sum_{r=0}^{\ell}
	\frac{ (2\pi ina_0^{-1})^{-k-1-\ell+r} \Gamma(-k) }
	{ \Gamma(\ell-r+1)\Gamma(-k-\ell+r) }
	\frac{B_{r}(x;a_0,a_1)}{r!}w^{-2+\ell}\:
	e^{2\pi inx/a_0} \notag\\
	&= \sum_{\ell=0}^{\infty} \sum_{r=0}^{\ell}
	\frac{ (2\pi ina_0^{-1})^{-k-1-\ell+r} (-1)^{\ell-r} \Gamma(k+1+\ell-r) }
	{ \Gamma(\ell-r+1)\Gamma(k+1) }
	\frac{B_{r}(x;a_0,a_1)}{r!}\:
	e^{2\pi inx/a_0}.
\end{align}
Thus
\begin{align}
	d_{\ell}
	&= \sum_{r=0}^{\ell}
	\frac{ (-1)^{\ell-r} \Gamma(k+1+\ell-r) }
	{ \Gamma(\ell-r+1)\Gamma(k+1) }
	\frac{B_r(x;a_0,a_1)}{r!}
	\left( \frac{a_0}{2\pi in} \right)^{k+1+\ell-r}
	e^{2\pi inx/a_0}.
\end{align}
Finally, we find that the residue around a pole of 2nd order is
\begin{align}
	d_1
	&= B_1(x;a_0,a_1)
	\left( \frac{a_0}{2\pi in} \right)^{k+1}
	e^{2\pi inx/a_0}
	- (k+1) B_0(x;a_0,a_1)
	\left( \frac{a_0}{2\pi in} \right)^{k+2}
	e^{2\pi inx/a_0} \notag\\
	&= \left(4x-\frac{5}{2}\right)
	\left( \frac{1/4}{2\pi in} \right)^{k+1}
	e^{2\pi in \cdot 4x}
	-4(k+1)
	\left( \frac{1/4}{2\pi in} \right)^{k+2}
	e^{2\pi in \cdot 4x}.
\end{align}
This agrees with the previous result.

\subsubsection{Saddle point approximation}
Similarly to the Hurwitz zeta function,
saddle point approximation is also useful to obtain Fourier expansion\eqref{eq:bzeta_fourier}.
Let us start from the integral representation \eqref{eq:bzeta_int} of the Barnes multiple zeta function.
When $s=-k\leq0$, the integral diverges around $t=0$.
In order to cure this problem, one may use the Hankel contour.
Instead we use a spiral integration contour following \cite{Witten:2013pra},
\begin{align}
	\zeta_N(-k,x; \mathbf{a})
	&= \frac{1}{\Gamma(-k-i\delta)}
	\int_{\mathcal{C}_{\delta}} \dd{t}\: t^{-k-i\delta-1}
	\frac{e^{-xt}}{ (1-e^{-a_1t}) \cdots (1-e^{-a_{N}t}) }
\end{align}
Here $\delta>0$ is a small regularization parameter to shift $k$,
which is an analog of the small constant of the Feynman prescription.
In the subsequent discussion,
we will omit the small regularization parameter $\delta$ as long as there is no risk of confusion.

If we write the integral in the form of
\begin{align}
	\zeta_N(-k,x; \mathbf{a})
	&=
	\frac{1}{\Gamma(-k)}
	\int_{\mathcal{C}} \dd{t}\: e^{S(t)},
\end{align}
the ``action'' is
\begin{align}
	S(t)
	&= -xt -\sum_{i=1}^{N}\ln(1-e^{-a_it}) - (k+1)\ln t.
\end{align}
Its saddle points are determined by an equation
\begin{align}
	0 &= S'(t)
	= -x -\sum_{i=1}^{N}\frac{a_i}{e^{a_it}-1} -\frac{k+1}{t}.
\end{align}
In the following we assume $N=2,\: \mathbf{a}=(a_1,a_2)=(1/4,1)$ for simplicity.
We can generalize it straightforwardly.
We also $k, x\gg 1$ so that saddle point approximation is valid.

We can find a real saddle by setting an ansatz $t = -1/\epsilon$.
The saddle point equation becomes
\begin{align}
	0 \sim
	%-\frac{x}{\epsilon} + N + (k+1).
	x -a_1-a_2-\epsilon(k+1)
\end{align}
Its solution is
\begin{align}
	\label{eq:bzeta_saddle_real}
	t_0 \sim
	-\frac{k+1}{x-a_1-a_2}.
\end{align}

Also we can find infinitely many complex saddles.
By setting an ansatz
\begin{align}
	a_2t=\epsilon+2\pi in, \quad
	n \in \integernum_{\neq0}, \;
	n \equiv 0 \: (\text{mod\:} a_2/a_1)
\end{align}
the saddle point equation becomes
\begin{align}
	0 \sim
	%a_2^{-1}x(2\pi in +\epsilon)
	%+ 2\frac{\epsilon+2\pi in}{\epsilon} + (k+1).
	x + \frac{2a_2}{\epsilon} + \frac{ (k+1)a_2 }{2\pi in+\epsilon}
\end{align}
Its solutions are
\begin{align}
	\label{eq:bzeta_saddle_cpx4-2}
	a_2t_n
	&\sim
	%2\pi in
	%- \frac{2}{ a_2^{-1}x + \frac{k}{2\pi in} }, \quad
	%n \in 4\mathbb{Z}_{\neq0}.
	2\pi in \left(
	1 - \frac{ 2a_2 }{ 2\pi inx +(k+1)a_2+2a_2 }
	\right), \quad
	n \in \integernum_{\neq0}, \;
	n \equiv 0 \: (\text{mod\:} a_2/a_1)
	%n \in \integernum_{\neq0}, \;
	%n \equiv0 \; (\text{mod} \: a_2/a_1), 
\end{align}

\begin{figure}[t]
	\centering
	\includegraphics[width=0.6\textwidth]{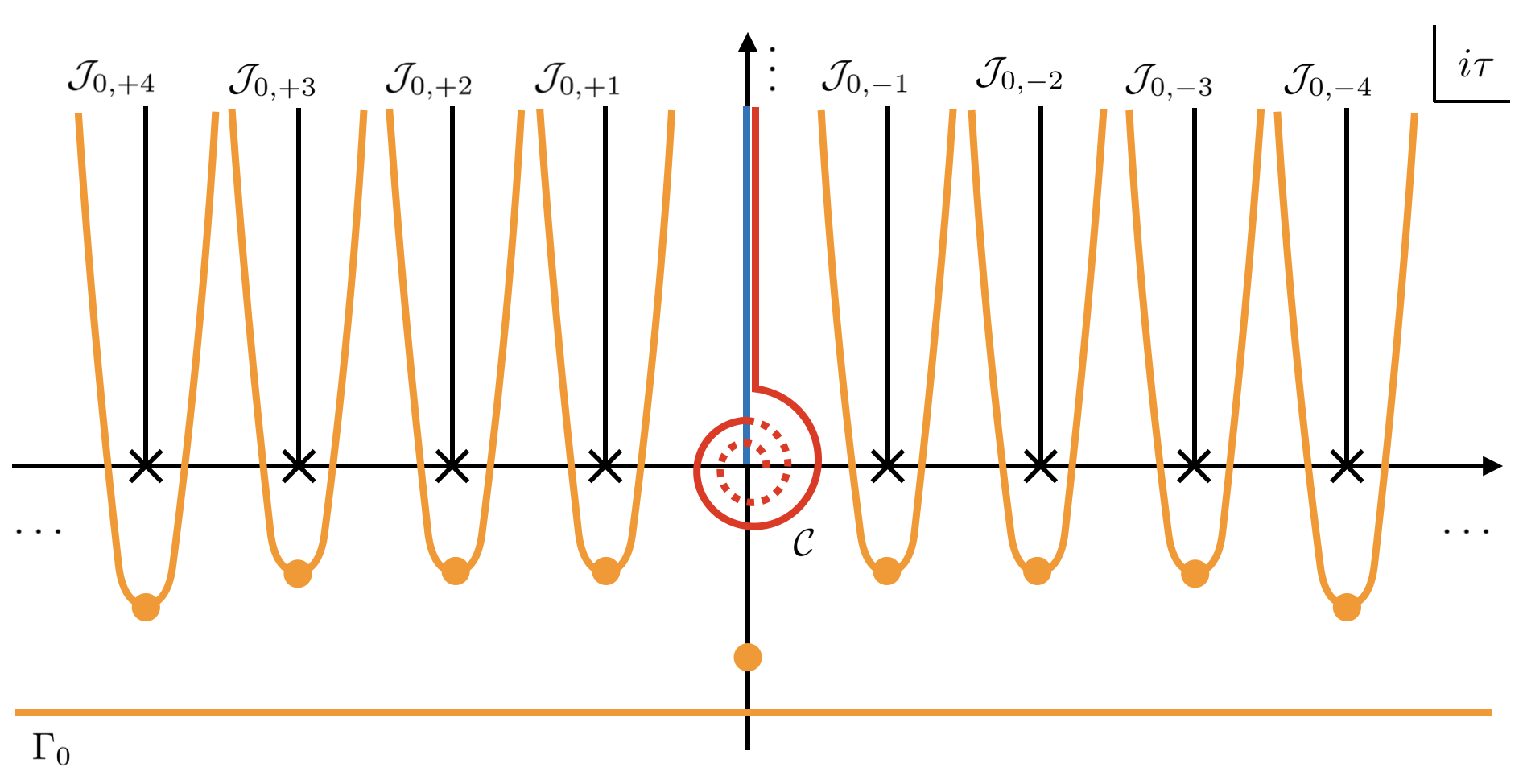}
	\caption{
		Thimble structure of the Barnes multiple zeta function with $(a_1,a_2)=(1/4,1)$.
		The blue line is the original formal integration contour.
		The red spiral contour is the redefined integration contour $\mathcal{C}$ to analytically continue the Hurwitz zeta function. Its dotted part indicates that the contour flows into the $l(<0)$-th Riemann sheets.
		The black lines from cross symbols are logarithmic branch cuts.
		The integration contour $\mathcal{C}$ is deformed to a combination of thimbles (orange curves) associated with complex saddles (orange bulbs).
		Note that there are two series of thimbles with $n \equiv, \not\equiv 0 \; (\text{mod} 4)$.
	}
	\label{fig:barnes_saddles}
\end{figure}

There is another series of complex saddles.
By setting an ansatz
\begin{align}
	a_2t=\epsilon+2\pi in, \quad
	n \in \integernum_{\neq0}, \;
	n \not\equiv 0 \: (\text{mod\:} a_2/a_1)
\end{align}
we have
\begin{align}
	0 \sim
	%a_2^{-1}x (2\pi in + \epsilon)
	%+ \frac{ (a_1/a_2)(\epsilon+2\pi in) }{ e^{(a_1/a_2)(\epsilon+2\pi in)}-1 }
	%+ \frac{\epsilon+2\pi in}{\epsilon} + (k+1).
	x + \frac{a_1}{e^{(a_1/a_2)(2\pi in +\epsilon)}-1}
	+ \frac{a_2}{\epsilon} + \frac{(k+1)a_2}{2\pi in+\epsilon}
\end{align}
Its solutions are
\begin{align}
	\label{eq:bzeta_saddle_cpx1}
	a_2t_n \sim
	%2\pi in
	%-\frac{1}{ a_2^{-1}x + \frac{k}{2\pi in} }, \quad
	%n \not\in 4\mathbb{Z}.
	2\pi in \left(
		1 - \frac{ a_2 }
		{ 2\pi inx + (k+1)a_2 + a_2 + \frac{2\pi in a_1}{ e^{(a_1/a_2)2\pi in}-1 } }.
	\right), \quad
	n \in \integernum_{\neq0}, \;
	n \not\equiv 0 \: (\text{mod\:} a_2/a_1).
	%n \in \integernum_{\neq0}, \;
	%n \not\equiv 0 \; (\text{mod}\: a_2/a_1)
\end{align}
The saddles and thimbles are illustrated in Fig.~\ref{fig:barnes_saddles}.
%\begin{align}
%	\mathcal{J}_{0,-4} \; \mathcal{J}_{0,-3} \;
%	\mathcal{J}_{0,-2} \; \mathcal{J}_{0,-1} \; \mathcal{J}_{0,+1} \; \mathcal{J}_{0,+2} \;
%	\mathcal{J}_{0,+4} \; \mathcal{J}_{0,+3}  \\
%	\Gamma_0 \\
%	\mathcal{C}
%\end{align}
On each $l$-th Riemann sheet,
we have two types of thimbles associated with complex saddles $a_2t \sim 2\pi in$
where $n \equiv0, \not\equiv 0$.
We denote them by $\mathcal{J}_{l,n}$.
Also, we denote a horizontal line by $\Gamma_{l}$.
The original integration contour $\mathcal{C}$ is deformed to a combination of these cureves as
\begin{align}
	\mathcal{C}
	\rightarrow
	\sum_{l=-\infty}^{0} \left[
	\sum_{n\neq0, n\equiv0 \: \text{mod}4} (-1)^{\sigma_n}\mathcal{J}_{l,n}
	+\sum_{n\neq0, n\not\equiv0 \: \text{mod}4} (-1)^{\sigma_n}\mathcal{J}_{l,n}
	+\Gamma_{l}
	\right].
\end{align}
where $(-1)^{\sigma_n}$ is the intersection number.
Note that the real saddle does not contribute to the integral.

At the complex saddles $t=t_{n}, \: n\equiv0$ on the $l$-th Riemann sheet, the action value is
\begin{align}
	e^{S(t_n)-2\pi i(k+1)l}
	&=
	\left( \frac{1}{t} \right)^{k+1}
	\frac{ e^{ -xt } }{ (1-e^{-a_1t})(1-e^{-a_2t}) } \notag\\
	&\sim
	\left( \frac{1}{2\pi in} \right)^{k+1}
	\frac{ e^{ -2\pi inx } }
	{ \left(1-e^{ +\frac{2\pi in \cdot 2a_1}{2\pi inx +(k+1)a_2+2a_2} }\right) \left(1-e^{ +\frac{2\pi in \cdot 2a_2}{2\pi inx +(k+1)a_2+2a_2} }\right) }
\end{align}
The fluctuation part, or the second derivative, is
\begin{align}
	S''(t)
	&=
	\sum_{i=1}^{N} \frac{a_i^2e^{a_it}}{(e^{a_it}-1)^2}
	+ \frac{k+1}{t^2},
\end{align}
thus
\begin{align}
	S''(t_n)
	&\sim
	2a_2^2
	\left(
	-\frac{ 2\pi inx+(k+1)a_2+\cdots }{ 2\pi in\cdot 2a_2 }
	\right)^2.
\end{align}
%$t=t_{4n}, \: n\neq0$ on the $l$-th Riemann sheet,
%the action value is
%\begin{align}
%	e^{ S(t_{4n}) - 2\pi i(k+1)l }
%	&\sim
%	%\left( -x-\frac{k}{2\pi i\cdot 4n} \right)^2
%	%\left( \frac{1/4}{2\pi in} \right)^{k+1}
%	%e^{ -2\pi in \cdot 4x - 2\pi i(k+1)l }.
%	\left( \frac{a_1}{2\pi n} \right)^{k+1}
%	\frac{ e^{ -2\pi in x/a_1 - 2\pi i(k+1)l } }
%	{ \left(1-e^{+\frac{2\pi in\cdot2a_2}{2\pi inx/a_1+(k+1)a_2+\cdots}}\right)^2 }
%\end{align}

At a complex saddle $t=t_n,\: n\not\equiv0$ on the $l$-th Riemann sheet,
the action value is
\begin{align}
	e^{S(t_n)-2\pi i(k+1)l}
	&\sim
	%\frac{ e^{2\pi in/4} }{ e^{2\pi in/4}-1 }
	%\left( -x-\frac{k}{2\pi in} \right)
	%\left( \frac{1}{2\pi in} \right)^{k+1}
	%e^{-2\pi inx - 2\pi i(k+1)l}.
	\frac{ 1 }{ 1-e^{-(a_1/a_2)2\pi in} }
	\left( \frac{a_2}{2\pi in} \right)^{k+1}
	\frac{ e^{ -2\pi inx/a_2 - 2\pi i(k+1)l } }
	{ \left(1-e^{+\frac{ 2\pi in\cdot a_2 }{2\pi nx+(k+1)a_2+\cdots}} \right) }
\end{align}
Its fluctuation part is
\begin{align}
	S''(t_n)
	&\sim
	%\left( -x-\frac{k}{2\pi in} \right)^2.
	a_2^2 \left(
		- \frac{2\pi inx +(k+1)a_2+\cdots }{ 2\pi in \cdot a_2 }
	\right)^2.
\end{align}

We sum up all these contributions from complex saddles.
We rewrite $n\equiv0 \rightarrow 4n$.
Using the gamma function reflection formula,
we get $k!$ from $1/\Gamma(-k)$ and $e^{-2\pi i(k+1)m}$.
Finally, rewriting $n\rightarrow-n$, we obtain
\begin{align}
	\label{eq:bzeta_fourier_saddle}
	&\zeta(-k,x) \notag \\
	&\sim
	\frac{1}{\Gamma(-k)}
	\sum_{m=-\infty}^{0} \sum_{n\neq0}
	(-1)^{\sigma_{n}}
	\sqrt{ \frac{2\pi}{-S''(t_n)} } \: e^{S(t_n) -2\pi i(k+1)m} \notag\\
	&\sim
	-k!
	\left[
	\sum_{ n\not\equiv0 }
	\frac{1}{e^{+(a_1/a_2)2\pi in}-1}
	\left( \frac{a_2}{2\pi in} \right)^{k+1}
	e^{ 2\pi inx/a_2}
	+\sum_{n\neq0}
	\left( -\frac{k+1}{2\pi in} +4x \right)
	\left( \frac{a_1}{2\pi in} \right)^{k+1}
	e^{ 2\pi in x/a_1 }
	\right].
\end{align}
This agrees with previous result \eqref{eq:hzeta_fourier} in the asymptotic region.

%========================
\section{Evaluating Truncated String Amplitudes}
\label{sec:evaluate_amp}
%========================

In the previous section Sec.~\ref{sec:barnes},
we reviewed poles and zeros of the Barnes multiple zeta function in asymptotic regions.
Typical form of integral expressions were, for example
\begin{align}
	\int_0^{\infty} \dd{t} \:
	t^{-k-1}\:
	\frac{ e^{-xt} }
	{ (1-e^{-a_1t})(1-e^{-a_2t}) }.
\end{align}
In the asymptotic region,
saddles were found around poles $a_1t, a_2t \sim 2\pi in$ of the integrand
where factors $\ln(1-e^{-a_it})$ in the action becomes large.
From the action value at these saddles,
we find that the integral is expanded by the following oscillating modes
\begin{align}
	\label{eq:evaluate_factors}
	\left( \frac{a_1}{2\pi in} \right)^{k+1} e^{ 2\pi in x/a_1 }, \quad
	\left( \frac{a_2}{2\pi in} \right)^{k+1} e^{ 2\pi in x/a_2 }.
\end{align}

The same structure appears in our truncated models.
For example in \eqref{eq:ann_amp_n2},
\begin{align}
	\int_0^\infty \dd{\tau}\:
	\tau^{-d/2}\:
	\frac{ \exp 2\pi\tau \left[ \frac{x_1x_2}{x_+x_2}(-\alpha'p^2)+x_1+x_2 \right] }
	{ (1-e^{-2\pi\tau})^{D-2-2\alpha'p^2} (1-e^{-2\pi x_1\tau})^{+\alpha'p^2}(1-e^{-2\pi x_2\tau})^{+\alpha'p^2} }.
\end{align}
We constructed further simplified model
by ignoring polarizations and momentum dependence of interacting vertices.
For example in\eqref{eq:trunc_oneloop},
\begin{align}
	\int_0^{\infty} \dd{\tau} \tau^{-d/2+2-1}\:
	\frac{ \exp 2\pi\tau \left[ \frac{x_1x_2}{x_1+x_2}(-\alpha'p^2)+x_1+x_2 \right]}
	{ (1-e^{-2\pi x_1\tau})(1-e^{-2\pi x_2\tau}) }.
\end{align}
Here the power of the denominator of \eqref{eq:trunc_oneloop} became a constant
in contrast to \eqref{eq:ann_amp_n2}.
Although these truncated models break modular invariance and momentum dependence of string interaction vertex,
the location of the poles of integrand is identical with the Barnes multiple zeta function.
In the following part Sec.~\ref{sec:evaluate_amp},
we will show that our truncated amplitudes are expanded by oscillatory factors in the same form as \eqref{eq:evaluate_factors}.

%--------------------------------
\subsection{One-loop diagram: Further truncated}
%--------------------------------
Let us start from the simplest one.
We further simplify our one-loop truncated model \eqref{eq:trunc_oneloop}
by fixing $M_1^2 = M_2^2$.
The amplitude becomes
\begin{align}
	\label{eq:trunc_oneloop_fixed}
	\tilde{\mathcal{A}}_1(p^2)
	&=
	\sum_{\alpha'M^2}
	\int_0^1 \dd{x_1}\dd{x_2} \delta(x_1+x_2-1)
	\left( \frac{1}{x_1+x_2} \right)^{d/2} \notag\\
	&\qquad\times
	\frac{1}{\Gamma(-d/2+2)}
	\int_0^{\infty} \dd{\tau} \tau^{-d/2+2-1}
	e^{ -2\pi\alpha'\tau\left( \frac{x_1x_2}{x_1+x_2}p^2+(x_1+x_2)M^2 \right) } \notag\\
	&=
	\int_0^1 \dd{x_1}\dd{x_2} \delta(x_1+x_2-1)
	\left( \frac{1}{x_1+x_2} \right)^{d/2} \notag\\
	&\qquad\times
	\frac{1}{\Gamma(-d/2+2)}
	\int_0^{\infty} \dd{\tau} \tau^{-d/2+2-1}\:
	\frac{ e^{ +2\pi\tau\left( \frac{x_1x_2}{x_1+x_2}(-\alpha'p^2)+x_1+x_2 \right) } }
	{ (1-e^{-2\pi (x_1+x_2)\tau}) } \notag\\
	&=
	\int_0^1 \dd{x} \:
	\frac{1}{\Gamma(-k)}
	\int_0^{\infty} \dd{\tau}\:
	\tau^{-k-1} \:
	\frac{ \exp 2\pi\tau\left( x(1-x)(-\alpha'p^2)+1 \right) }
	{ 1-e^{-2\pi\tau} } \notag\\
	&=
	(2\pi)^k
	\int_0^1 \dd{x}\:
	\zeta(-k, -x(1-x)(-\alpha'p^2)-1).
\end{align}
It reduces to the Hurwitz zeta function.

Let us evaluate this integral by saddle point approximation.
Let us temporally write $\alpha'p^2/4 = \xi, \: 2\pi\tau \rightarrow\tau$.
The saddle equations are
\begin{subequations}
\begin{align}
	0 &= -4(1-2x) \: \xi\tau \\
	0 &= -4x(1-x) \: \xi+1 - \frac{1}{e^{\tau}-1} - \frac{k+1}{\tau}
\end{align}
\end{subequations}
The first equation gives the unique solution
\begin{align}
	x = \frac{1}{2}.
\end{align}
In the asymptotic region $p^2, k \gg 1$,
the second equation has solutions
\begin{align}
	\tau
	\sim
	%\dfrac{ d/2 }{ p^2/4 }, \quad
	%\frac{1}{ \frac{p^2}{4}-\frac{d/2-1}{2\pi in} } + 2\pi in
	-\frac{k+1}{\xi-2}, \;
	2\pi in \left( 1-\frac{1}{2\pi in(\xi-1)+k+2} \right).
\end{align}
where $n \in \integernum_{\neq0}$.
Using the results in Sec.~\ref{sec:barnes},
we have contributions from the thimbles of these complex saddles $\mathcal{J}_{l,n}$,
\begin{align}
	\left. \tilde{\mathcal{A}}_1(p^2) \right|_{\mathcal{J}}
	\propto
	(2\pi)^k
	\Gamma(k+1)
	\sum_{n\neq0}
	\left( \frac{1}{2\pi in} \right)^{k+1}
	e^{ 2\pi in \xi }.
\end{align}
This result is interpreted as follows. %Fig.~\ref{fig:one-loop_interpret}.
The saddle values $\tau_1 = x_1\tau, \: \tau_2 = x_2\tau$ are proper time of excitations traveling on the first/second loop propagator respectively.
In this one-loop case, two loop propagators are equivalent.
Thus, the dominant contribution should be around $\tau_1\sim\tau_2$, thus, $x_1\sim x_2$.
Correspondingly, loop momentum is equally distributed $\ell\sim p/2$.
Evaluating the integrand around these points,
we have a oscillatory factor $e^{ 2\pi in (p/2)^2 }$.
%\begin{figure}[t]
%	\centering
%	\includegraphics[height=60mm]{fig/depth-1_interpret.pdf}
%	%\caption{}
%	\label{fig:one-loop_interpret}
%\end{figure}

\begin{figure}[H]
	\centering
	\includegraphics[width=0.45\textwidth]{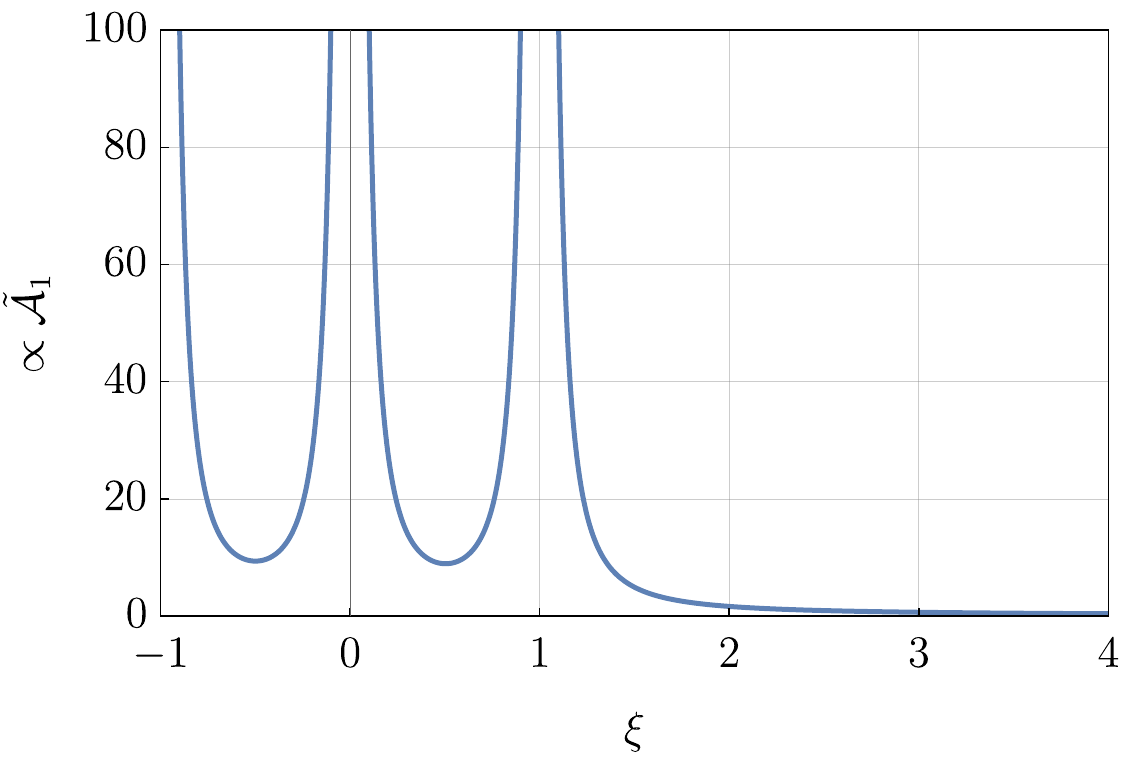} \hspace{5mm}
	\includegraphics[width=0.45\textwidth]{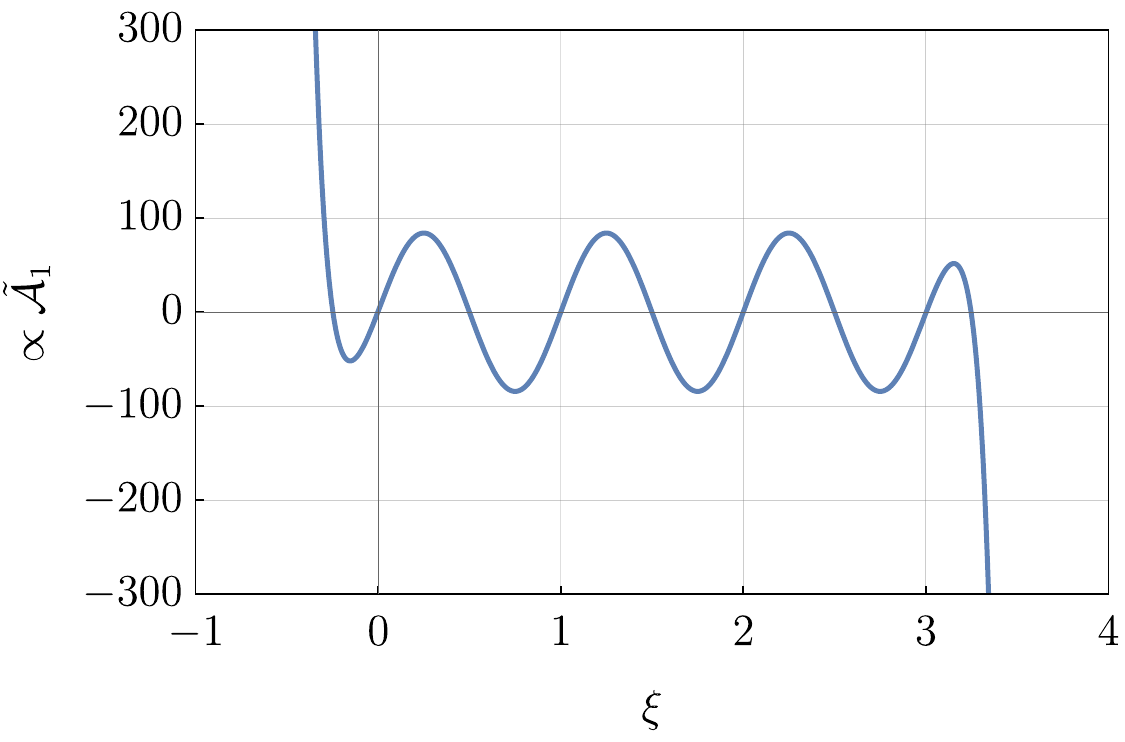}
	\caption{
		Plots of one-loop model \eqref{eq:trunc_oneloop_fixed} evaluated at the saddle point
		$\tilde{\mathcal{A}}_1(p^2) \propto \zeta(-k,\xi-1)$
		with respect to $\xi = \alpha'p^2/4$.
		The left figure is for positive $s=-k=2$
		while the right figure is for negative $s=-k=-20$.
		They show single periodic structure.
	}
	\label{fig:one-loop_fix_plots}
\end{figure}

As we saw in Sec.~\ref{sec:barnes},
thimbles of complex saddles $\mathcal{J}_{l,n}$ give the oscillatory part of the Hurwitz zeta function.
The remaining part is given by the vertical integration contour $\Gamma_l$.
Sum of these contributions are simply written by the Hurwitz zeta function evaluated at the saddle $x=1/2$.
\begin{align}
	\mathcal{A}_1(p^2)
	\propto
	(2\pi)^k\zeta(-k,\xi-1)
\end{align}
This result is plotted in Fig.~\ref{fig:one-loop_fix_plots}.

%--------------------------------
\subsection{One-loop diagram}
\begin{figure}[H]
	\centering
	\includegraphics[width=0.45\textwidth]{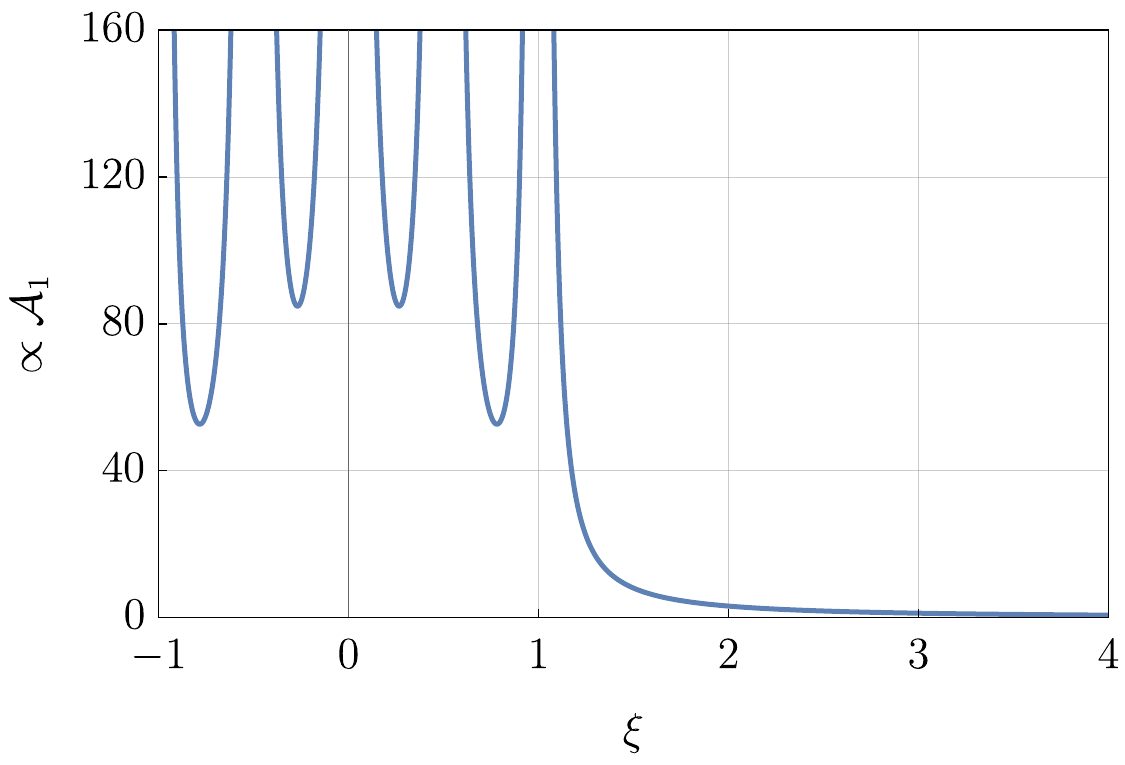} \hspace{5mm}
	\includegraphics[width=0.45\textwidth]{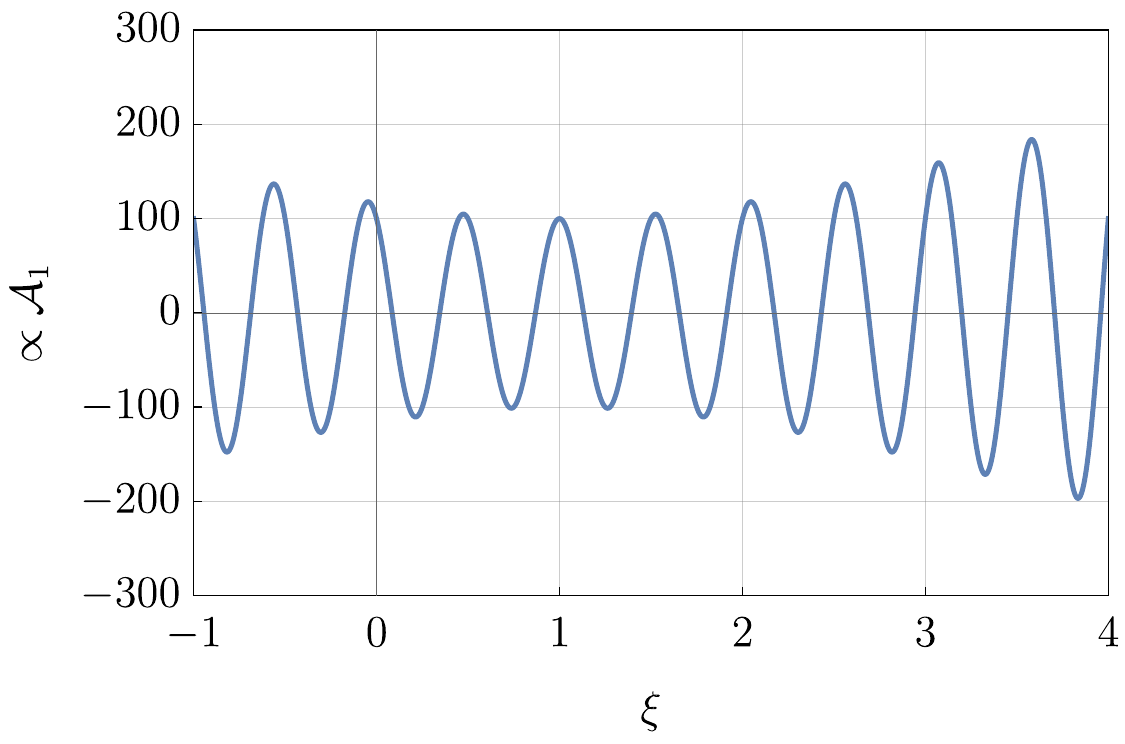}
	\caption{
		Plots of one-loop model \eqref{eq:trunc_oneloop} evaluated at the saddle point
		$\mathcal{A}_1(p^2) \propto \zeta(-k,\xi-1;1/2,1/2)$
		with respect to $\xi = \alpha'p^2/4$.
		The left figure is for positive $s=-k=2$
		while the right figure is for negative $s=-k=-20$.
		All of these blue curves are truncation of infinite series expression at the first term $n=1$.
		They show single periodic structure.
	}
	\label{fig:one-loop_plots}
\end{figure}

%--------------------------------
Let us proceed to the one-loop truncated model \eqref{eq:trunc_oneloop}.
Introducing a Lagrange multiplier $\lambda$ to impose a condition $x_1+x_2-1=0$, we have
\begin{align}
	\mathcal{A}_1(p^2)
	&\propto
	\frac{(2\pi)^k}{\Gamma(-k)}
	\int_{-i\infty}^{i\infty} \dd{\lambda}
	\int_0^1 \dd{x_1}\dd{x_2}
	\int_0^{\infty} \dd{\tau}\:
	e^{ S(\lambda,x_1,x_2,\tau) },
\end{align}
where
\begin{align}
	S(\lambda,x_1,x_2,\tau)
	&=
	-4\frac{x_1x_2}{x_1+x_2} \: \xi\tau
	+(x_1+x_2)\tau \notag\\
	&\qquad
	-\ln (1-e^{-x_1\tau})(1-e^{-x_2\tau})
	-(k+1)\ln\tau \notag\\
	&\qquad\qquad
	-(k+2)\ln(x_1+x_2)
	-\lambda(x_1+x_2-1).
\end{align}
Its saddle point equations are
\begin{subequations}
\begin{align}
	0 &= \pdv{S}{x_1}
	= -4\left(\frac{x_2}{x_1+x_2}\right)^2 \xi\tau + \tau
	-\frac{ \tau }{ e^{x_1\tau}-1 }
	-\frac{k+2}{x_1+x_2}
	-\lambda \\
	0 &= \pdv{S}{x_2}
	= -4\left(\frac{x_1}{x_1+x_2}\right)^2 \xi\tau + \tau
	-\frac{ \tau }{ e^{x_2\tau}-1 }
	-\frac{k+2}{x_1+x_2}
	-\lambda \\
	0 &= \pdv{S}{\tau}
	= -4\frac{x_1x_2}{x_1+x_2} \xi + (x_1+x_2)
	-\frac{ x_1 }{ e^{x_1\tau}-1 } -\frac{ x_2 }{ e^{x_2\tau}-1 }
	-\frac{k+1}{\tau} \\
	0 &= \pdv{S}{\lambda}
	= -x_1-x_2+1
\end{align}
\end{subequations}
From the first two equations
\begin{align}
	4\frac{ x_1-x_2 }{ x_1+x_2 } \xi\tau
	= \frac{ \tau }{ e^{x_1\tau}-1 } - \frac{\tau}{ e^{x_2\tau}-1 }.
\end{align}
This has a solution
\begin{align}
	x_1 = x_2
\end{align}
And from the last equation
\begin{align}
	x\coloneqq x_1 = x_2 = \frac{1}{2}.
\end{align}
Substituting this into the third equation
\begin{align}
	0 &=
	-(\xi-1)
	-2\frac{x}{ e^{x\tau}-1 }
	-\frac{k+1}{\tau}.
\end{align}
It has solutions
\begin{align}
	\tau
	&\sim
	-\frac{k+1}{(\xi-1)-2x}, \:
	\frac{ 2\pi in }{ x } \left( 1-\frac{2x}{2\pi in(\xi-1)+(k+1)x+2x} \right)
\end{align}
As we saw in Sec~\ref{sec:barnes}, complex saddles contribute to the integral.
The action value is
\begin{align}
	e^{ S(x_1,x_2,\tau) }
	&=
	\left( \frac{1}{x_1+x_2} \right)^{k+2}
	\left( \frac{1}{\tau} \right)^{k+1}
	\frac{ \exp\left[ -4\frac{x_1x_2}{x_1+x_2}\xi\tau + (x_1+x_2)\tau \right] }
	{ (1-e^{-x_1\tau}) (1-e^{-x_2\tau}) } \notag\\
	&\sim
	\left( \frac{x}{2\pi in} \right)^{k+1}
	\frac{ e^{ -2\pi in(\xi-1)/x } }
	{ \left( 1-e^{+\frac{2\pi in \cdot 2x}{2\pi in(\xi-1)+(k+1)x+2x}} \right)^2 }
\end{align}
The fluctuation part is
\begin{align}
	\pdv[2]{S(x_1,x_2,\tau)}{\tau}
	&\sim \frac{ x_1^2 e^{x_1\tau} }{ (e^{x_1\tau}-1)^2 }
	+\frac{ x_2^2 e^{x_1\tau} }{ (e^{x_2\tau}-1)^2 }
	+\frac{k+1}{\tau^2} \notag\\
	&\sim
	2x^2 \left( -\frac{2\pi in(\xi-1)+(k+1)x+2x}{ 2\pi in\cdot 2x } \right)^2
\end{align}
Combining these,
we obtain contributions from the thimbles $\mathcal{J}_{l,n}$
\begin{align}
	\left. \mathcal{A}_1(p^2) \right|_{\mathcal{J}}
	&\sim
	\frac{ (2\pi)^k }{ \Gamma(-k) }
	\sum_{ l=-\infty }^{0}
	\sum_{ n\neq0 }
	(-1)^{\sigma_n} \sqrt{ \frac{2\pi}{-S''(\tau_n)} } \:
	e^{ S(\tau_n)-2\pi i(k+1)l } \notag\\
	&\sim
	(2\pi)^k \Gamma(k+1)
	\sum_{n\neq0}
	\left( -\frac{k+1}{2\pi in} + (\xi-1)/x \right)
	\left( \frac{x}{2\pi in} \right)^{k+1}
	e^{ 2\pi in (\xi-1)/x }
\end{align}
The remaining part comes from the vertical contour.
Summing all of these contributions, we have
\begin{align}
	\mathcal{A}_1(p^2)
	\propto
	(2\pi)^k \zeta_2(-k,\xi-1; 1/2,1/2).
\end{align}
The result is plotted in Fig.~\ref{fig:one-loop_plots}.

%--------------------------------
\subsection{Two-loop diagrams}
%--------------------------------

We can generalize the previous results straightforwardly.
For example in the reducible case \eqref{eq:trunc_twoloop_red},
the action is
\begin{align}
	S(\lambda,x_i,\tau)
	&=
	\left( \frac{x_1x_2}{x_1+x_2}+x_3+\frac{x_4x_5}{x_4+x_5} \right)(-\alpha'p^2)\tau
	+(x_1+\cdots+x_5)\tau \notag\\
	&\qquad
	-\ln(1-e^{-x_1\tau}) -\cdots -\ln(1-e^{-x_5\tau})
	-(k+1)\ln\tau \notag\\
	&\qquad\qquad
	-\frac{d}{2}\ln(x_1+x_2) -\frac{d}{2}\ln(x_4+x_5)
	-\lambda(x_1+\cdots+x_5-1).
\end{align}
The saddle point equations are
\begin{subequations}
\begin{align}
	0&= \pdv{S}{x_1}
	= \left( \frac{x_2}{x_1+x_2} \right)^2(-\alpha'p^2)\tau+\tau
	-\frac{\tau}{e^{x_1\tau}-1}
	-\frac{d/2}{x_1+x_2} -\lambda \\
	0&= \pdv{S}{x_2}
	= \left( \frac{x_1}{x_1+x_2} \right)^2(-\alpha'p^2)\tau+\tau
	-\frac{\tau}{e^{x_2\tau}-1}
	-\frac{d/2}{x_1+x_2} -\lambda \\
	0&= \pdv{S}{x_3}
	= (-\alpha'p^2)\tau+\tau
	-\frac{\tau}{e^{x_3\tau}-1}
	-\lambda \\
	0&= \pdv{S}{x_4}
	= \left( \frac{x_5}{x_4+x_5} \right)^2(-\alpha'p^2)\tau+\tau
	-\frac{\tau}{e^{x_4\tau}-1}
	-\frac{d/2}{x_4+x_5} -\lambda \\
	0&= \pdv{S}{x_5}
	= \left( \frac{x_4}{x_4+x_5} \right)^2(-\alpha'p^2)\tau+\tau
	-\frac{\tau}{e^{x_5\tau}-1}
	-\frac{d/2}{x_4+x_5} -\lambda
\end{align}
\end{subequations}
and
\begin{align}
	0 &= \pdv{S}{\tau}
	= \left( \frac{x_1x_2}{x_1+x_2}+x_3+\frac{x_4x_5}{x_4+x_5} \right)(-\alpha'p^2)
	+(x_1+\cdots+x_5)
	-\frac{x_1}{e^{x_1\tau}-1} -\cdots -\frac{x_5}{e^{x_5\tau}-1}
	-\frac{k+1}{\tau}
\end{align}
and
\begin{align}
	0 &= \pdv{S}{\lambda}
	= x_1 +\cdots +x_5 -1.
\end{align}
Solving these equations, we can find infinitely many complex saddles.
From the symmetry of the diagram and the uncertainty principle,
we can guess the location of saddles.
The momentum $p$ flowing into the diagram is equally distributed to the $1$-st, $2$-nd propagator and flows into the $3$-rd propagator
\begin{align}
	\frac{p}{2}\sqrt{x_1\tau} \sim \frac{p}{2}\sqrt{x_2\tau}
	\sim p \sqrt{x_3\tau}
	\sim \frac{p}{2}\sqrt{x_4\tau} \sim \frac{p}{2}\sqrt{x_5\tau}.
\end{align}
Thus, we find saddles around
\begin{subequations}
\begin{align}
	&x_1 \sim x_2 \sim 4x_3 \sim x_4 \sim x_5, \\
	&x_1\tau, x_3\tau \sim 2\pi in.
\end{align}
\end{subequations}
Using these saddle values, the amplitude is
\begin{align}
	\mathcal{A}_2
	\sim
	(2\pi)^k
	\zeta_5(-k,\xi-1;x_1,x_2,x_3,x_4,x_5).
\end{align}
where $k=d-5$.
Its oscillatory part consists of two types of modes
\begin{align}
	\sim
	\left( \frac{x_1}{2\pi in} \right) e^{2\pi in \xi /x_1}, \;
	\left( \frac{x_3}{2\pi in} \right) e^{2\pi in \xi /x_3}.
\end{align}

Similarly, in the irreducible case \eqref{eq:trunc_twoloop_irred},
we can solve saddle point equations.
From the symmetry of the diagram, we find saddles around
\begin{subequations}
\begin{align}
	&x_2 \sim x_5, \; x_3\sim x_4, \\
	&x_1\tau, x_2\tau, x_3\tau \sim 2\pi in.
\end{align}
\end{subequations}
Using these saddle values, the amplitude is
\begin{align}
	\mathcal{A}_2
	\sim
	(2\pi)^k \zeta_5(-k,\xi-1;x_1,x_2,x_3,x_4,x_5)
\end{align}
where $k=d-5$.
Its oscillatory part consists of three types of modes
\begin{align}
	\sim
	\left( \frac{x_1}{2\pi in} \right) e^{2\pi in \xi /x_1}, \;
	\left( \frac{x_2}{2\pi in} \right) e^{2\pi in \xi /x_2}, \;
	\left( \frac{x_3}{2\pi in} \right) e^{2\pi in \xi /x_3}.
\end{align}

%--------------------------------
\subsection{Nested melon diagrams}
%--------------------------------
Let us proceed to a special class of higher-loop diagrams \eqref{eq:trunc_nested}.
The bipartite depth-$n$ amplitude can be written as
\begin{align}
	\label{eq:n-loop_zeta_saddle}
	\mathcal{A}^{(n)}
	&\propto
	\frac{ (2\pi)^k }{ \Gamma(-k^{(n)}) }
	\int_{-i\infty}^{+i\infty} \dd{\lambda}
	\int_0^1 \prod_{a,\overline{a},b} \dd{x_a}\dd{x_{\overline{a}}}\dd{x_b}
	\int_0^{\infty} \dd{\tau}\:
	e^{ S^{(n)}(x_a,x_{\overline{a}},x_b, \tau, \lambda) },
\end{align}
where
\begin{align}
	S^{(n)}(x_a,x_{\overline{a}},x_b,\tau,\lambda)
	&=
	-\left( \frac{x^{(n)}_0x^{(n)}_1}{x^{(n)}_0+x^{(n)}_1} p^2 -1 \right)\tau
	- \frac{d}{2} \sum_{i=0}^{n}\sum_{a\in\mathbb{B}_{n-i}}
	\ln \left( x_{a0}^{(i)}+x_{a1}^{(i)} \right) \notag\\
	&\qquad
	-\sum_{a,\overline{a},b}
	\ln (1-e^{-x_a\tau}) (1-e^{-x_{\overline{a}}\tau}) (1-e^{-x_b\tau})
	-(k^{(n)}+1) \ln \tau \notag\\
	&\qquad\qquad
	-\lambda \left(
	\sum_a x_a + \sum_{\overline{a}} x_{\overline{a}} + \sum_b x_b -1
	\right).
\end{align}
The indices $a,\overline{a},b$ run over $\mathbb{B}_1^n,\overline{\mathbb{B}}_1^n, \mathbb{B}_{n+1}$.
We will omit them unless there is no risk of confusion.
The following computations may be tedious,
but just straightforward generalization of the previous computations.

The saddle point equations are
\begin{subequations}
	\label{eq:der_1dig}
	\begin{align}
		\pdv{S^{(n)}}{x_0}
		&= -\left[
		p^2\tau \left( \frac{x_1^{(n)}}{x_0^{(n)}+x_1^{(n)}} \right)^2
		+ \frac{d/2}{x_0^{(n)}+x_1^{(n)}}
		\right]
		-\frac{\tau}{e^{x_0\tau}-1}
		-\lambda, \\
		\pdv{S^{(n)}}{x_1}
		&= -\left[
		p^2\tau \left( \frac{x_0^{(n)}}{x_0^{(n)}+x_1^{(n)}} \right)^2
		+ \frac{d/2}{x_0^{(n)}+x_1^{(n)}}
		\right]
		-\frac{\tau}{e^{x_1\tau}-1}
		-\lambda, \\
		\pdv{S^{(n)}}{x_{\overline{0}}}
		&= -\left[
		p^2\tau \left( \frac{x_1^{(n)}}{x_0^{(n)}+x_1^{(n)}} \right)^2
		+ \frac{d/2}{x_0^{(n)}+x_1^{(n)}}
		\right]
		-\frac{\tau}{e^{x_{\overline{0}}\tau}-1}
		-\lambda, \\
		\pdv{S^{(n)}}{x_{\overline{1}}}
		&= -\left[
		p^2\tau \left( \frac{x_1^{(n)}}{x_0^{(n)}+x_1^{(n)}} \right)^2
		+ \frac{d/2}{x_0^{(n)}+x_1^{(n)}}
		\right]
		-\frac{\tau}{e^{x_{\overline{1}}\tau}-1}
		-\lambda,
	\end{align}
\end{subequations}
\begin{subequations}
	\label{eq:der_2dig}
	\begin{align}
		\pdv{S^{(n)}}{x_{00}}
		=& -\left[
		p^2\tau \left( \frac{x_1^{(n)}}{x_0^{(n)}+x_1^{(n)}} \right)^2
		+ \frac{d/2}{x_0^{(n)}+x_1^{(n)}}
		\right]
		\left( \frac{x_{01}^{(n-1)}}{x_{00}^{(n-1)}+x_{01}^{(n-1)}} \right)^2 \notag\\
		&- \frac{d/2}{x_{00}^{(n-1)}+x_{01}^{(n-1)}}
		-\frac{\tau}{e^{x_{00}\tau}-1}
		-\lambda, \\
		\pdv{S^{(n)}}{x_{01}}
		=& -\left[
		p^2\tau \left( \frac{x_1^{(n)}}{x_0^{(n)}+x_1^{(n)}} \right)^2
		+ \frac{d/2}{x_0^{(n)}+x_1^{(n)}}
		\right]
		\left( \frac{x_{00}^{(n-1)}}{x_{00}^{(n-1)}+x_{01}^{(n-1)}} \right)^2 \notag\\
		&- \frac{d/2}{x_{00}^{(n-1)}+x_{01}^{(n-1)}}
		-\frac{\tau}{e^{x_{01}\tau}-1}
		-\lambda, \\
		\pdv{S^{(n)}}{x_{10}}
		=& -\left[
		p^2\tau \left( \frac{x_0^{(n)}}{x_0^{(n)}+x_1^{(n)}} \right)^2
		+ \frac{d/2}{x_0^{(n)}+x_1^{(n)}}
		\right]
		\left( \frac{x_{11}^{(n-1)}}{x_{10}^{(n-1)}+x_{11}^{(n-1)}} \right)^2 \notag\\
		&- \frac{d/2}{x_{10}^{(n-1)}+x_{11}^{(n-1)}}
		-\frac{\tau}{e^{x_{10}\tau}-1}
		-\lambda, \\
		\pdv{S^{(n)}}{x_{11}}
		=& -\left[
		p^2\tau \left( \frac{x_0^{(n)}}{x_0^{(n)}+x_1^{(n)}} \right)^2
		+ \frac{d/2}{x_0^{(n)}+x_1^{(n)}}
		\right]
		\left( \frac{x_{10}^{(n-1)}}{x_{10}^{(n-1)}+x_{11}^{(n-1)}} \right)^2 \notag\\
		&- \frac{d/2}{x_{10}^{(n-1)}+x_{11}^{(n-1)}}
		-\frac{\tau}{e^{x_{11}\tau}-1}
		-\lambda, \\
		\vdots\qquad&
	\end{align}
\end{subequations}
%\begin{align}
%	\vdots \notag
%\end{align}
\begin{subequations}
	\label{eq:der_n1dig}
	\begin{align}
		\pdv{S^{(n)}}{x_{00\dots00}}
		=& -\left[
		p^2\tau \left( \frac{x_1^{(n)}}{x_0^{(n)}+x_1^{(n)}} \right)^2
		+ \frac{d/2}{x_0^{(n)}+x_1^{(n)}}
		\right]
		\left( \frac{x_{01}^{(n-1)}}{x_{00}^{(n-1)}+x_{01}^{(n-1)}} \right)^2
		\cdots
		\left( \frac{x_{00\dots01}^{(0)}}{x_{00\dots00}^{(0)}+x_{00\dots01}^{(0)}} \right)^2 \notag\\
		&-\frac{d/2}{x_{00}^{(n-1)}+x_{01}^{(n-1)}}
		\left( \frac{x_{001}^{(n-2)}}{x_{000}^{(n-2)}+x_{001}^{(n-2)}} \right)^2
		\cdots
		\left( \frac{x_{00\dots01}^{(0)}}{x_{00\dots00}^{(0)}+x_{00\dots01}^{(0)}} \right)^2 \notag\\
		& -\cdots \notag\\
		& -\frac{d/2}{x_{00\dots00}^{(0)}+x_{00\dots01}^{(0)}}
		-\frac{\tau}{e^{x_{00\dots00}\tau}-1}
		-\lambda, \notag\\
		\vdots\qquad&
	\end{align}
\end{subequations}
and
\begin{align}
	\label{eq:der_tau}
	\pdv{S^{(n)}}{\tau}
	&= -\left( \frac{x^{(n)}_0x^{(n)}_1}{x^{(n)}_0+x^{(n)}_1} p^2 -1 \right)
	-\sum_a \frac{x_a}{e^{x_a\tau}-1}
	-\sum_{\overline{a}} \frac{x_{\overline{a}}}{e^{x_{\overline{a}}\tau}-1}
	-\sum_b \frac{x_b}{e^{x_b\tau}-1}
	-\frac{k^{(n)}+1}{\tau},
\end{align}
and
\begin{align}
	\label{eq:der_lam}
	\pdv{S^{(n)}}{\lambda}
	&=
	-\sum_a x_a
	-\sum_{\overline{a}} x_{\overline{a}}
	-\sum_b x_b
	+1.
\end{align}
From \eqref{eq:der_1dig}, we have
\begin{align}
	p^2\tau
	\frac{x_0^{(n)}-x_1^{(n)}}{x_0^{(n)}+x_1^{(n)}}
	= \frac{\tau}{e^{x_0\tau}-1} - \frac{\tau}{e^{x_1\tau}-1}
	= \frac{\tau}{e^{x_{\overline{0}}\tau}-1} - \frac{\tau}{e^{x_{\overline{1}}\tau}-1}.
\end{align}
These equations have a solution
\begin{align}
	x_0^{(n)} = x_{1}^{(n)}, \quad
	x_0 = x_1, \quad
	x_{\overline{0}} = x_{\overline{1}}.
\end{align}
Substituting these back into \eqref{eq:der_1dig}, we have
\begin{align}
	0 = \frac{\tau}{e^{x_0\tau}-1} - \frac{\tau}{e^{x_{\overline{0}}\tau}-1}.
\end{align}
Then it has a solution
\begin{align}
	x_0 = x_{\overline{0}}.
\end{align}
Similarly, from \eqref{eq:der_2dig}, we have
\begin{subequations}
	\begin{align}
		\left[
		p^2\tau \left( \frac{x_1^{(n)}}{x_0^{(n)}+x_1^{(n)}} \right)^2
		+ \frac{d/2}{x_0^{(n)}+x_1^{(n)}}
		\right]
		\frac{x_{00}^{(n-1)}-x_{01}^{(n-1)}}{x_{00}^{(n-1)}+x_{01}^{(n-1)}}
		&= \frac{\tau}{e^{x_{00}\tau}-1} - \frac{\tau}{e^{x_{01}\tau}-1} \notag\\
		&= \frac{\tau}{e^{x_{\overline{00}}\tau}-1} - \frac{\tau}{e^{x_{\overline{01}}\tau}-1}, \\
		\left[
		p^2\tau \left( \frac{x_0^{(n)}}{x_0^{(n)}+x_1^{(n)}} \right)^2
		+ \frac{d/2}{x_0^{(n)}+x_1^{(n)}}
		\right]
		\frac{x_{10}^{(n-1)}-x_{11}^{(n-1)}}{x_{10}^{(n-1)}+x_{11}^{(n-1)}}
		&= \frac{\tau}{e^{x_{10}\tau}-1} - \frac{\tau}{e^{x_{11}\tau}-1} \notag\\
		&= \frac{\tau}{e^{x_{\overline{10}}\tau}-1} - \frac{\tau}{e^{x_{\overline{11}}\tau}-1}.
	\end{align}
\end{subequations}
These equations have a solution
\begin{subequations}
	\begin{align}
		&x_{00}^{(n-1)} = x_{01}^{(n-1)}, \quad
		x_{00}=x_{01}, \quad
		x_{\overline{00}}=x_{\overline{01}}, \\
		&x_{10}^{(n-1)} = x_{11}^{(n-1)}, \quad
		x_{10}=x_{11}, \quad
		x_{\overline{10}}=x_{\overline{11}}.
	\end{align}
\end{subequations}
Substituting these back into \eqref{eq:der_2dig}, we have
\begin{subequations}
	\begin{align}
		0 &= \frac{\tau}{e^{x_{00}\tau}-1} - \frac{\tau}{e^{x_{\overline{00}}\tau}-1}, \\
		0 &= \frac{\tau}{e^{x_{10}\tau}-1} - \frac{\tau}{e^{x_{\overline{10}}\tau}-1},
	\end{align}
\end{subequations}
and
\begin{align}
	-\frac{d/2}{2x_{00}^{(n-1)}} + \frac{d_2}{2x_{10}^{(n-1)}}
	&= \frac{\tau}{e^{x_{00}\tau}-1} - \frac{\tau}{e^{x_{10}\tau}-1} \notag\\
	&= \frac{\tau}{e^{x_{\overline{00}}\tau}-1} - \frac{\tau}{e^{x_{\overline{10}}\tau}-1}.
\end{align}
These have a solution
\begin{align}
	x_{00} = x_{\overline{00}}, \quad
	x_{10} = x_{\overline{10}},
\end{align}
and
\begin{align}
	x_{00}^{(n-1)} = x_{10}^{(n-1)}, \quad
	x_{00} = x_{10}, \quad
	x_{\overline{00}} = x_{\overline{10}}.
\end{align}
Repeating these processes, we find a solution
\begin{subequations}
	\begin{align}
		&y_i \coloneqq x_{a_i} = x_{\overline{a}_i}, \quad a_i\in\mathbb{B}_{i},\: i=1,\dots,n \\
		&y_{n+1} \coloneqq x_b, \quad b\in\mathbb{B}_{n+1},
	\end{align}
\end{subequations}
where $y_1,\dots,y_n,y_{n+1}$ are some constants.

Substituting these back into \eqref{eq:der_1dig}, \eqref{eq:der_2dig}, \eqref{eq:der_n1dig}, we have a series of equations
\begin{align}
	\label{eq:sdl_eq_y}
	0 &=
	\left[ \frac{p^2\tau}{4}+\frac{d/2}{x_0^{(n)}} \right]\frac{1}{4^{i-1}}
	+\frac{d/2}{2x_{00}^{(n-1)}}\frac{1}{4^{i-2}}
	+\cdots
	+\frac{d/2}{2x_{00\dots0}^{(n-i+1)}}\frac{1}{4^{0}}
	+\frac{\tau}{e^{y_i\tau}-1}
	+\lambda
\end{align}
where $i=1,\dots,n+1$, and where
\begin{subequations}
	\begin{align}
		x_{00\dots00}^{(0)}
		&= y_{n+1} \\
		x_{00\dots0}^{(1)}
		&= 2y_n + \frac{x_{00\dots00}^{(0)}}{2}
		= 2y_n + \frac{y_{n+1}}{2} \\
		\vdots& \notag\\
		x_{0}^{(n)}
		&= 2y_1 + \frac{x_{00}^{(n-1)}}{2}
		= 2y_1 + \frac{
			2y_2 + \frac{
				2y_3 +\frac{
					\dots+\frac{
						2y_n+\frac{y_{n+1}}{2}
					}{\cdots}
				}{2}
			}{2}
		}{2}.
	\end{align}
\end{subequations}
Subtracting the $i,(i+1)$-th equations of \eqref{eq:sdl_eq_y}, we have
\begin{subequations}
	\begin{align}
		\label{eq:sdl_eq_y_red}
		0 &=
		\left[ \frac{p^2\tau}{4}+\frac{d/2}{x_0^{(n)}} \right] \frac{3}{4^i}
		-\frac{d/2}{x_{00}^{(n-1)}} \frac{3}{4^{i-1}}
		-\cdots
		-\frac{d/2}{x^{(n-i+1)}_{00\dots}} \frac{3}{4^1}
		-\frac{d/2}{x^{(n-i)}_{00\dots}} \cdot1 \notag\\
		&\qquad\qquad
		+\frac{\tau}{e^{y_i\tau}-1} - \frac{\tau}{e^{y_{i+1}\tau}-1}
	\end{align}
	for $i=1,\dots,n$.
	From \eqref{eq:der_tau} and \eqref{eq:der_lam}, we also have
	\begin{align}
		\label{eq:sdl_eq_tau_red}
		0 &=
		\left( \frac{x^{(n)}_0}{2} p^2 -1 \right)
		+\sum_{i=1}^{n} \frac{2\cdot2^iy_i}{e^{y_i\tau}-1}
		+\frac{2^{n+1}y_{n+1}}{e^{y_{n+1}\tau}-1}
		+\frac{k^{(n)}+1}{\tau}, \\
		\label{eq:sdl_eq_lam_red}
		0 &=
		\sum_{i=1}^{n} 2\cdot2^iy_i
		+ 2^{n+1}y_{n+1}
		-1.
	\end{align}
\end{subequations}

Now, we can find saddles by solving \eqref{eq:sdl_eq_y_red}, \eqref{eq:sdl_eq_tau_red} and \eqref{eq:sdl_eq_lam_red} simultaneously.
Although these equations are complicated, they are greatly simplified in the high energy limit $\alpha'p^2 \gg 1$.
The first two equations \eqref{eq:sdl_eq_y_red}, \eqref{eq:sdl_eq_tau_red} reduce to
\begin{align}
	\mqty(
	1 & -1 & 0 & \cdots & 0 & 0 \\
	0 & 1 & -1 &  & 0 & 0 \\
	\vdots &&&& \ddots & \vdots \\
	0 & 0 & 0 &  & 1 & -1 \\
	2^0y_1 & 2^1y_2 & 2^2y_3 & \cdots & 2^{n-1}y_{n} & 2^{n-1}y_{n+1}
	) \mqty(
	\frac{1}{e^{y_1\tau}-1} \\
	\frac{1}{e^{y_2\tau}-1} \\
	\vdots \\
	\frac{1}{e^{y_{n}\tau}-1} \\
	\frac{1}{e^{y_{n+1}\tau}-1}
	) \simeq
	\frac{-p^2}{4}
	\mqty(
	3/4^1 \\
	3/4^2 \\
	\vdots \\
	3/4^{n} \\
	x_0^{(n)}/2
	).
\end{align}
We can find the inverse of the matrix on the left-hand side of the equation by the Gauss-Jordan elimination and by using \eqref{eq:sdl_eq_lam_red}.
The result is
\begin{align}
	\mqty(
	1-2^2y_1 & 1-2^2y_1-2^3y_2 & \cdots & 1-2^2y_1-2^3y_2-\cdots-2^{n+1}y_n & 4 \\
	\ \ \ -2^2y_1 & 1-2^2y_1-2^3y_2 &  & 1-2^2y_1-2^3y_2-\cdots-2^{n+1}y_n & 4 \\
	\vdots & &&& 4 \\
	-2^2y_1 & \ \ \ -2^2y_1-2^3y_2 & & 1-2^2y_1-2^3y_2-\cdots-2^{n+1}y_n & 4 \\
	-2^2y_1 & \ \ \ -2^2y_1-2^3y_2 & & \ \ \ -2^2y_1-2^3y_2-\cdots-2^{n+1}y_n & 4
	).
\end{align}
%Noting that
%\begin{align}
%	\frac{x_0^{(n)}}{2}
%	&= \sum_{i=1}^{n} \frac{y_i}{2^{i-1}} + \frac{y_{n+1}}{2^{n+1}},
%\end{align}
Multiplying the inverse matrix to the previous equation, we obtain a simple expression
\begin{align}
	\frac{1}{e^{y_i\tau}-1}
	& \simeq \frac{-p^2}{4^i}, \quad
	i=1,\dots,n.
\end{align}
This equation combined with \eqref{eq:sdl_eq_lam_red} allows a series of complex solutions labeled by integers
\begin{align}
	\label{eq:amp_sdl}
	&y_i \simeq \frac{y_{n+1}}{4^{n-i+1}} \ (i = 1,\dots,n), \quad
	y_{n+1}\tau \simeq -\frac{4^{n+1}}{p^2}+2\pi i \cdot 4^{n}\mathbb{Z}, \notag\\
	&y_{n+1} \simeq \left( \frac{8^n-1}{8-1}\frac{1}{4^{n-1}}+2^{n+1} \right)^{-1}.
\end{align}

The values of $\{y_i\}$ have natural physical interpretation.
In the high energy limit, the momentum flowing along the $a(\in\mathbb{B}_i)$-th propagator may equally split into the $a0$-th and the $a1$-th propagators. %as illustrated in Fig.~\ref{fig:saddle_interpret}.
Then, the momentum integrals of \eqref{eq:amp_depn_int} are dominated by contributions around
\begin{align}
	\ell_{a0} \simeq \ell_{a1} \simeq \frac{\ell_{a}}{2}.
\end{align}
According to the uncertainty principle,
particles of half momentum travel twice long time.
The integral representation \eqref{eq:amp_depn_int} and dimensional analysis tell us that $\sqrt{y_i\tau}$ is interpreted as the proper time for the particle.
Here note that $\tau$ has mass dimension $[\tau]=-2$.
Thus, the integrals over Feynman parameters of \eqref{eq:amp_depn_int} are dominated by contributions around
\begin{align}
	\sqrt{y_{i+1}\tau} \sim 2\sqrt{y_i\tau}.
\end{align}
These agrees with a solution of the saddle point equations \eqref{eq:amp_sdl}.
%\begin{figure}[t]
%	\centering
%	\includegraphics[height=60mm]{fig/saddle_interpret.pdf}
%	\caption{
%		Saddle interpretation
%	}
%	\label{fig:saddle_interpret}
%\end{figure}

Here we must note that we have \textit{not} proven that these saddles are the \textit{all} of the contributing saddles.
For example, we may find saddles $y_i$ far away from the real axis.
Nevertheless, these saddles we found allows natural physical interpretation and consistent with the symmetry of the diagram.
We leave this problem for future works.

Substituting these values, the action reduces to
\begin{align}
	S^{(n)}(y_i,y_{n+1};\tau)
	&=
	- \xi\tau
	- \sum_{i=1}^n 2^{i+1} \ln (1-e^{-\tau/4^{n-i+1}})
	-2^{n+1} \ln (1-e^{-\tau})
	- (k+1)\ln\tau.
\end{align}
Here we introduced a variable to rescale the momentum as
\begin{align}
	\xi \simeq \frac{x_0^{(n)}/2}{y_{n+1}} \: \alpha'p^2
	= \frac{3-2^{-n+1}}{2^{n+1}} \: \alpha'p^2.
\end{align}
The saddle point equation in terms of $\tau$ becomes
\begin{align}
	0
	&= S'(y_i,y_{n+1};\tau) \notag\\
	&=
	-\xi
	-\sum_{i=1}^{n} 2^{i+1} \frac{1/4^{n-i+1}}{e^{\tau/4^{n-i+1}}-1}
	- 2^{n+1} \frac{ 1 }{ e^{\tau}-1 }
	-\frac{k+1}{\tau}.
\end{align}
We assume $k, \xi \gg 1$ where saddle point approximation is valid.

Now let us find complex saddles.
The first series of complex saddles are found by setting an ansatz
\begin{align}
	\tau = 2\pi i\cdot 4^{n-i+1}m +\epsilon, \quad
	m\notin 4\mathbb{Z},
\end{align}
for $i = 1,\dots,n$.
Substituting this ansatz into the saddle point equation, we have
\begin{align}
	0
	&\simeq
	2\pi im \cdot 4^{n-i+1}\xi
	+ \sum_{j=1}^{i-1} 2^{j+1} \frac{ 2\pi im / 4^{i-j} }{ e^{2\pi im/4^{i-j}}-1 }
	+ \sum_{j=i}^{n} 2^{j+1} \frac{ 2\pi im / 4^{i-j} }{ \epsilon/4^{n-j+1} } \notag\\
	&
	-2^{n+1} \frac{ 2\pi im / 4^{i-n-1} }{ \epsilon }
	- (k+1).
\end{align}
Its solution is
\begin{align}
	\tau_{ 4^{n-i+1}m }
	&\simeq
	2\pi i\cdot 4^{n-i+1}m
	- \frac{ \sum_{j=i}^{n} 2^{j+1} +2^{n+1} }
	{ \xi+\frac{k}{2\pi im\cdot 4^{n-i+1}} }, \quad
	m\notin 4\mathbb{Z}.
\end{align}

Another series of complex saddles can be found by setting
\begin{align}
	\tau = 2\pi im + \epsilon, \quad
	m\notin 4\mathbb{Z}.
\end{align}
Substituting this ansatz into the saddle point equation, we have
\begin{align}
	0
	&\simeq
	2\pi im \cdot 4^{n-i+1}\xi
	+ \sum_{j=1}^{n} 2^{j+1} \frac{ 2\pi im / 4^{i-j} }{ e^{2\pi im/4^{i-j}}-1 }
	-2^{n+1} \frac{ 2\pi im }{ \epsilon }
	- (k+1).
\end{align}
Its solution is
\begin{align}
	\tau_{m}
	\simeq
	2\pi im
	- \frac{2^{n+1}}{ \xi+\frac{k}{2\pi im} }, \quad
	m\notin 4\mathbb{Z}.
\end{align}

%The saddles and thimbles are illustrated in Fig.~\ref{fig:saddle_thimble_ampn}.
%\begin{figure}[t]
%	\centering
%	\includegraphics[height=60mm]{fig/saddle_thimble_ampn.pdf}
%	\caption{
%		$\tau$-plane
%	}
%	\label{fig:saddle_thimble_ampn}
%\end{figure}
The thimble structure is a straightforward generalization of Fig.~\ref{fig:barnes_saddles}.
The original contour $\mathcal{C}$ can be deformed to the sum of the thimbles and a horizontal line on each Riemann sheet.
\begin{align}
	\mathcal{C}
	\rightarrow
	\sum_{l=-\infty}^{0} \left[
	\sum_{m\notin 4\mathbb{Z}} \sum_{i=1}^{n+1}
	\mathcal{J}_{l,4^{n-i+1}m}
	+\Gamma_l
	\right].
\end{align}
From the figure, we find that the complex saddles at $\tau=\tau_n, \: n\neq0$ on each Riemann sheet contribute to the integral with the same intersection number $(-1)^{\sigma_n}=1$, whereas the real saddle does not contribute.

At the complex saddles $\tau=\tau_{4^{n-i+1}m}, \: m\notin4\mathbb{Z}$ on the $l$-th Riemann sheet,
the action values are
\begin{align}
	&e^{S(y_i,y_{n+1};\tau_{4^{n-i+1}m}) -2\pi i(k+1)l } \notag\\
	&\simeq
	\prod_{j=1}^{i-1} \left( 1-e^{-2\pi im/4^{i-j}} \right)^{-2^{j+1}}
	\prod_{j=i}^{n} \left( -\frac{ (2^{i+1}+\cdots+2^{n+1}+2^{n+1})/4^{n-j+1} }
	{ \xi+\frac{k}{2\pi im\cdot 4^{n-i+1}} } \right)^{ -2^{j+1} } \notag\\
	&\times
	\left( -\frac{ 2^{i+1}+\cdots+2^{n+1}+2^{n+1} }
	{ \xi+\frac{k}{2\pi im} } \right)^{ -2^{n+1} }
	\left( \frac{1/4^{n-i+1}}{ 2\pi im } \right)^{k+1}
	e^{ -2\pi im\cdot 4^{n-i+1}\xi -2\pi i(k+1)l }.
\end{align}
Since the second derivative of $S(y_i,y_{n+1};\tau)$ is
\begin{align}
	S''(y_i,y_{n+1};\tau)
	&=
	\sum_{i=1}^{n} 2^{i+1}
	\frac{ (1/4^{n-i+1})^2 }{ (e^{\tau/4^{n-i+1}}-1)^2 }
	+2^{n+1} \frac{1}{ (e^{\tau}-1)^2 }
	+\frac{k+1}{\tau^2},
\end{align}
Its fluctuation around the saddle is
\begin{align}
	S''(y_i,y_{n+1};\tau_{4^{n-i+1}m})
	\simeq
	\frac{1}{ 2^{i+1}+\cdots+2^{n+1}+2^{n+1} }
	\left( -\xi-\frac{k}{2\pi im\cdot 4^{n-i+1}} \right)^2.
\end{align}
Thus
\begin{align}
	&\frac{1}{\sqrt{ S''(y_i,y_{n+1};\tau_{4^{n-i+1}m}) }}
	e^{ S(y_i,y_{n+1};\tau_{4^{n-i+1}m}) -2\pi i(k+1)l } \notag\\
	&\simeq
	\frac{ (4^{n-i+1})^{2^{i+1}}\cdots (4^1)^{2^{n+1}} }
	{ (2^{i+1}\cdots+2^{n+1}+2^{n+1})^{ 2^{i+1}+\cdots+2^{n+1}+2^{n+1}-1/2 } } \notag\\
	&\times
	\frac{ \left( -\xi-\frac{k}{2\pi im\cdot 4^{n-i+1}} \right)^{2^{i+1}+\cdots+2^{n+1}+2^{n+1}-1} }
	{ \prod_{j=1}^{i-1} \left( 1-e^{-2\pi im/4^{i-j}} \right)^{2^{j+1}} }
	\left( \frac{1/4^{n-i+1}}{2\pi im} \right)^{k+1}
	e^{ -2\pi im\cdot 4^{n-i+1}\xi }.
\end{align}

At the other complex saddles $\tau=\tau_{m}, \: \notin4\mathbb{Z}$,
the action values are
\begin{align}
	&e^{ S(y_i,y_{n+1};\tau_m) -2\pi i(k+1)l } \notag\\
	&\simeq
	\prod_{j=1}^{n} \left( 1-e^{-2\pi im/4^{n-j+1}} \right)^{-2^{j+1}}
	\left( -\frac{ 2^{n+1} }
	{ \xi+\frac{k}{2\pi im} } \right)^{ -2^{n+1} } \notag\\
	&\times
	\left( \frac{1}{ 2\pi im } \right)^{k+1}
	e^{ -2\pi im\xi -2\pi i(k+1)l }.
\end{align}
Its fluctuation part is
\begin{align}
	S''(y_i,y_{n+1};\tau_m)
	&\simeq
	\frac{1}{2^{n+1}}
	\left( -\xi-\frac{k}{2\pi im} \right)^2.
\end{align}
Thus
\begin{align}
	&\frac{1}{ \sqrt{S''(y_i,y_{n+1};\tau_m)} }\:
	e^{ S(y_i,y_{n+1};\tau_m) -2\pi i(k+1)l } \notag\\
	&\simeq
	\frac{1}{ (2^{n+1})^{2^{n+1}-1/2} }
	\frac{ \left( -\xi-\frac{k}{2\pi im} \right)^{2^{n+1}-1} }
	{ \prod_{j=1}^{n} \left( 1-e^{-2\pi im/4^{n-j+1}} \right)^{2^{j+1}}  } \notag\\
	&\times
	\left( \frac{1}{ 2\pi im } \right)^{k+1}
	e^{ -2\pi im\xi -2\pi i(k+1)l }.
\end{align}
Combining these contributions,
we obtain the fluctuating part of the amplitude
\begin{align}
	\left. \mathcal{A}^{(n)} \right|_{\mathcal{J}}
	&\propto
	\frac{ (2\pi)^k }{ \Gamma(-k^{(n)}) }
	\sum_{l=-\infty}^{0}
	\sum_{ m\notin 4\mathbb{Z} }
	\left[
	\sqrt{ \frac{2\pi}{ -S''(y_i,y_{n+1};\tau_m)} }\:
	e^{ S(y_i,y_{n+1};\tau_m) -2\pi i(k+1)l } \right. \notag\\
	&\qquad\qquad \left.
	+ \sum_{i=1}^{n}
	\sqrt{ \frac{2\pi}{ -S''(y_i,y_{n+1};\tau_{4^{n-i+1}m})} }\:
	e^{ S(y_i,y_{n+1};\tau_{4^{n-i+1}m}) -2\pi i(k+1)l }
	\right]  \notag\\
	&\propto
	(2\pi)^k \Gamma(k+1)
	\sum_{ m\notin 4\mathbb{Z} }
	\sum_{ i=1 }^{n+1}
	c_{4^{n-i+1}m}(\xi,k)
	\left( \frac{1/4^{n-i+1}}{2\pi im} \right)^{k+1}
	e^{ -2\pi im \cdot 4^{n-i+1}\xi }.
\end{align}
Adding the remaining part from $\Gamma_l$, the full expression is
\begin{align}
	\mathcal{A}^{(n)}
	&\propto
	(2\pi)^k \:
	\zeta_{N^{(n)}} \left(
	-k^{(n)}, \xi;
	\underbrace{ \frac{1}{4^n} }_{2^{1+1}},
	\underbrace{ \frac{1}{4^{n-1}} }_{2^{2+1}},
	\dots,
	\underbrace{ \frac{1}{4^1} }_{2^{n+1}},
	\underbrace{ \frac{1}{4^0} }_{2^{n+1}}
	\right).
\end{align}
where
\begin{align}
	N^{(n)}
	= 3\cdot2^{n+1}-4, \quad
	k^{(n)}
	= (2^{n+1}-1)\frac{d}{2} - (3\cdot2^{n+1}-4).
\end{align}

\begin{figure}[t]
	\centering
	\includegraphics[width=0.31\textwidth]{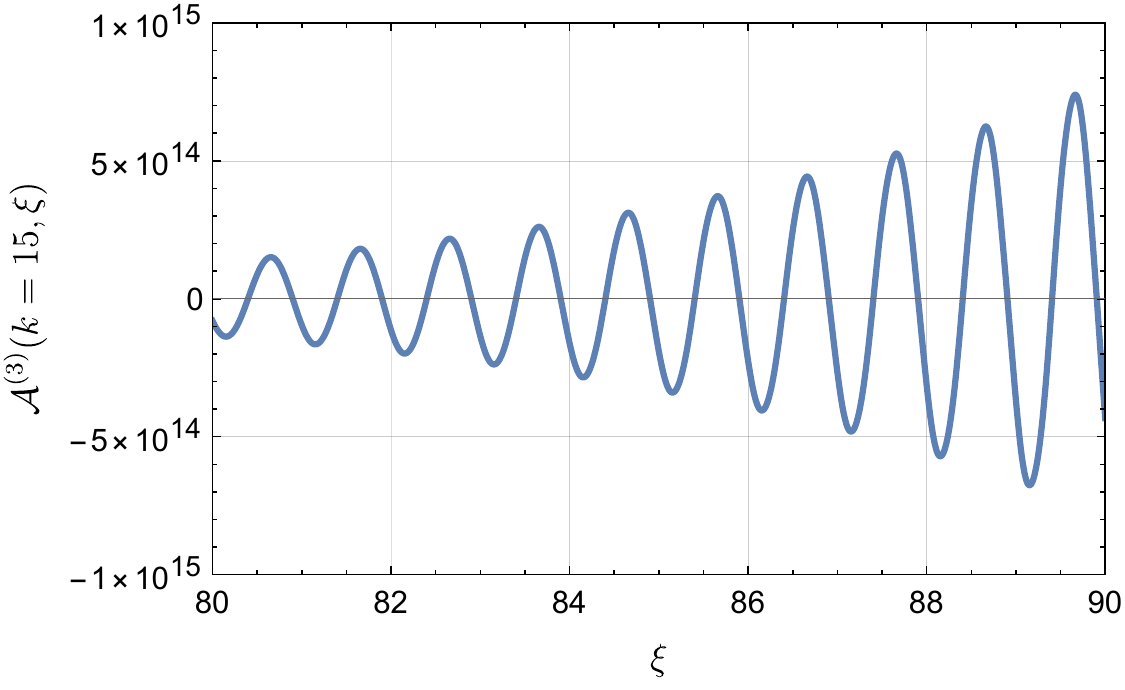}
	\includegraphics[width=0.31\textwidth]{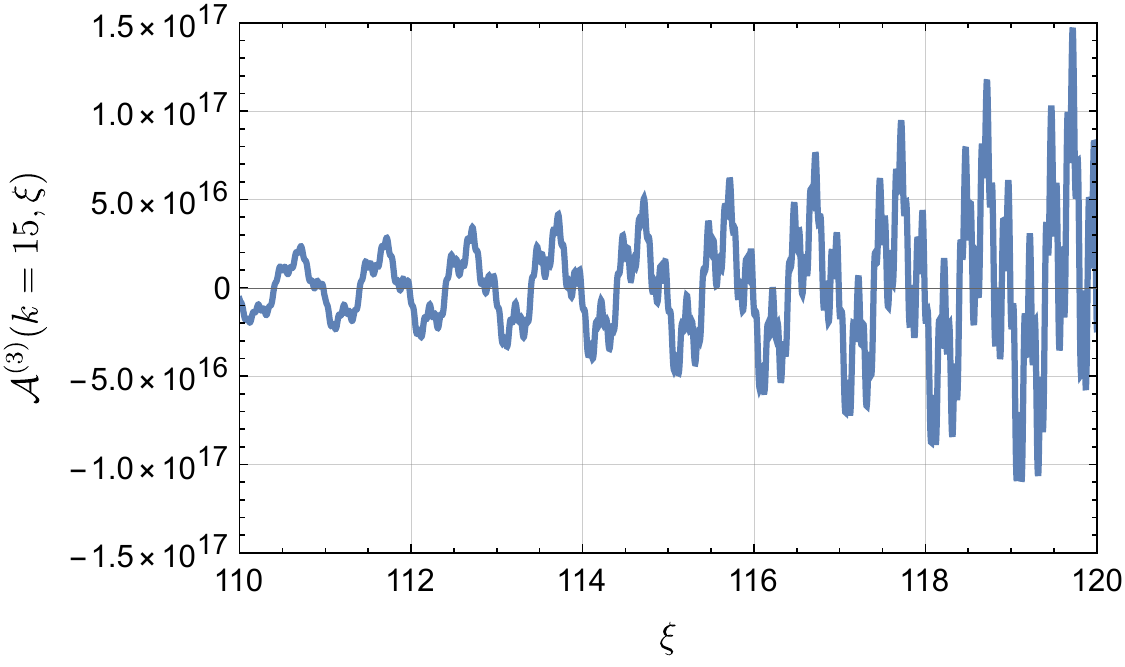}
	\includegraphics[width=0.31\textwidth]{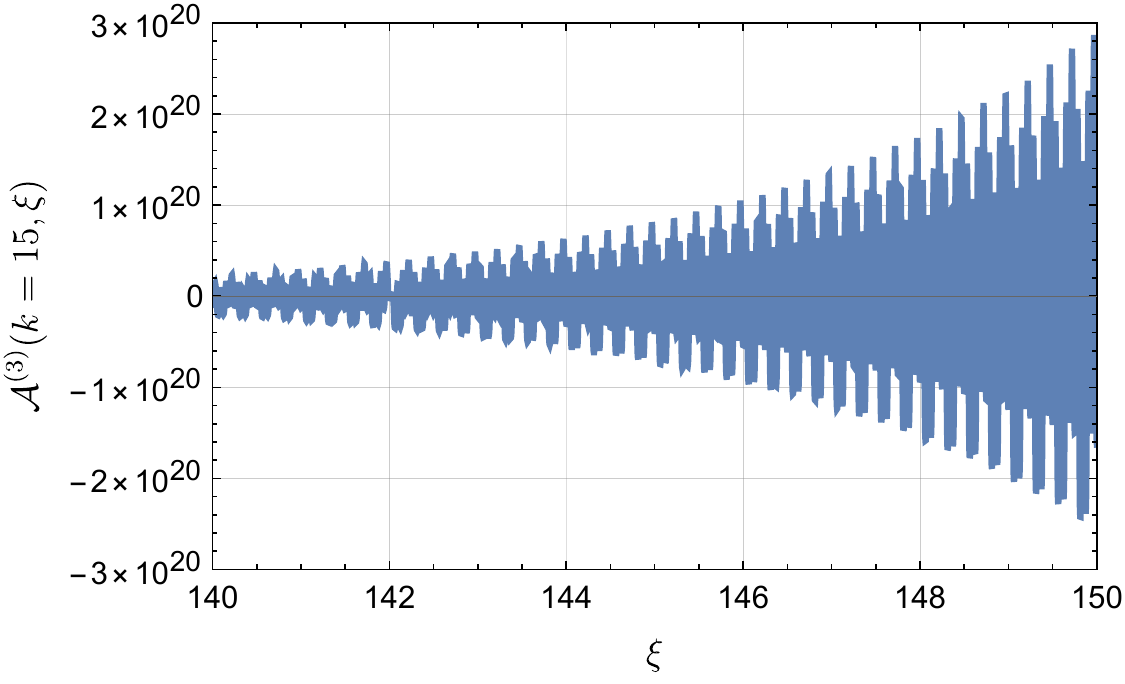} \\
	\includegraphics[width=0.31\textwidth]{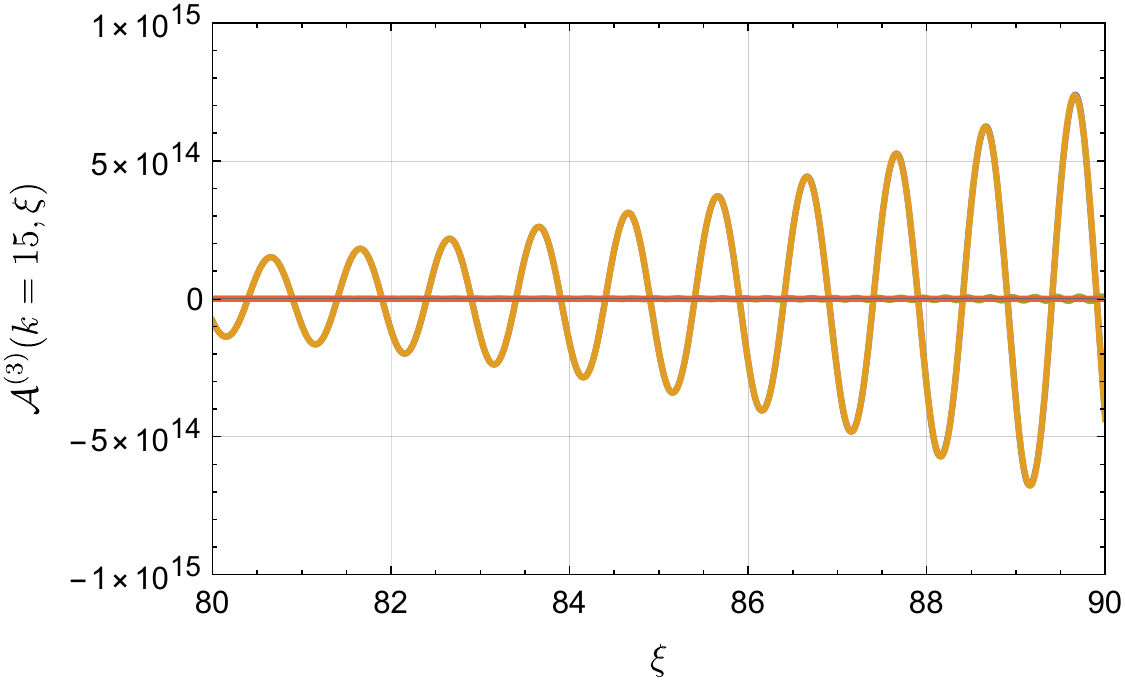}
	\includegraphics[width=0.31\textwidth]{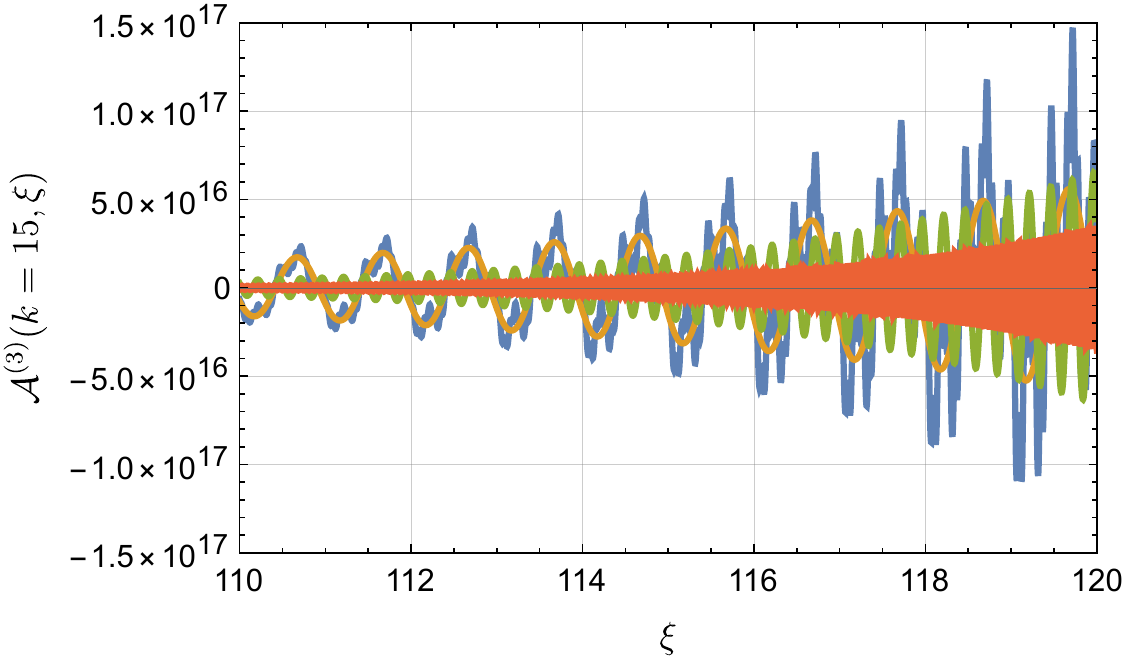}
	\includegraphics[width=0.31\textwidth]{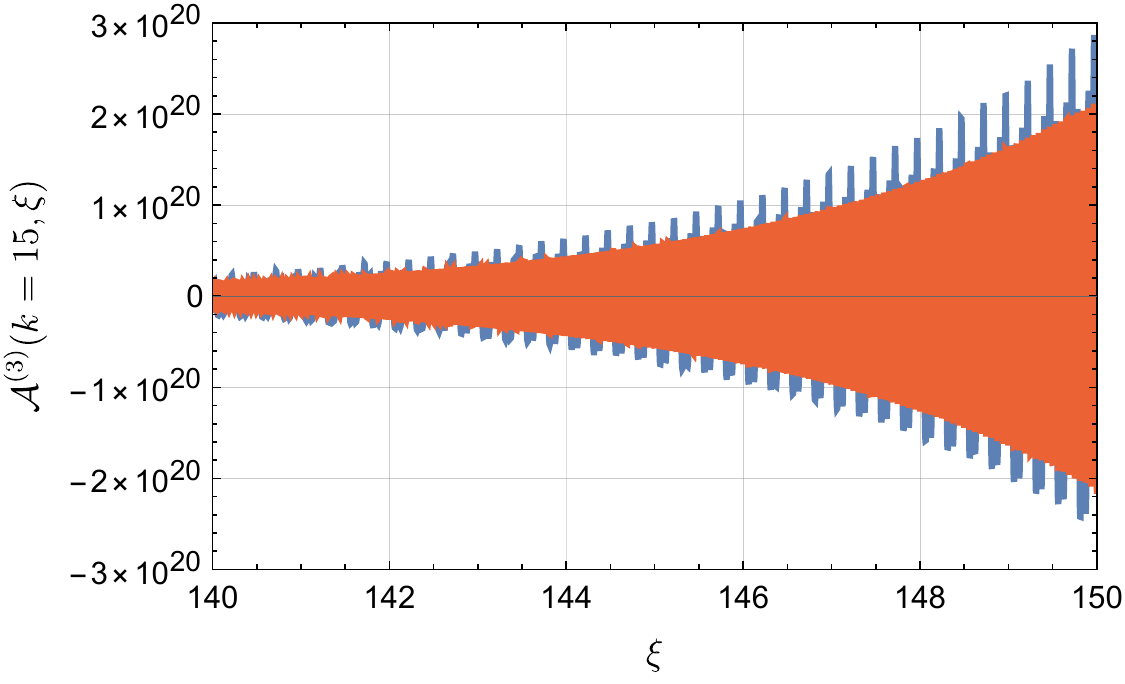}
	\caption{
		Plots of the nested melon diagrams $n=3$ in typical parameter region $s=-k=-15$.
		The left figures are around $\xi \propto p^2 \sim 85$,
		the middle are around $\xi \propto p^2 \sim 115$,
		and the right are around $\xi \propto p^2 \sim 145$.
		The blue curves are the sum of all contributions
		while the orange curves are contributions from $i=n+1,\: m=1,2,\dots$ modes,
		green are from $i=n,\: m=1,2,\dots$ modes,
		and red are from $i=n-1,\: m=1,2,\dots$ modes.
		As momentum $p^2$ becomes larger, or equivalently $k$ becomes smaller,
		the size of higher frequency modes becomes larger.
		Then the amplitude becomes more erratic.
	}
	\label{fig:nested_plot}
\end{figure}

When $s=-k^{(n)} > N^{(n)}$, the amplitude has poles at
\begin{align}
	\xi \sim -(1/4^n)\integernum_{\geq} - (1/4^{n-1})\integernum_{\geq} - \cdots
	-(1/4^1)\integernum_{\geq} - (1/4^{0})\integernum_{\geq}.
\end{align}
When $s=-k^{(n)} \ll -1$, its oscillatory part is expanded by a series of Fourier modes
\begin{align}
	\sim
	\left( \frac{1/4^{n-i+1}}{2\pi im} \right)^{k+1}
	e^{ -2\pi im \cdot 4^{n-i+1}\xi }.
\end{align}
Its typical oscillatory behaviors are plotted in Fig~\ref{fig:nested_plot}.
For each $i$ mode, $m=1,2,\dots$ modes are suppressed.
When $\xi$ is small, the lowest frequency mode $(i,m)=(n+1,1)$ dominates the amplitude.
When $\xi$ becomes larger, higher frequency modes $(i,m)=(n,1),(n-1,1)$ become compatible with other modes.

%--------------------------------
\subsection{Momentum dependent coupling}
%--------------------------------

Finally, we evaluate \eqref{eq:ann_amp_n2} to check that the momentum dependence of string interaction vertex does not affect qualitative behaviors.
In this case the action is
\begin{align}
	S(\lambda,x_1,x_2,\tau)
	&=
	-4\frac{x_1x_2}{x_1+x_2}\xi\tau + (x_1+x_2)\tau \notag\\
	&\qquad -4\xi \ln(1-e^{-x_1\tau}) -4\xi \ln(1-e^{x_2\tau}) -(D-2-8\xi)\ln(1-e^{-\tau}) \notag\\
	&\qquad\qquad -(k+1)\ln\tau - (d/2)\ln(x_1+x_2) -\lambda(x_1+x_2-1),
\end{align}
where $\xi = \alpha'p^2/4,\: k=d/2-1$.
Its saddle equations are
\begin{subequations}
\begin{align}
	0 &= \pdv{S}{x_1}
	= -4\left( \frac{x_2}{x_1+x_2} \right)^2 {\xi\tau} + \tau
	-4\xi \frac{\tau}{e^{x_1\tau}-1} -\frac{d/2}{x_1+x_2} -\lambda \\
	0 &= \pdv{S}{x_2}
	= -4\left( \frac{x_1}{x_1+x_2} \right)^2 {\xi\tau} + \tau
	-4\xi \frac{\tau}{e^{x_2\tau}-1} -\frac{d/2}{x_1+x_2} -\lambda \\
	0 &= \pdv{S}{\tau}
	= -4\frac{x_1x_2}{x_1+x_2}\xi + (x_1+x_2)
	-4\xi \frac{x_1}{e^{x_1\tau}-1} -4\xi \frac{x_2}{e^{x_2\tau}-1}
	-\frac{D-2-8\xi}{e^{\tau}-1}
	-\frac{k+1}{\tau}, \\
	0 &= \pdv{S}{\lambda}
	= x_1+x_2-1.
\end{align}
\end{subequations}
From the first two equations
\begin{align}
	4\frac{x_1-x_2}{x_1+x_2}\xi\tau
	= 4\xi \left( \frac{\tau}{e^{x_1\tau}-1}-\frac{\tau}{e^{x_2\tau}-1} \right)
\end{align}
This has a solution
\begin{align}
	x \coloneqq x_1 = x_2 = \frac{1}{2}.
\end{align}
Substituting this into the third equation
\begin{align}
	0 &= -(\xi-1)
	-2\frac{4\xi x}{e^{x\tau}-1} -\frac{D-2-8\xi}{e^{\tau}-1}
	-\frac{k+1}{\tau}.
\end{align}

A series of complex saddles are found by setting an ansatz
\begin{align}
	x\tau = 2\pi in + \epsilon, \quad n\neq0.
\end{align}
Then we find
\begin{align}
	x\tau
	\sim 2\pi in\left(
		1 - \frac{ (D-2)x }{ 2\pi in(\xi-1) + (k+1)x + (D-2)x }
	\right).
\end{align}
At this saddle, the action value is
\begin{align}
	e^{S}
	&= \left( \frac{1}{\tau} \right)^{k+1}
	\frac{ e^{ -4\frac{x_1x_2}{x_1+x_2}\xi\tau + (x_1+x_2)\tau } }
	{ (1-e^{-\tau})^{D-2-8\xi} (1-e^{-x_1\tau})^{4\xi} (1-e^{-x_2\tau})^{4\xi} } \notag\\
	&\sim
	x^{-8\xi}
	\left( -\frac{2\pi in(\xi-1)+(k+D)x}{2\pi in(D-2)} \right)^{D-2}
	\cdot
	\left( \frac{x}{2\pi in} \right)^{k+1}
	e^{ -2\pi in(\xi-1)/x }
\end{align}
The fluctuation part is
\begin{align}
	\pdv[2]{S}{\tau}
	&= 4\xi \sum_{i=1,2} \frac{x_i^2 e^{x_i\tau}}{(e^{x_i\tau}-1)^2}
	+(D-2-8\xi)\frac{1}{(e^{\tau}-1)^2}
	+\frac{k+1}{\tau^2} \notag\\
	&\sim
	4\xi \cdot 2x^2
	\left( -\frac{2\pi in(\xi-1)+(k+D)x}{2\pi in(D-2)x} \right)^2
\end{align}
Combining these
\begin{align}
	\frac{1}{\sqrt{S''}} \: e^S
	\propto
	\left(
		- \frac{k+D}{2\pi in \cdot \frac{D-2}{2}}
		+\frac{ (\xi-1)x }{ \frac{D-2}{2} }
	\right)
	\xi^{-1/2} x^{-8\xi+1}
	\cdot
	\left( \frac{x}{2\pi in} \right)^{k+1}
	e^{ 2\pi in(\xi-1)/x }
\end{align}
As long as $k \gg \xi \gg 1$, the oscillatory behavior $\sim e^{2\pi in(\xi-1)/x}$ is qualitatively the same.

%Another series of complex saddles are found by setting an ansatz
%\begin{align}
%	\tau = 2\pi in+\epsilon, \quad n\neq 0.
%\end{align}
%Then we find
%\begin{align}
%	\tau \sim
%	2\pi in \left(
%		1 - \frac{ D-2-8\xi }{ 2\pi in(\xi-1) + (k+1) + (D-2-8\xi) + \frac{2\pi in\cdot 2x\cdot 4\xi}{e^{2\pi inx}-1} }
%	\right).
%\end{align}
%At this saddle the action value is
%\begin{align}
%	e^{S}
%	\sim
%	\frac{1}{ (1-e^{-2\pi inx})^{8\xi} }
%	\left( -\frac{2\pi in(\xi-1)+(k+2)+\cdots}{D-2-8\xi} \right)^{ D-2-8\xi }
%	\cdot
%	\left( \frac{1}{2\pi in} \right)^{k+1}
%	e^{ -2\pi in(\xi-1) }.
%\end{align}
%The fluctuation part is
%\begin{align}
%	\pdv[2]{S}{\tau}
%	\sim
%	(D-2-8)
%	\left( -\frac{ 2\pi in(\xi-1)+(k+1)+\cdots }{ 2\pi in(D-2-8\xi) } \right)^2.
%\end{align}
%Combining these
%\begin{align}
%	\frac{1}{S''}e^{S}
%	\propto
%\end{align}

%========================
\section{Conjecture}
\label{sec:conjec}
%========================

%----------------------------------
\subsection{Oscillatory behaviors}
%----------------------------------
In the previous section, we have seen that our truncated model reduces to the Barnes multiple zeta function.
Depending on parameter region, it has multiple poles or zeros.

In particular, the oscillatory part of our nested melon diagram consists of Fourier modes
\begin{align}
	\label{eq:nested_scale}
	\sim
	\left( \frac{1/4^{j}}{2\pi im} \right)^{k+1}
	e^{ 2\pi im \cdot 4^j \xi }
\end{align}
The plot Fig.~\ref{fig:nested_plot} showed that
there is a sweet parameter region where $(j,m)=(0,1), (1,1), (2,1),\dots$ modes dominantly contributes to the amplitude.
Other modes with larger $m$ are suppressed by a factor $(1/m)^{k+1}$.
The size of higher $j$ modes are compatible with each other.
Then, the curve $\mathcal{A}^{(n)}(\xi)$ becomes more erratic.

%----------------------------------
\subsection{Weierstrass function}
%----------------------------------
Such highly oscillatory behaviors resembles a famous function, the Weierstrass function.
It is defined by
\begin{align}
	\label{eq:weierstrass_def}
	W(x)
	= \sum_{n=0}^{\infty} a^n \cos(2\pi b^nx),
\end{align}
where $a,b$ are real parameters.
Each term of this infinite series is a smooth function of $x$.
However, this function behaves pathologically.
If the parameters satisfy
\begin{align}
	\label{eq:weier_nondif_cond}
	b>1, \: \frac{1}{b}<a<1,
\end{align}
the Weierstrass function is nowhere differentiable.
Otherwise the infinite series diverges or becomes differentiable.

Such a pathological behavior originates in its self-similar structure,
\begin{align}
	a W(bx) = W(x) - \cos(2\pi x).
\end{align}
If we zoom into the curve, we see similar structure.
We can never find a smooth slope to define derivative.
The fractal dimension\footnote{
	There are various types of measures to define its fractal dimension.
	In this paper, we do not discuss details.
} of the curve is give by
\begin{align}
	D_{\text{F}} = 2 + \frac{\ln a}{\ln b}.
\end{align}

%----------------------------------
\subsection{Fractality}
%----------------------------------
From the spiky behaviors in Fig~\ref{fig:nested_plot}
and from comparison between \eqref{eq:nested_scale} and \eqref{eq:weierstrass_def},
it is expected that our amplitude becomes fractal in a certain asymptotic region.
Motivated by this, let us estimate the fractal dimension of the amplitude.

In the asymptotic region, our nested model scales as \eqref{eq:nested_scale}.
Its sub-series $(j,m)=(0,1),(1,1),(2,1), \dots$ resembles the Weierstrass function
\begin{align}
	a \rightarrow (1/4)^{k}, \quad
	b \rightarrow 4.
\end{align}
Our approximation was valid for $k\gg1$.
In this region, \eqref{eq:weier_nondif_cond} is not satisfied.
This is consistent with the fact that amplitude is an analytic function.
However, motivated by the spiky behaviors in Fig.~\ref{fig:nested_plot}, let us extrapolate our result to smaller $k$ region, and observe outcomes.
For smaller $k$, parameters $a,b$ approach the fractal region \eqref{eq:weier_nondif_cond}.
We can estimate its fractal dimension
\begin{align}
	D_{\text{F}} \sim 2 + \frac{ \ln 4^{-k} }{ \ln 4 } = 2-k.
\end{align}
It can be fit into
\begin{align}
	1 < D_{\text{F}} < 2.
\end{align}
Now let us estimate the spacetime dimension where the curve approach a fractal one.
Recalling the definition of $k^{(n)}$,
in the large depth limit $n\gg1$,
\begin{align}
	k^{(n)}
	\sim
	2^{n+1} \left( \frac{d}{2}-3 \right).
\end{align}
The curve approach a fractal one if $k^{(n)}$ approach $k^{(n)}\sim1$.
This means that
\begin{align}
	d \sim 6 + \mathcal{O}(2^{-n-1}).
\end{align}

It is remarkable that the value $d\sim 6$ is a special spacetime dimension where our three-point interaction coupling $g$ becomes dimensionless $[g]=0$.
Our truncated string model
\begin{align}
	\mathcal{A}
	\sim
	g^{\#} 
	\int \dd[d]\ell
	\sum_{\alpha'M^2} \frac{1}{ (p-\ell)^2+M^2 }\cdots
\end{align}
consists of variables $\alpha'p^2, \alpha'M^2, g$.
The mass $\alpha'M^2$ is summed over all stringy mass tower.
If we focus on the asymptotic region $\alpha'p^2\gg1$, its mass gap becomes negligible.
Thus, the left dimension-full parameter is only coupling $g$.
If $d\sim6$, there is no typical scale to suppress higher frequency modes.
Then, the curve can approach a fractal curve.

In summary, we have observed that our truncated string amplitude has a tendency to approach a fractal curve
if the following conditions are satisfied.
\begin{itemize}
	\item High energy asymptotic limit $\alpha' p^2 \gg 1$
	\item Higher loop limit $n \gg 1$
	\item Without dimension-full coupling constants
\end{itemize}

It is expected that this feature is universal for other types of diagrams.
Although we focused on a clean example, bifurcating nested melon diagrams,
the key point of such oscillatory behaviors were the form of the $\tau$ integrand.
As explained in the beginning of Sec.~\ref{sec:evaluate_amp},
our truncated models share the same pole structure of $\tau$ integral.
We can find saddles around the pole, then the saddles give oscillatory terms in the form of \eqref{eq:nested_scale}.
Even if we recover the momentum dependence of coupling, and increase the truncation level,
the form of the $\tau$ integrand is the same as checked at the end of Sec.~\ref{sec:evaluate_amp}.

We must comment on analyticity.
String amplitudes should be analytic functions.
Contrary fractal functions are not differentiable.
This point will be understood as follows.
Tree level amplitudes have organized pole/zero structure as the Veneziano amplitude is expressed by the Euler beta function.
If we go to higher genus amplitudes, their pole/zero structure will become more complicated.
Accordingly the curve of amplitudes become more spiky.
Amplitudes are still analytic.
It will be analogous to increasing the truncation level $N$ of the Weierstrass function
\begin{align}
	\sum_{n=0}^{N} a^n \cos (2\pi b^nx)
\end{align}
As long as $N$ is finite, it is differentiable.
If $N$ is sent to infinity, it becomes nowhere differentiable.
In the same way, higher genus amplitudes will approach fractal functions.
If we believe that black hole-string correspondence is valid for their amplitudes,
such messy string amplitudes are identical to black hole amplitudes.
It will give microscopic understanding of black hole thermal radiation.

Also, we must note that fractal behaviors of string amplitudes, even if they exist, should not be as clean as the Weierstrass function.
In Sec.~\ref{sec:evaluate_amp}, we focused on a bifurcating model.
Then we obtained Fourier modes $e^{2\pi im\cdot 4^j \xi}$.
If we consider trifurcating model, we will obtain $e^{2\pi im\cdot 9^j \xi}$.
Furthermore we can consider their combinations.
Each family of diagrams will have different fractal dimension.
Observable string amplitudes are superposition of such diagrams.
Thus, fractal behaviors should not be clean only with the one fractal dimension.
Rather, it should be multi-fractal.

Based on these observations, we state our conjecture.
We conjecture that
\textit{
\begin{itemize}
	\item String amplitudes, as functions of momenta, approach multi-fractal functions in high energy asymptotic region if higher genus contributions are fully included.
	\item Their fractal dimensions are determined by the type of string theory and by the spacetime dimension where scatterings occur.
\end{itemize}
}

%========================
\section{Conclusion and Discussions}
\label{sec:conc}
%========================

%----------------------------------
\subsection{Summary}
%----------------------------------
Motivated by the rich structure of string amplitudes,
and by the black hole-string correspondence,
we introduced stringy toy amplitudes in Sec.~\ref{sec:truncate_string_amp}
by truncating a sub-class of infinitely many stringy excitations.
We sacrificed the modular invariance, thus interactions needed to be introduced by hand.
Instead, we preserved the form of the integral over the Schwinger proper time reviewed in Sec.~\ref{sec:string_amp} (The form of integrals were compared in the beginning of Sec.~\ref{sec:evaluate_amp}).
This implies that the oscillatory behaviors of our models are universal for other types of diagrams and for other models with increased truncation level.

The integrals over the Schwinger proper time were expressed by the Barnes multiple zeta function reviewed in Sec.~\ref{sec:barnes}.
The integrals have infinite series of complex saddles.
Evaluating the integrals by saddle point approximation,
we have obtained their Fourier mode expansions in Sec.~\ref{sec:evaluate_amp}.

We have observed that the oscillatory behaviors of nested multi-loop diagrams resembles the fractal behaviors of the Weierstrass function in Sec.~\ref{sec:conjec}.
Estimated fractal dimension $D_{\text{F}}$ approached $1<D_{\text{F}}<2$ when our model loses dimension full parameters.
Finally, we conjectured that
string amplitudes should approach multi-fractal functions in the high energy asymptotic region
if higher genus contributions are fully included.
Their fractal dimensions should be determined by the type of string theory and by the spacetime dimension where scatterings occur.

%----------------------------------
\subsection{Future prospects}
%----------------------------------
It will be interesting to seek physical interpretation for the complex saddles of our model.
The Schwinger proper time $\tau$ of our model was originally the Euclidean time of propagating stringy excitations.
Thus the complex saddles $a_i\tau \sim 2\pi in$ should have Lorentzian time interpretation.
One possible candidate is the Lorentzian time delay of particles traveling along loops.
Such interpretation will be justified by creating wave packets and evaluating Lorentzian time-delay explicitly.
Here we must note that the highly oscillatory behaviors were found in large $\alpha'p^2\gg1$ region where string momentum is imaginary.
We still need to find relations between such oscillatory behaviors and physical observables such as time-delay.
Once Lorentzian time-delay is evaluated, it is compared directly to the dime-delay of black holes.
Black hole time-delay is associated with the redshift near horizon.
Thus, it will be expected that our models give microscopic understanding of black hole horizons.

%It will also be interesting to try to evaluate Lyapunov exponent of string scatterings,
%motivated by a series of works\cite{Gross:2021gsj,Rosenhaus:2021xhm}.
%In classical scatterings, it is know that the Lyapunov exponent is determined from the time-delay and the fractal dimension of phase space.
%\cite{Shenker:2013pqa,Kitaev14,Polchinski:2015cea}
%\cite{Gross:2021gsj,Rosenhaus:2021xhm,Firrotta:2022cku,Bianchi:2022mhs,Firrotta:2023wem,Bianchi:2023uby,10.1093/ptep/ptad045,Hashimoto:2022bll}.
%\cite{Hashimoto:2022bll}

%It will be interesting to quantify erraticity of string amplitudes and compare to the one of perturbative QFTs\cite{Henn:2024qwe}.

%Spacetime uncertainty

%Fractality of amplitudes remind us of some quantum gravity researches on dynamical triangulation (see a review e.g. \cite{Kawamoto:1993an}).
%It might be interesting to discuss relations to them.

Nevertheless, we have to note that our model have caveats.
One of them is modular invariance breaking.
Our model preserved the form of the Schwinger proper time integral.
It is expected to be universal even if we increase truncation level.
However, there is a fairly reasonable chance that highly oscillatory behaviors we have found will vanish if truncation level goes to infinity to recover the original string amplitudes.
Also, we have not renormalized our diagrams in order to study erratic behaviors of the original string amplitudes.
We have not provided quantitative arguments that how our model is similar to string amplitudes.

Another point is the limitation of saddle point approximation.
Our approximation was valid only of asymptotic regions $k, p^2 \gg 1$.
We extrapolated our results to discuss possible fractality.
If we try to discuss fractality more precisely, we need to go smaller $k$ regions.
However, around $k\sim1$, fluctuations around saddles become larger and it makes convergence rate of the Fourier mode expansion much slower.
Although saddle point approximation have led us to our conjecture,
we may need other powerful tools to study fractality more precisely.
We leave these problems for future works.

%===========================
\subsection*{Acknowledgments}
%===========================
The author would like to thank Koji Hashimoto for valuable discussions and comments for a manuscript
and Tamiaki Yoneya for inspiring discussions on time-delay and spacetime uncertainty.
The work of T. Y. was supported in part by JSPS KAKENHI Grant No. JP22H05115 and
JP22KJ1896.

%================================================================================
\newpage
\appendix

%========================
\section{Analytic Continuation of Euler Gamma Function}
\label{sec:ac_gamma}
%========================

In this appendix, we review a method of analytic continuation in \cite{Witten:2013pra} from a slightly different perspective.
This approach can be generalized straightforwardly to analytically continue the Barnes multiple zeta function.

A modern way to define the Euler gamma function is using the Weierstrass product form
\begin{align}
	\frac{1}{\Gamma(z)}
	&= z e^{\gamma z}
	\prod_{n=1}^{\infty}
	\left(1+\frac{z}{n}\right)e^{-z/n}
\end{align}
where
\begin{align}
	\gamma = \lim_{k\rightarrow\infty} 1+\frac{1}{2}+\cdots+\frac{1}{k}-\ln k.
\end{align}
This is analytic for all $z<\infty$.
Classical ways are the Euler infinite product
\begin{align}
	\Gamma(z)
	= \lim_{n\rightarrow\infty} \frac{1\cdot 2\cdot 3\cdots n}{z(z+1)(z+2)\cdots(z+n)} n^{z}, \quad
	z\neq0,-1,-2,\dots
\end{align}
and the Euler integral
\begin{align}
	\Gamma(z)
	= \int_0^{\infty} \dd{t}
	t^{z-1} e^{-t}, \quad
	\Re z > 0.
\end{align}

A standard way to analytically continue the Euler integral is replacing the integration contour to the Hankel contour
\begin{align}
	C: \infty e^{2\pi i\cdot0}
	\rightarrow \epsilon e^{2\pi i\cdot0}
	\rightarrow \epsilon e^{2\pi i}
	\rightarrow \infty e^{2\pi i}.
\end{align}
Then we have
\begin{align}
	\Gamma(z)
	&= \frac{1}{ e^{2\pi iz}-1 }
	\int_{C} \dd{t} t^{z-1} e^{-t}
\end{align}
While the original Euler integral converges only for $\Re z>0$,
the new integral expression converges except $z\not\in\integernum$.
From the identity theorem, the new integral gives the analytic continuation of the original definition.

\begin{figure}[t]
	\centering
	\includegraphics[height=60mm]{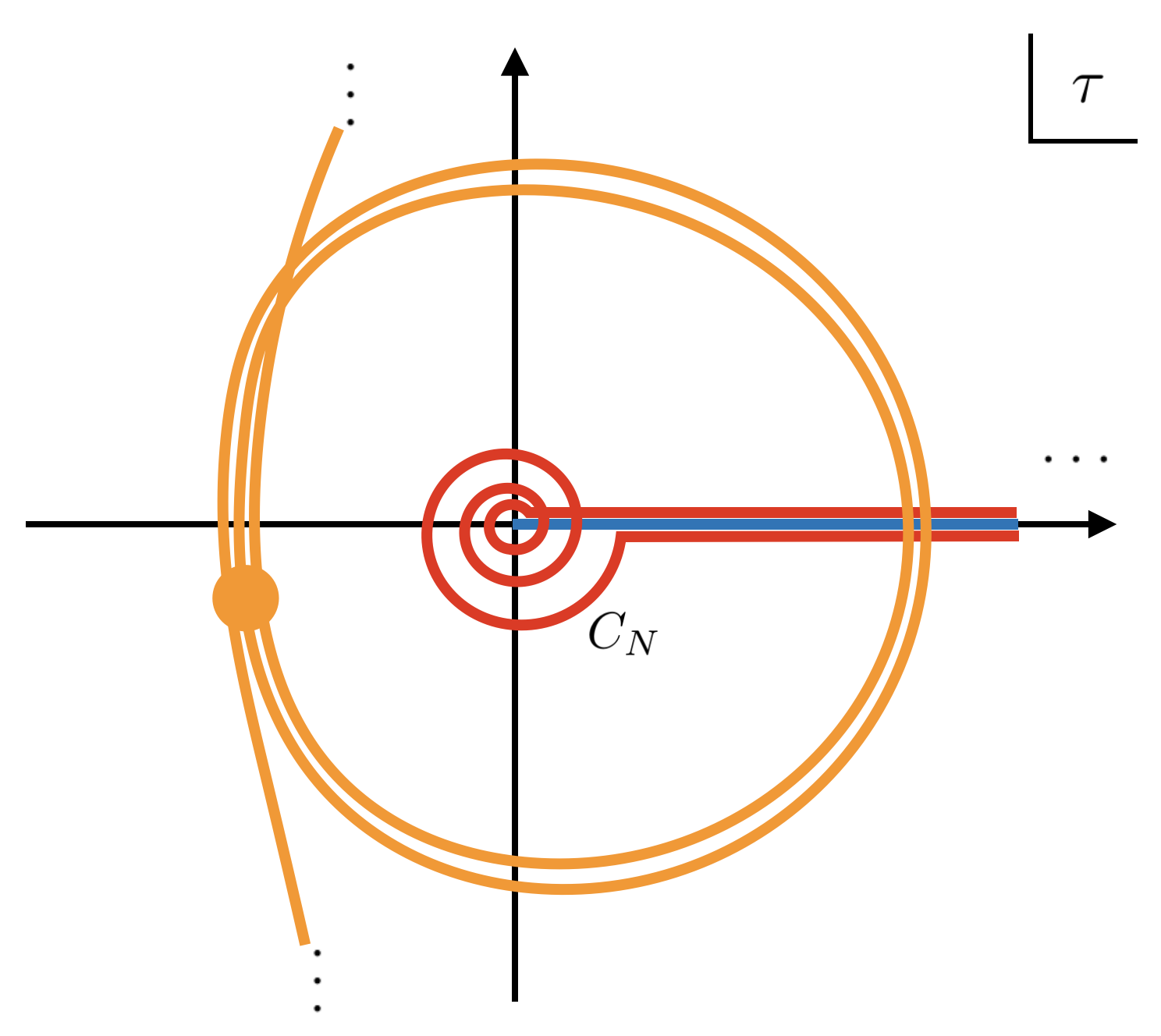}
	\caption{
		Integration contours for analytically continuing the Euler gamma function.
		The blue line is the original integration contour,
		which may be formal contour depending on the sign of $\Re z$.
		The red spiral contour is the redefined integration contour.
		When $\Re z>0,\: \Im z<0$, it reduces to the original blue contour
		while when $\Re z<0,\: \Im z<0$, it is deformed to the orange contour.
		Thimble analysis tells us that in the latter case,
		all of the orange saddles on each Riemann sheet contributes to the integral.
	}
	\label{fig:gamma_contour}
\end{figure}

Motivated by the Hankel contour,
let us use a spiral integration contour which encircles around the origin $N$ times
\begin{align}
	C_N : \;
	\infty e^{-2\pi iN} \rightarrow \epsilon e^{-2\pi iN}
	\rightarrow \epsilon e^{-2\pi i0} \rightarrow \infty e^{-2\pi i0}
\end{align}
as illustrated in Fig.~\ref{fig:gamma_contour} as a red contour.
When $\Re z>0$, using the Euler integral expression, we have
\begin{align}
	\int_{C_N} \dd{t} t^{z-1} e^{-t}
	= (1-e^{-2\pi iN(z-1)}) \Gamma(z).
\end{align}
The left hand side converges for all $z\not\in\integernum$ as long as $C_N$ is deformed to avoid the singularity at the origin.
From the identity theorem, now we have obtained analytic continuation of the Euler integral
\begin{align}
	\Gamma(z)
	&= \frac{1}{1-e^{-2\pi iN(z-1)}}
	\int_{C_N} \dd{t} t^{z-1} e^{-t}, \quad
	z\not\in \mathbb{Z}
\end{align}

Particularly when\footnote{
	The condition $\Im z <0$ matches the Feynman-$i\varepsilon$ prescription.
} $\Re z>0, \: \Im z <0$,
the integral is decomposed into
\begin{align}
	\Gamma(z)
	&= \frac{1}{1-e^{-2\pi iN(z-1)}} \left[
	\int_{\infty e^{-2\pi iN}}^{\epsilon e^{-2\pi iN}}
	+\int_{\epsilon e^{-2\pi iN}}^{\epsilon e^{-2\pi i0}}
	+\int_{\epsilon e^{-2\pi iN}}^{\infty e^{-2\pi i0}}
	\right].
\end{align}
The first two terms satisfy
\begin{align}
	\abs{ \int_{\infty e^{-2\pi iN}}^{\epsilon e^{-2\pi iN}} }
	&\leq e^{-2\pi N(-\Im z)} \int_{\infty}^{\epsilon} \dd{r} r^{\Re z-1} e^{-r} \notag\\
	\abs{ \int_{\epsilon e^{-2\pi iN}}^{\epsilon e^{-2\pi i0}} }
	&\leq \epsilon^{\Re z}e^{+\epsilon} \int_{-2\pi N}^{0} \dd{\theta} e^{-(-\theta)(-\Im z)}
\end{align}
Thus in the limit $N\rightarrow\infty, \epsilon^{\Re z} N \rightarrow 0$,
it reduces to the original integral
\begin{align}
	\Gamma(z)
	\rightarrow
	\int_0^{\infty} \dd{r} r^{z-1} e^{-r},
\end{align}
as illustrated in Fig.~\ref{fig:gamma_contour} as a blue contour.
In this case, the integral can be evaluated at the unique saddle near the integration contour.
It gives the Stirling formula
\begin{align}
	\Gamma(z)
	\sim e^{ (z-1/2)\ln z -z + \ln\sqrt{2\pi} }.
\end{align}

On the other hand when $\Re z <0, \Im z <0$, the integral is decomposed into
\begin{align}
	\Gamma(z)
	&=  \frac{1}{1-e^{-2\pi iN(z-1)}} \left[
	\int_{\downarrow}
	+ \sum_{n=-N}^{0} \int_{\bigcirc\; \text{on $n$-th sheet}}	+ \int_{\downarrow}
	\right]
\end{align}
as illustrated in Fig.~\ref{fig:gamma_contour} as an orange contour.
In this case, the contour runs on $N+1$ Riemann sheets.
On each Riemann sheet, we find a complex saddle.
Deforming the new contour to the thimbles associated with these complex saddles,
we find
\begin{align}
	\Gamma(z)
	&\rightarrow
	\sum_{n=-\infty}^{0}
	\int_{\mathcal{J}_{n}} \dd{t} t^{z-1}e^{-t} \notag\\
	&\sim \left( \sum_{n=-\infty}^{0} e^{2\pi in(z-1)} \right)
	e^{ (z-1/2)\ln z-z+\ln\sqrt{2\pi} } \notag\\
	&\sim
	\frac{\pi}{\sin\pi z} \frac{1}{\Gamma(1-z)}.
\end{align}
This result is consistent with another way to analytically continue the gamma function by the reflection formula
\begin{align}
	\Gamma(z) \Gamma(1-z)
	= \frac{\pi}{\sin \pi z}.
\end{align}
Thus the new spiral integration contour gives appropriate analytical continuation of the gamma function.
Deforming the redefined contour, we can find appropriate poles (and zeros) of the original function.
%\begin{align}
%	\tau \quad
%	C_N
%\end{align}

%================================================================================
\newpage
%========================
% References
%========================

\bibliographystyle{utphys}
\bibliography{ref.bib}

\end{document}